\RequirePackage{fix-cm}
\documentclass[smallextended]{svjour3}       % onecolumn (second format)
\smartqed  % flush right qed marks, e.g. at end of proof
\usepackage{graphicx}
\usepackage{natbib,amsmath,amssymb}
\usepackage{mathabx}
\usepackage{hyperref}
\begin{document}

\title{Low energy cosmic rays%\thanks{Grants or other notes
%about the article that should go on the front page should be
%placed here. General acknowledgments should be placed at the end of the article.}
}
\subtitle{Regulators of the dense interstellar medium}

%\titlerunning{Short form of title}        % if too long for running head

\author{Stefano Gabici         %\and
        %Second Author %etc.
}

%\authorrunning{Short form of author list} % if too long for running head

\institute{S. Gabici \at
	     Universit\'e de Paris, CNRS, Astroparticule et Cosmologie,  F-75006 Paris, France \\
              %Tel.: +123-45-678910\\
              %Fax: +123-45-678910\\
              \email{gabici@apc.in2p3.fr}           %  \\
%             \emph{Present address:} of F. Author  %  if needed
        %   \and
          % S. Author \at
            %  second address
}

\date{Received: date / Accepted: date}
% The correct dates will be entered by the editor

\maketitle

\begin{abstract}
Low energy cosmic rays (up to the GeV energy domain) play a crucial role in the physics and chemistry of the densest phase of the interstellar medium.
Unlike interstellar ionising radiation, they can penetrate large column densities of gas, and reach molecular cloud cores.
By maintaining there a small but not negligible gas ionisation fraction, they dictate the coupling between the plasma and the magnetic field, which in turn affects the dynamical evolution of clouds and impacts on the process of star and planet formation.
The cosmic-ray ionisation  of molecular hydrogen in interstellar clouds also drives the rich interstellar chemistry revealed by observations of spectral lines in a broad region of the electromagnetic spectrum, spanning from the submillimetre to the visual band.
Some recent developments in various branches of astrophysics provide us with an unprecedented view on low energy cosmic rays.
Accurate measurements and constraints on the intensity of such particles are now available both for the very local interstellar medium and for distant interstellar clouds. 
The interpretation of these recent data is currently debated, and the emerging picture calls for a reassessment of the scenario invoked to describe the origin and/or the transport of low energy cosmic rays in the Galaxy.

% \PACS{PACS code1 \and PACS code2 \and more}
% \subclass{MSC code1 \and MSC code2 \and more}
\end{abstract}

\setcounter{tocdepth}{2}

\tableofcontents

\section{Introduction}
\label{sec:intro}

The formation of stars is a central question in astrophysics \citep{shu1987,maclow2004,mckee2007,krumholz2014}.
While it is certain that star formation takes place inside interstellar molecular clouds (MCs) as the result of the gravitational collapse of their dense cores, the details of such process remain quite uncertain.

MCs are cold, dense, magnetised, and turbulent \citep{crutcher2012,hennebelle2012,heyer2015}. While the first two features in the list tend to favour the formation of stars, as they imply low thermal pressure support and short gravitational free-fall time, respectively, the latter two oppose to it, as they provide a non-thermal pressure support against gravity. 
The relevance of the magnetic pressure support and the impact of magnetohydrodynamical turbulence on the dynamical evolution of clouds depend on how tightly the gas and the magnetic field are coupled.

Ultimately, the level of coupling depends on the gas ionisation fraction: a neutral gas would not be affected at all by the presence of a magnetic field and, conversely, a magnetic field would be frozen into a fully ionised gas.
The ionisation fraction found in dense MCs is at the level of $\approx 10^{-7}$, which is quite small, but nevertheless much larger than what one would expect in a very cold ($\sim 10$ K) gas which, due to its large cloud column density, is protected by external sources of ionising radiation \citep{mckee1989}.
It follows that an additional source of ionisation must be present inside MCs.

A similar line of reasoning can be pushed forward also in connection with the formation of planetary systems, which is intimately connected to star formation. 
This is because the dynamics and evolution of protoplanetary disks depend, again, on the gas ionization level, which dictates the effectiveness of mechanisms of magnetic field transport such as magneto-rotational instability or magneto-centrifugally launched winds \citep{wardle2007,armitage2011}.

Finally, a surprisingly complex chemistry has been revealed by spectroscopic observations of interstellar clouds from the submillimetre to the visible band.
Such chemistry is made possible by the large gas column density of clouds, which protects molecules from interstellar radiation and allow them to survive and proliferate.
However, the rate of neutral-neutral chemical reactions is way too slow in the diluted and cold interstellar medium (ISM), implying that some level of ionization of the gas is mandatory in order to allow faster ion-neutral reactions to build up the molecules we observe \citep{bergin2007,caselli2012,tielens2013}. Likewise, the chemical richness of planetary systems is inherited from the processes taking place in the parent MC, and is further affected by the ionization state of protoplanetary disks \citep{vandishoeck2020}.

What said so far clearly indicates that a central question in astrophysics is: \textit{what keeps the densest phase of the ISM slightly ionised?}

\begin{figure*}
% Use the relevant command to insert your figure file.
% For example, with the graphicx package use
\center
  \includegraphics[width=0.7\textwidth]{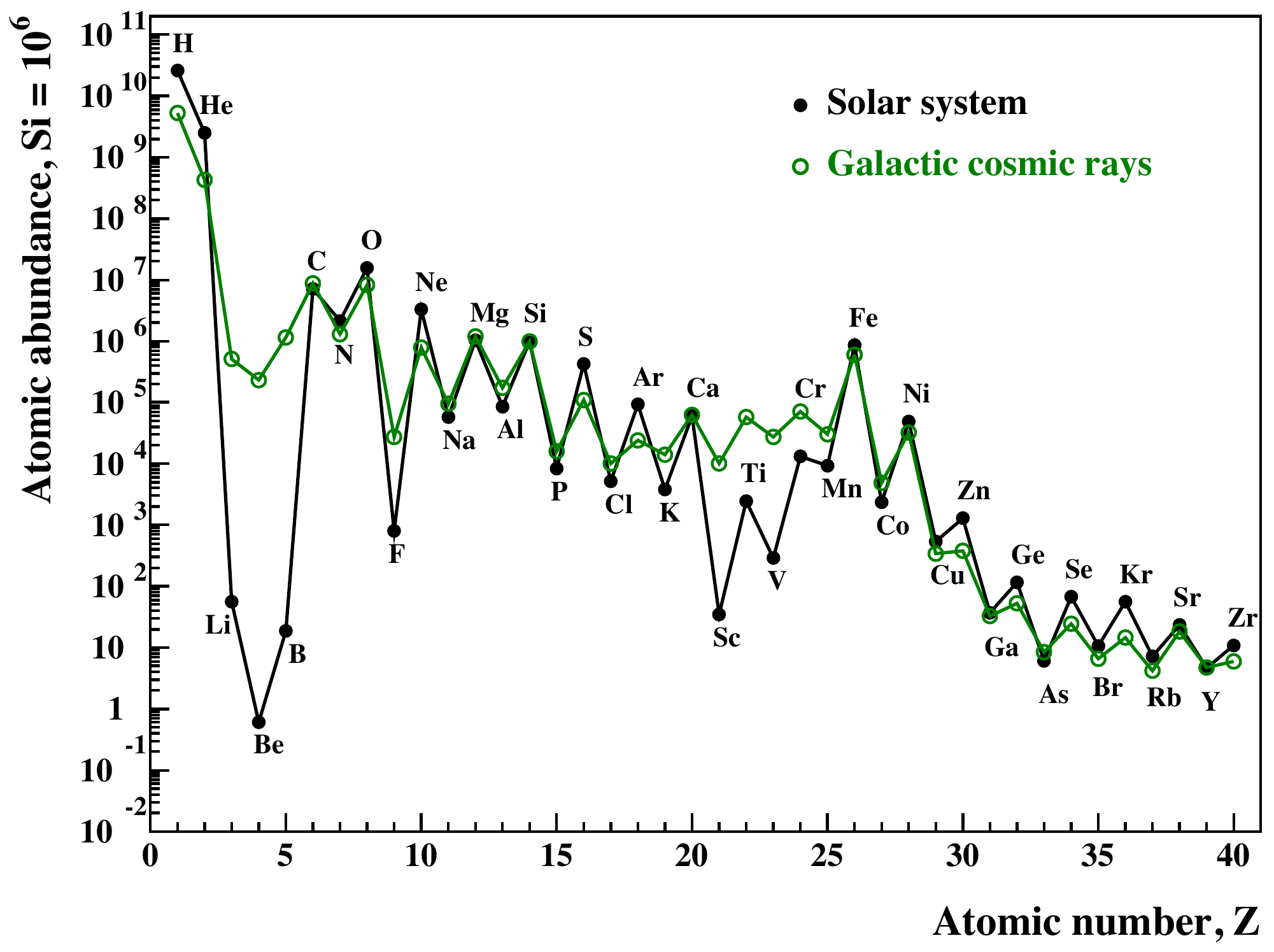}
%  \includegraphics[width=0.80\textwidth]{boeziop.jpg}
 % \includegraphics[width=0.80\textwidth]{boeziohe.jpg}
  %\includegraphics[width=0.80\textwidth]{boezioBCO.jpg}
% figure caption is below the figure
\caption{Cosmic abundances of elements in the solar system (black) and in cosmic rays (green). Figure from \citet{tatischeff2018}.}
\label{fig:vincent}       % Give a unique label
\end{figure*}

Once stars are formed from the gravitational collapse of MC cores, nuclear fusion reactions begin to operate in their hot interiors, and to create elements heavier than hydrogen in a process called stellar nucleosynthesis \citep{burbidge1957,cameron1957,trimble1991,wallerstein1997,woosley2002}. 
The very early Universe emerging from Big Bang nucleosynthesis was only made of hydrogen, some helium, and a very small fraction of lithium \citep{coc2017}. 
It is thanks to stellar nucleosynthesis that the great variety of elements that we observe today was generated. % later, in the nuclear reactions taking place in stellar interiors, during stellar explosions, and, as recently confirmed observationally, in the explosions following double neutron star mergers \citep{pian2017}.
Notwithstanding the great success of both Big Bang and stellar nucleosynthesis theories in predicting the observed abundances of elements, it became soon clear that a substantial fraction of the lithium and the totality of beryllium and boron found in the present day Universe must have been produced in another way.
A peculiar origin of lithium, beryllium and boron (LiBeB) was suggested by the fact that their cosmic (solar system) abundances are much smaller (many orders of magnitude) than those of their neighbours in the periodic table (see Fig.~\ref{fig:vincent}). %\citep{lodders2009}.
\citealt{burbidge1957} pointed out that LiBeB are fragile, and are destroyed very effectively in the thermonuclear reactions taking place in the hot and dense stellar interiors.
Their origin, then, had to be searched elsewhere.
In the absence of a viable mechanism, they invoked an unspecified {\it x-process}, operating in an environment where both temperature and density must be low, as the responsible for the synthesis of light elements.

A second fundamental questions then emerges: \textit{what produces the light elements (LiBeB) observed in the Universe?}

Remarkably, both the crucial questions asked above have the very same answer, that is, {\it the interactions of low energy cosmic rays (LECRs) with interstellar matter}.

Cosmic rays (CRs) are mostly made of energetic atomic nuclei (mainly protons). 
They fill the entire Galaxy and reach the Earth as an isotropic flux of particles.
Unlike ultraviolet radiation, the intensity of CRs is not completely attenuated by large column densities of matter.
Therefore, CRs are %believed to be 
the only ionising agents able to penetrate the depths of MCs and maintain there the gas ionisation fraction at the observed level of $\approx 10^{-7}$.
Moreover, secondary electrons produced in ionisation events cool by transferring their energy to the ambient gas, providing the source of heating needed to explain the gas temperatures measured in clouds.
Not all CRs contribute significantly to the ionisation and heating of the gas, but only low energy ($\lesssim 1$ GeV) ones \citep{hayakawa1961,spitzer1968,field1969}.

The link between LECRs and the synthesis of light elements became evident after the intensity of CR nuclei heavier than hydrogen was measured, revealing a striking difference with respect to cosmic abundances.
The ratio between the abundance of CNO and LiBeB nuclei is very large ($\approx 10^6$) in the solar system but only of the order of a few for CR particles (see Fig.~\ref{fig:vincent}). %in \citealt{tatischeff2018}).
This puzzle was solved in the early seventies, when it was understood that the synthesis of LiBeB mainly results from spallation of CNO nuclei of interstellar gas by LECRs. 
Spallation occurs when a nucleus in the interstellar gas (for example carbon), struck by a CR particle (for example a proton), ejects a number of lighter particles and transforms into a LiBeB isotope.
To this {\it direct} channel one should also add the contribution from the {\it reverse} process, i.e. spallation of CNO LECRs by interstellar nuclei, and from $\alpha + \alpha$ reactions, where two helium nuclei collide to synthesize lithium.
These three main reactions involving LECRs and interstellar matter account for most of the LiBeB isotopes produced by the x-process, which is now called spallogenic nucleosynthesis \citep{reeves1970,meneguzzi1971}. 

%Beside ionizing and heating dense MCs and synthesizing light elements, 
LECRs also impact on the diffuse interstellar gas on large scales.
This follows from the fact that the interstellar energy densities of CRs, magnetic field and thermal and turbulent gas are in rough equipartition, with LECRs providing a sizeable contribution to the total energy density of CRs.
Therefore, CRs may affect the dynamics and the characteristics of diffuse interstellar matter in various ways, such as providing the pressure support needed to launch galactic winds \citep{recchia2020} or exciting magnetoydrodynamical turbulence as they stream across magnetized plasmas \citep{wentzel1974,zweibel2017}.

Finally, understanding where and how LECRs are accelerated, how they are transported from their sources to the Earth, and what is their final fate is important {\it per se}: revealing the origin of CRs is one of the main open questions in high energy astrophysics.
Supernova remnant shocks, that form in the ISM as the result of supernova explosions, are often invoked as the sites where the bulk of galactic CRs are accelerated \citep{drury2012,blasi2013}.
However, two things should be kept in mind.
First of all, the supernova remnant origin remains to date an hypothesis \citep[see][for a critical review]{gabici2019}, and alternative acceleration sites have been proposed, including stellar wind termination shocks, stellar clusters, or superbubbles, i.e. the cavities inflated in the ISM by the collective effect of stellar winds and recurrent supernova explosions in star clusters \citep{cesarsky1983,bykov2014,lingenfelter2018,aharonian2019}. 
Second, at present it is not quite clear whether LECRs, which are characterized by sub-GeV particle energies, have the same origin as higher energy ones \citep[see for example the discussion on protostellar shocks as LECR accelerators in][]{padovani2020}, nor if the intensity of LECRs measured within the solar system is representative of the entire Galaxy or simply reflects some very local ambient conditions (for example the presence or absence of nearby CR sources, see \citealt{phan2021}, or the fact that we live inside an interstellar cavity called the {\it local bubble}, see e.g. \citealt{silsbee2019}).
%\citep[for example the fact that we live inside an interstellar cavity called the {\it local bubble}, see e.g.][]{silsbee2019}.
All in all, the origin of galactic CRs is still not well understood, and this is particularly true for LECRs.
 
Addressing all the issues raised above goes beyond the scope of this review, which will be focussed on the impact that LECRs have on the densest phase of the ISM.
This choice is motivated by a number of recent developments in space exploration (the Voyager probes crossing the heliopause, \citealt{stone2013,stone2019}), laboratory astrophysics (the measurement of the dissociative recombination rate of H$_3^+$, a pivotal molecule in interstellar cloud chemistry, \citealt{mccall2003}) and CR astrophysics (the advent of a precision era in high energy CR measurements from space, \citealt{boezio2020,aguilar2021}).
These activities triggered a renewed interest in both observational and theoretical studies of interstellar clouds irradiated by LECRs.
An attempt to review this field seems therefore timely.

Before proceeding, I provide here a list, certainly incomplete, of review articles covering aspects of the physics of LECRs which are not treated here.
The reader interested in the spallogenic nucleosythesis of light elements is referred to the reviews by \citet{reeves1994}, \citet{vangioni2000}, and \citet{tatischeff2018}.
The origin of the isotopic composition of CRs is discussed in \citet{ptuskin1998} and \citet{wiedenbeck2007}.
The extended review by \citet{padovani2020} treats, among others, topics which are not covered by the present review, including the impact of LECRs in circumstellar disks, stellar cosmic rays, LECR acceleration in protostellar jets, and extragalactic studies of LECRs.
Finally, a discussion on the effect that CRs might have on the origin of life can be found in \citet{dartnell2011}.

The remaining of the paper is structured as follows.
After an introduction on CRs (Sec.~\ref{sec:CRs}), we will review the difficulties encountered by direct (Sec.~\ref{sec:direct}) and indirect (Sec.~\ref{sec:indirect} and \ref{sec:integral}) attempts to measure the intensity and spectrum of LECRs. In these two Sections, we will also highlight the recent observational breakthroughs that radically changed or improved our knowledge of LECRs.
Sec.~\ref{sec:transport} will be devoted to a discussion on the transport of LECRs in and around MCs. 
To this purpose, we will consider both isolated MCs and clouds located in the proximity of powerful CR sources.
A list of open issues in LECR astrophysics will be provided in Sec.~\ref{sec:questions}, and we will conclude in Sec.~\ref{sec:conclusions}, where future perspectives will be also outlined.

%{\bf The multiphase interstellar medium (ISM) consists of a mixture of dense and cold clouds, either atomic or molecular \citep{snow2006,heyer2015}, and a diffuse gas, either warm or hot \citep{mckee1995,ferriere2001,cox2005}.
%MCs (hereafter MCs), though occupying a very small fraction of the volume of the Galactic disk, have been the subject of extensive observational and theoretical investigations for at least two reasons.}

\section{What are cosmic rays?}
\label{sec:CRs}

CRs are energetic charged particles that reach the Earth's atmosphere from outer space.
The distribution in the sky of the arrival direction of CRs is remarkably isotropic.
The vast majority of CR particles are atomic nuclei, with a contribution from electrons at the percent level, and an even smaller one from antimatter (positrons and antiprotons).
Amongst CR nuclei, protons largely dominate, with helium and heavier nuclei (metals) amounting to $\sim$10\% and $\sim$1\% of the total number of particles, respectively \citep[for reviews see][]{gaisser2016,gabici2019,boezio2020}.

With the exception of anomalous CRs\footnote{Anomalous CRs are neutral atoms in the local ISM that, due to the motion of the Sun, enter the heliosphere, become partially (mainly singly) ionised due to charge-exchange, solar radiation, or electron impact, and are eventually accelerated in the solar wind up to energies of $\lesssim 100$ MeV/nucleon \citep{reames1999,potgieter2013}.}, they are originated outside of the heliosphere. 
%As I will discuss in Sec.~\ref{sec:gamma}, constraints coming from gamma-ray observations showed that 
The vast majority of them are in fact originated within our Galaxy, and only the highest energy particles, that won't be discussed here, likely have an extragalactic origin as they can hardly be confined by the Galactic magnetic field.
Understanding the origin of the bulk of CRs is one of the most fundamental and long standing questions in high energy astrophysics.
The CR elemental composition and energy spectra are now measured with great accuracy and carry crucial information about where CRs are accelerated and how they traveled from their sources to us. 

\begin{figure*}
% Use the relevant command to insert your figure file.
% For example, with the graphicx package use
\center
  \includegraphics[width=0.7\textwidth]{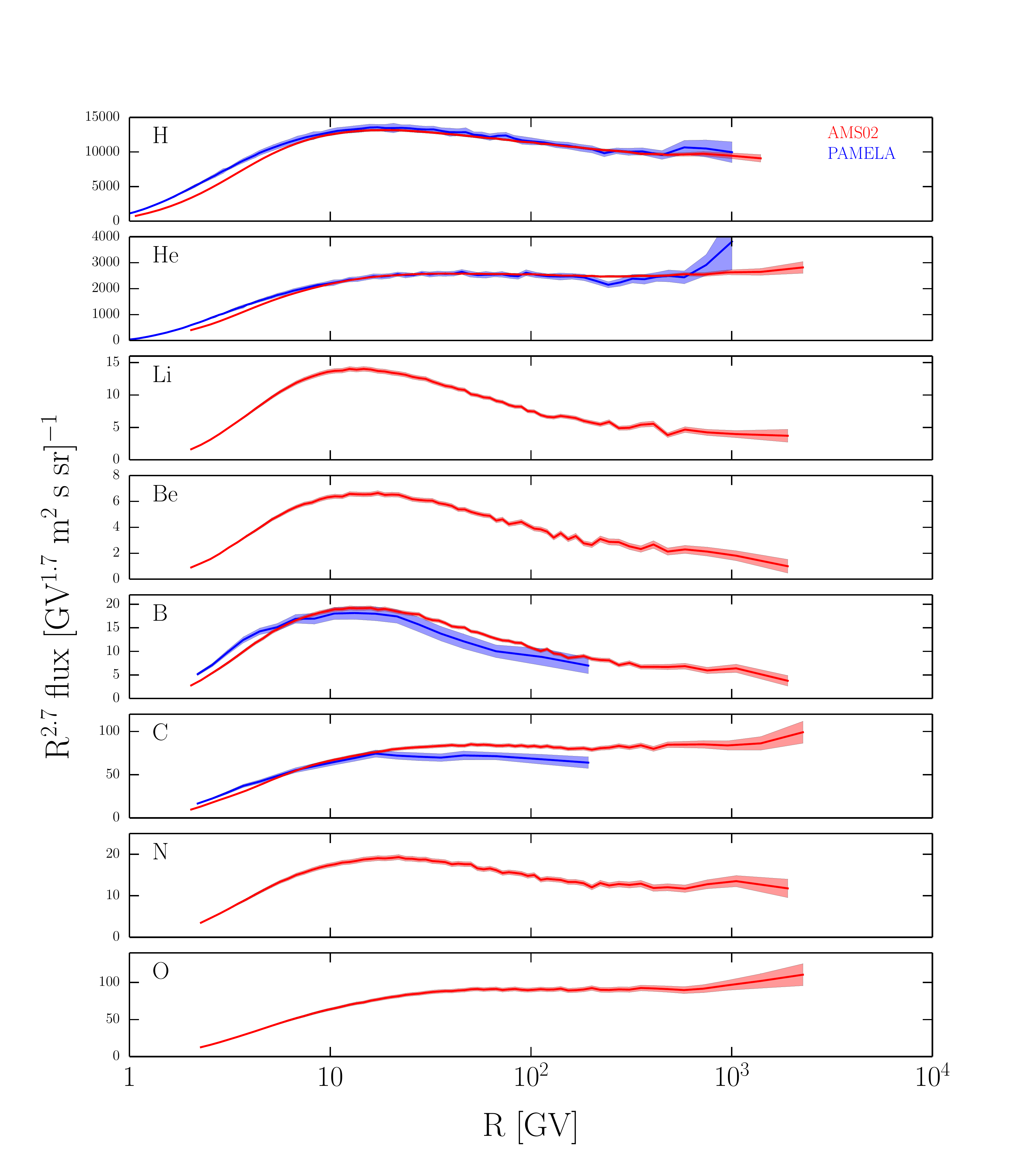}
%  \includegraphics[width=0.80\textwidth]{boeziop.jpg}
 % \includegraphics[width=0.80\textwidth]{boeziohe.jpg}
  %\includegraphics[width=0.80\textwidth]{boezioBCO.jpg}
% figure caption is below the figure
\caption{Spectra of CR nuclei of atomic number from $Z = 1$ to 8 (top to bottom) as a function of rigidity. Data from AMS-02 (red) and PAMELA (blue).  Figure from \citet{gabici2019}.}
%protons (top panel), He (middle panel) and B, C, and O nuclei (bottom panel) measured by various space instruments, as indicated in the Figure labels, or by AMS-02 if not otherwise indicated. Spectra are plotted as a function of the kinetic energy per nucleon (top and middle panelsa) or of the particle rigidity (bottom panel). All spectra have been multiplied by $E^{-2.7}$. Figures from \citet{boezio2020}, where references to the datasets can also be found.}
\label{fig:carmelo}       % Give a unique label
\end{figure*}

Fig.~\ref{fig:carmelo} shows the local spectra of CR nuclei of atomic number from $Z = 1$ to 8 (top to bottom).
Data have been collected by detectors operating at the top of the Earth's atmosphere: the satellite borne PAMELA (blue) and the Alpha Magnetic Spectrometer (AMS-02, red shaded regions) mounted on the International Space Station \citep{adriani2014,aguilar2002}.
For a better representation, spectra have been multiplied by the particle rigidity\footnote{The rigidity of a fully ionised nucleus of momentum $p$ is $R = p c/Z e$, where $e$ is the elementary charge. If $p c$ is expressed in eV and the particle charge in natural units ($e = 1$) the rigidity has units of Volts. Particles of equal rigidity have the same gyration radius around a magnetic field of strength $B$, $r_g = R/B$, and therefore follow the same trajectory.} $R$ to the power 2.7.
The suppression observed in all spectra at rigidities smaller than a few tens of GV is due to the effect of the solar wind on low energy particles, and will be discussed in detail in Sec.~\ref{sec:local}.

Two things should be noted.
First of all, for particle rigidities above $\sim$ 10 GV, the spectra of lithium, berillium, and boron are markedly steeper than those of the other elements, which are roughly flat once multiplied by $R^{2.7}$.
Second, while the Solar abundances of LiBeB are many orders of magnitude smaller than that of carbon, such a difference is way smaller for CRs (see Fig.~\ref{fig:vincent}).
As seen in the Introduction, this peculiarity in the abundances of light elements fits well within a scenario where such elements are produced as the result of the interactions of LECRs in the ISM.

\subsection{The confinement of cosmic rays in the Galaxy: boron-over-carbon ratio}
\label{sec:B/C}

To understand more quantitatively the peculiarities in the intensity and spectra of light elements, consider the production of CR boron due to spallation of heavier CR nuclei by interstellar matter.
In such process, an energetic nucleus (most likely of carbon or oxygen, which are the most abundant elements heavier than boron, see Fig.~\ref{fig:vincent}) hits a nucleus of interstellar gas (most likely an hydrogen nucleus), loses one or few nucleons in the impact and transforms into boron.
%This is indeed the dominant channel for the production of CR boron.
Let us call $n_C(E)$ and $n_O(E)$ the average number density of CR carbon and oxygen nuclei in the Galaxy having an energy per nucleon $E$, and $\sigma_{C \rightarrow B}$ and $\sigma_{O \rightarrow B}$ the appropriate spallation cross sections for boron production.
In first approximation, the energy per nucleon is conserved in spallation reactions, and at large enough energies the spallation cross sections are roughly energy independent \citep[see e.g. Fig.~4 in][]{tatischeff2018}.
Therefore, for a fixed value of $E$, the production rate of CR boron in the Galaxy is $q_B \sim n_H^{ISM} v ( \sigma_{C \rightarrow B} n_C + \sigma_{O \rightarrow B} n_O )$, where $n_H^{ISM}$ is the hydrogen density of the interstellar gas and $v$ is the velocity of the incident nucleus\footnote{Nuclei characterised by the same energy per nucleon move at the same speed.}.
Noting from Fig.~\ref{fig:carmelo} that the measured abundances and spectral slopes of CR C and O are almost identical, $n_C(E) \approx n_O(E)$, the expression for the production rate of CR boron can be simplified to $q_B \approx n_H^{ISM} v n_C ( \sigma_{C \rightarrow B} + \sigma_{O \rightarrow B})$.
By doing so, we are implicitly assuming that the spectrum of CRs observed locally is representative of the entire Galaxy, i.e. that there are no large spatial variations of the CR intensity in the Galactic disk.

After being produced, CR boron nuclei %diffuse in the turbulent ambient magnetic field 
spend some time $\tau_{ISM}$ wandering in the ISM before escaping the Galaxy, or they spallate on interstellar hydrogen to transform into lighter nuclei in a typical time $\tau_B = (n_H^{ISM} \sigma_B v)^{-1}$.
Here, $\sigma_B$ is the {\it total} spallation cross section for CR boron.
The balance between production and escape/destruction leads to an equilibrium density of CR boron equal to $n_B = q_B \tau_{eff}$, where $\tau_{eff}^{-1} = \tau_{ISM}^{-1} + \tau_B^{-1}$ is an effective timescale.

To proceed further, it is convenient to introduce a quantity called {\it grammage}, defined as the mean amount of matter traversed by CRs in a given time $\tau_i$, as ${\rm X}_i = m_p n_H^{ISM} v \tau_i$, where $m_p$ is the hydrogen (proton) mass, and where the presence of helium and heavier elements in the ISM is neglected.
Then, ${\rm X}_{ISM}(E)$ represents the mean amount of matter traversed by CRs before escaping the Galaxy, while ${\rm X}_B$ corresponds to that traversed by CR boron before spallating to transform in a lighter element. 
Note that while the former may be an energy dependent quantity, the latter is not, as long as particles of large enough energy are considered (as $\sigma_B$ is roughly constant).
Rearranging all the equations derived above one finally gets an expression for the CR {\it Boron-over-Carbon ratio}  (B/C), which is an observable quantity \citep{gaisser2016}:
\begin{equation}
\label{eq:B/C}
\frac{n_B}{n_C} \sim \frac{{\rm X}_{ISM}}{1+\frac{{\rm X}_{ISM}}{{\rm X}_B}} \frac{\sigma_{C \rightarrow B}+\sigma_{O \rightarrow B}}{m_p} ~ .
\end{equation}
Eq.~\ref{eq:B/C} illustrates well how the measurements of the CR B/C ratio provide us with an estimate of the residence time of CRs in the ISM.
In particular, the energy dependence of the B/C ratio is fully determined by that of  ${\rm X}_{ISM} \propto \tau_{ISM}$, as all the other quantities on the right hand side of the equation are roughly energy independent.

% For two-column wide figures use
\begin{figure*}
% Use the relevant command to insert your figure file.
% For example, with the graphicx package use
\center
  \includegraphics[width=0.7\textwidth]{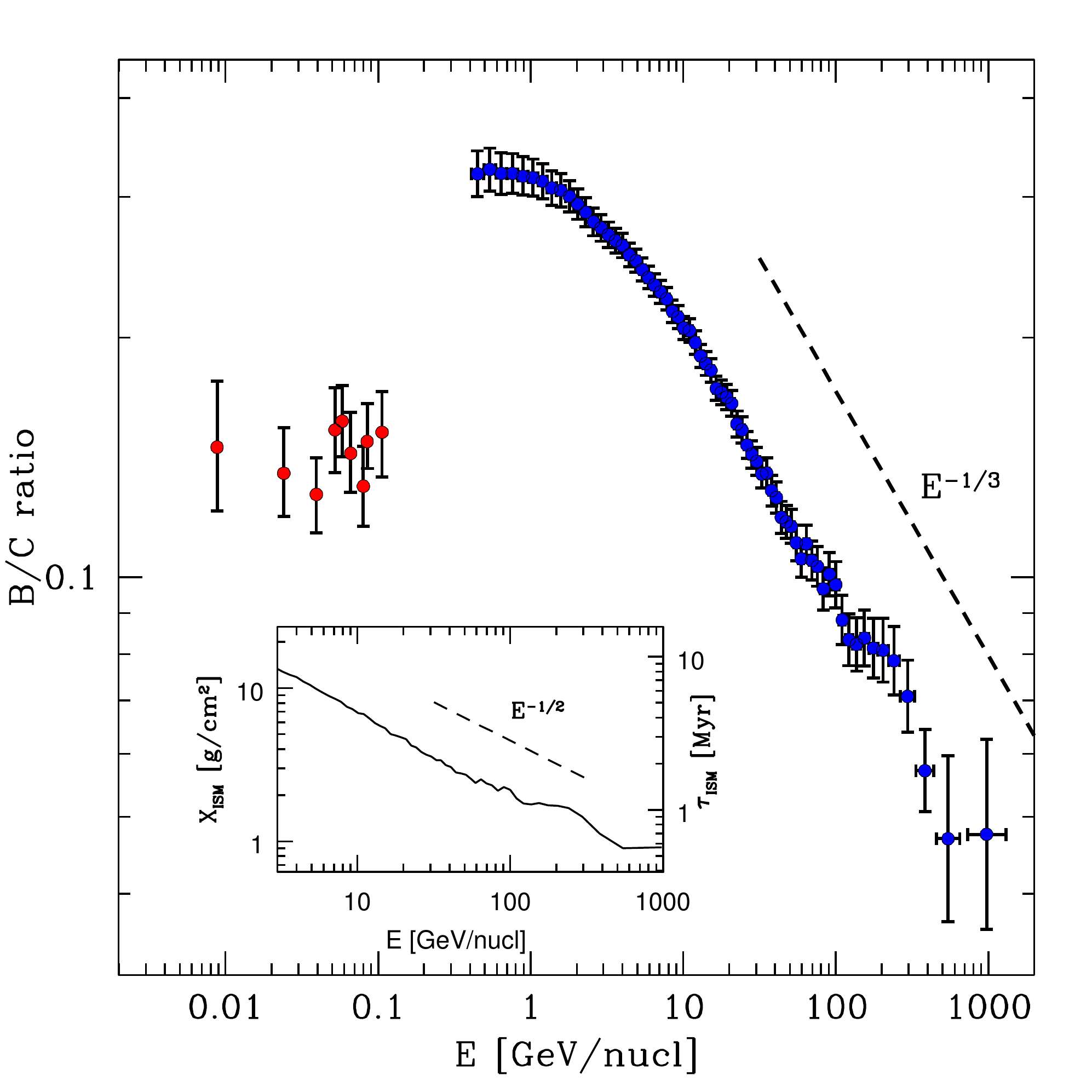}
% figure caption is below the figure
\caption{Boron over carbon (B/C) ratio. Data from AMS-02 (blue points, \citealt{aguilar2016}) and Voyager 1 (red points, \citealt{cummings2016}). The dashed line represents the power law $E^{-1/3}$.
The grammage derived by data and the corresponding CR residence time in the ISM are shown in the inset (solid line) together with the power law $E^{-1/2}$ (dashed line).}
\label{fig:B/C}       % Give a unique label
\end{figure*}

The observed B/C ratio is shown in Fig.~\ref{fig:B/C}, as measured at the top of the Earth's atmosphere by AMS-02 (blue points), and in the local ISM by the Voyager 1 probe (red points).
The two datasets do not seem to connect smoothly, and exhibit quite different behaviours.
The B/C ratio measured by AMS-02, after a plateau, decreases steadily with particle energy, while that measured by Voyager 1 is virtually energy independent.
Here, we will limit the analysis to the higher energy dataset (AMS-02), which can be interpreted in a rather straightforward way, and we avoid a discussion on low energy data, whose interpretation is very uncertain (see e.g. Fig.~1 in \citealt{tatischeff2021} or Fig.~9 in \citealt{cummings2016}).

The first thing that can be inferred from the high energy observation of the B/C ratio is that the residence time of CRs in the ISM is indeed energy dependent, and it decreases with energy.
The absolute value of the grammage ${\rm X}_{ISM}$ can be read directly from Fig.~\ref{fig:B/C}.
This can be done by using Eq.~\ref{eq:B/C} with the measured values $\sigma_{C \rightarrow B} + \sigma_{O \rightarrow B} \approx 100$ mb and ${\rm X}_B \approx 7$ g/cm$^2$ \citep{gaisser2016}, which gives ${\rm X}_{ISM} \approx 13$ and $\approx 7$ g/cm$^2$ for a particle energy of $\sim 3$ and 10 GeV/nucleon, respectively.
Such an estimate is quite close to results from accurate studies (\citealt{jones2001} obtained ${\rm X}_g \sim 12$~g/cm$^2$ at $\sim$5 GeV/nucleon).
The grammages derived here correspond to residence times for CRs in the ISM equal to $\tau_{ISM} \sim$ 8 and 4~Myr for particle energies of 3 and 10 GeV/nucleon, respectively, for a typical density of the ISM $n_H^{ISM} \sim 1$ cm$^{-3}$.

The decrease of the observed B/C ratio with particle energy implies that {\it well above} 10 GeV/nucleon, ${\rm X}_{ISM}$ becomes much smaller than ${\rm X}_B$ and Eq.~\ref{eq:B/C} asymptotically reduces to $n_B/n_C \propto {\rm X}_{ISM}$.
\cite{aguilar2016} found that B/C data can be fitted with a single power law in energy $E^{-\delta}$ with $\delta \sim 1/3$ for particle energies larger than $\sim 30$ GeV/nucleon.
It is sometimes erroneously claimed that such measured value of $\delta$ also describes the energy scaling of the grammage ${\rm X}_{ISM} \propto E^{-\delta}$. 
However, such claim is incorrect because the particle energies probed by AMS-02 are not large enough to satisfy the condition ${\rm X}_{ISM} \gg {\rm X}_B$.
This is shown in the inset of Fig.~\ref{fig:B/C}, where the value of the grammage inferred from data and from Eq.~\ref{eq:B/C} is plotted for particle energies exceeding 3 GeV/nucleon\footnote{At smaller particle energies the effect of Solar modulation becomes significant.}.
The decrease of the grammage with particle energy is in fact faster than $E^{-1/3}$, and rather closer to $\approx E^{-1/2}$. This is in broad agreement with more sophisticated studies (the reader interested in an accurate derivation of the grammage is referred to the recent works by \citealt{evoli2019} and \citealt{genolini2019}).

Now that the grammages traversed by CRs of different energies have been estimated from data, it is worth comparing them with the gas surface density of the Galactic disk ${\rm X}_{disk}$, which is of the order of few milligrams per squared centimeter \citep{ferriere2001}.
The latter is a proxy for the minimum possible average grammage, i.e., that accumulated by a CR produced in the disk and moving away from the production site along a straight line.
Strikingly, the accumulated grammages are about three orders of magnitude larger than the disk surface density.
This remarkable difference can be interpreted in two ways: either CRs remain somehow (diffusively?) confined in the gaseous disk for a time $\tau_{ISM}$ before escaping the Galaxy, or they are confined in a much larger volume for a time $\tau_{esc} \gg \tau_{ISM}$ and during this time they cross the disk $\approx {\rm X}_{ISM}/{\rm X}_{disk} = \mathcal{O}(10^3-10^4)$ times.
The problem is that the measurements of the CR B/C ratio can only constrain the product between the typical confinement time and the {\it average} gas density felt by CRs (i.e. the grammage) and not the two quantities separately.
Therefore, in order to distinguish between the two scenarios, an additional observable is needed.
As the mean density of the Galactic disk is very well measured \citep[e.g.][]{ferriere2001}, what is much needed is an independent measurement of the confinement time.

\subsection{Radioactive isotopes as cosmic ray clocks}

Short lived radioactive isotopes are present in the cosmic radiation, being produced in the spallation of CR nuclei by interstellar matter.
They can be used as {\it cosmic ray clocks}, as their abundance relative to stable isotopes may depend quite strongly on the escape time from the Galaxy $\tau_{esc}$.
This is true for isotopes whose decay time $\tau_{rad}$  does not exceed $\tau_{esc}$, otherwise they would behave exactly as stable isotopes.
Also, $\tau_{rad}$ should be long enough to allow the intensity of radioactive isotopes to be detectable by available instruments.
$^{10}$Be, with a lifetime (at rest) of $\tau_{rad} \approx$ 2 Myr (half-life equal to $\tau_{1/2} \sim 1.4$ Myr) is therefore an ideal cosmic ray clock \citep{ptuskin1998}.
In the observer rest frame, the lifetime of a $^{10}$Be CR isotope has to be multiplied by its Lorentz factor $\gamma$ due to relativistic time dilation.
The measured abundance of $^{10}$Be with respect to its stable companion $^9$Be provides, to date, the most stringent constrain on the escape time of CRs from the Galaxy.

Unfortunately, at least three complications arise when studying $^{10}$Be in CRs.
First of all, measurements of such isotope are available only at particle energies smaller than $\sim 2$ GeV/nucleon \citep[e.g.][and references therein]{nozzoli2021}\footnote{At the International Cosmic Ray Conference that took place (virtually) in Berlin in July 2021, the AMS-02 collaboration presented preliminary results on the measurements of the isotopic ratio $^{10}{\rm Be}/^9{\rm Be}$ up to energies of $\sim$~10 GeV/nucleon.}.
At these energies, ionisation losses in the ISM cannot be neglected, as it was done, implicitly, in writing Eq.~\ref{eq:B/C} (see Sec.~\ref{sec:losses} for a discussion on energy losses).
Second, besides spallation of CR C and O by interstellar matter, 
%also the reaction $^{11}$B + p $\rightarrow ^{10}$Be + X 
other reactions produce a significant amount of $^{10}$Be \citep[e.g.][]{genolini2018}.
Predicting the abundance of such isotope in CRs therefore would require the solution of a network of reactions. %, and is affected by uncertainties in the knowledge of production cross sections \citep[see e.g.][]{simon2003}.
Third, at low particle energies, where measurements are available, the particles' Lorentz factor is $\gamma \lesssim 2$ and the lifetime of $^{10}$Be in the observer frame is a factor of a few shorter than $\tau_{ISM}$, which is in turn smaller (or at most equal, in the limiting case when the confinement volume coincides with the Galactic disk) than $\tau_{esc}$.
Assuming that the confinement of CRs in the Galaxy is diffusive, and that the Galaxy is a flattened system of height $H$, one can estimate the particle diffusion coefficient as $D \sim H^2/\tau_{esc}$.
It follows that the CR isotopes of $^{10}$Be observed at the Earth must have been produced within a distance $l \sim \sqrt{D \tau_{rad}}$, i.e. within a volume significantly smaller than the entire confining volume, as $l \sim H \sqrt{\tau_{rad}/\tau_{esc}} < H$.
To interpret correctly the measurements of $^{10}$Be, then, it is mandatory to know with some accuracy the spatial distribution of matter in the local ISM.
This is of particular relevance as the Solar system is known to be embedded in a large cavity in the ISM, called {\it local bubble} \citep[see][and the discussion in Sec.~\ref{sec:questions}]{streitmatter2001,donato2002}.

Despite all these complications, the way in which the measurements of the abundance ratio $^{10}$Be/$^9$Be in CRs can be used to constrain their confinement time in the Galaxy can be understood in a qualitative way if one considers only the most important difference between the behaviour of the two isotopes $^{9}$Be and $^{10}$Be: the former is stable, the latter is not \citep{hayakawa1958}.
Assume then that the CR isotopes $^{10}$Be and $^9$Be are produced at a rate $q_{10}$ and $q_9$ and removed from the Galaxy in a typical time $\tau_{rad}$ and $\tau_{esc}$, respectively. In the absence of any other effect, at equilibrium one would get $n(^{10}{\rm Be})/n(^9 {\rm Be}) = (q_{10}/q_9) (\gamma \tau_{rad}/\tau_{esc})$.
The isotopic ratio at production $q_{10}/q_9$ is known, as it depends mainly on the production cross sections.
Therefore, the observed abundance ratio $n(^{10}{\rm Be})/n(^9 {\rm Be})$ provides an estimate for the CR escape time from the Galaxy $\tau_{esc}$.

The comparison of model predictions with the observed $n(^{10}Be)/n(^9 Be)$ ratio indicates that the residence time of CRs in the Galaxy $\tau_{esc}$ is significantly longer than the time spent into the ISM $\tau_{ISM}$, and is found to be of the order of $\approx 100$ Myr for particle energies $\lesssim 1$ GeV/nucleon \citep[e.g.][]{ptuskin1998}. 
This implies that the confinement volume of CRs is larger than the Galactic disk (where grammage is accumulated), and that CRs spend most of the time outside of the disk, in a larger region called Galactic halo, before leaving the Galaxy.
The existence of an extended Galactic halo is also required by radio observation of the Galaxy at high latitudes \citep{beuermann1985}. 

\subsection{Diffusive models for the transport of cosmic rays} %, and the derivation of the source injection spectrum}
\label{sec:diffusive}

The transport of CRs in a disk-plus-halo system is often modelled in a simplified way by taking the Galaxy as a box inside which CRs move at a speed $v$ and undergo a random walk in space with mean free path $\lambda$. 
In this scenario CRs diffuse with a diffusion coefficient  $D \sim \lambda v$, and occasionally cross an infinitely thin disk where interstellar matter is concentrated \citep[e.g.][]{jones2001}.
If $H$ is the height of the halo, it can be shown that a CR particle of velocity $v$ will cross the disk $N_{cross} \sim H v/D \sim H/\lambda \gg 1$ times before leaving the Galaxy.
The grammage determined from the B/C ratio is then $\sim N_{cross} {\rm X}_{disk}$ and therefore constrains the ratio $D/H$ only, and not $D$ and $H$ separately. 
On the other hand, the measurement of the abundance ratio $^{10}$Be/$^9$Be is sensitive to the escape time only, which is $\tau_{esc} \sim H^2/D$.
Then, the simultaneous knowledge of these two abundance ratios (B/C and $^{10}$Be/$^9$Be) allows us to estimate both the diffusion coefficient ($D \approx 10^{28}$ cm$^2$/s at $\approx 1$ GeV/nucleon) and the size of the Galactic halo (several kiloparsecs).
For an accurate derivation of these two quantities see e.g. \citet{strong2007} and references therein.

Finally, the energy dependence of the diffusion coefficient can also be constrained within the framework of the simple diffusive scenario outlined above. 
Indeed, for relativistic particle energies ($v \sim c$) the scaling simplifies to $D \, \propto \, 1/{\rm X}_{ISM} \propto 1/\tau_{esc} \propto E^{\delta}$ with $\delta \approx 0.5$, as determined from Fig.~\ref{fig:B/C}.
Accurate studies provide a range of values $\delta \sim 0.3 ... 0.6$ compatible with observations \citep{strong2007}. 
In fact, it would be more appropriate to think in terms of particle rigidity $R$ rather energy per nucleon $E$ (see footnote 2), but for relativistic particles one simply has $R \approx (A/Z) E \approx 2 E$, and $D \propto R^{\delta}$.
This shows that particles of larger rigidity diffuse faster and leave the Galaxy earlier than particles of lower rigidity, and this fact can be used to constrain the spectrum that CR sources, whatever their nature, must inject in the ISM.

To do so, let us assume that CR sources located in the Galactic disk inject in the ISM {\it primary} nuclei (H, He, C, etc.) characterised by an energy spectrum $q_{p,inj} \propto R^{-\alpha}$.
The energy dependent escape of CRs from the Galaxy will steepen the spectrum towards an equilibrium particle distribution function $n_p \propto ~q_{p,inj} \tau_{esc} \propto ~q_{p,inj}/D \propto R^{-(\alpha+\delta)} \equiv R^{-s}$, where $s \approx 2.7$ is a measured quantity (Fig.~\ref{fig:carmelo}).
{\it Secondary} CR nuclei (e.g. LiBeB) are produced by spallation of heavier primaries by interstellar matter, and their spectrum is steep as it mimics the equilibrium spectrum of primaries: $q_{s,inj} \propto n_p$.
A further spectral steepening is induced by CR escape, to give an equilibrium spectrum of secondaries $n_s \propto q_{s,inj} \tau_{esc} \propto n_p/D \propto R^{-(\alpha+ 2 \delta)}$.
In fact, this latter result applies only at very large energies (larger than those shown in Fig.~\ref{fig:carmelo} and \ref{fig:B/C}).
The steepness of $n_s$ is less pronounced, but still clearly visible at lower energies.
These simple considerations explain the difference in the spectra of primary and secondary CR nuclei shown in Fig.~\ref{fig:carmelo}. 
Moreover, knowing that $s \approx 2.7$ and $\delta \approx 0.3 ... 0.6$ one can deduce that at relativistic particle energies the injection spectrum of CRs in the ISM must have a slope $\alpha \approx 2.1 ... 2.4$.
This is a remarkable result, as it has been obtained without making any assumption on the nature of CR sources, except for the fact that they have to be located in the Galactic disk.

It is now well understood that the confinement of CRs in the Galaxy is due to the scattering off the turbulent interstellar magnetic field, which deflects the trajectory of CRs and prevents them to escape quickly from the Galaxy \citep[e.g.][]{zweibel2013}. 
The scattering is so effective to keep the CR particle distribution function very close to isotropy, in agreement with what is observed at the Earth's location.
The isotropization of the trajectories of CRs is the main obstacle in the search for the sources of such particles, as the arrival direction of a CR does not point to the position of its accelerator.
As a consequence, the production of CRs at many discrete Galactic sources results in an almost isotropic flux of particles at the Earth.
As we will see in the following (Sec.~\ref{sec:gamma}), indirect observations of CRs have to be employed to overcome this problem.
Instead of observing CRs, one can search for neutral particles, most notably photons, which are produced by CR interactions with matter or radiation at (or in the vicinity of) the acceleration site.
As neutral particles are not deflected by magnetic fields their arrival direction pinpoints the position of the CR source.
Unfortunately, despite many decades of indirect searches, we still do not have a conclusive answer to the problem of the origin of CRs.
We postpone a discussion of this crucial issue to Sec.~\ref{sec:SNR}.

%\section{Difficulties in the direct and indirect observations of low energy cosmic rays}
%\label{sec:problems}
\section{Difficulties in the direct observations of the local interstellar spectrum of low energy cosmic rays}
\label{sec:direct}

The brief introduction on CRs provided in the previous section was mainly focussed on particles of energy $\gtrsim 1$ GeV/nucleon.
We discuss now the main observational difficulties encountered in the study of particles of comparable and lower energy, which here are referred to as LECRs. To do so, it is mandatory to distinguish between measurements of the {\it local} (near Earth) and {\it remote} (at some given location in the Galactic disk) intensity of low energy particles.
As we will see in the following, recent progresses in diverse branches of astrophysics and space sciences allowed us to overcome, or at least strongly mitigate such difficulties and to obtain an unprecedented view on LECRs. 

%\subsection{Difficulties in the measurement of the {\it local} interstellar spectrum of low energy cosmic rays: solar modulation}
\subsection{Solar modulation}
\label{sec:local}

%Cosmic Rays (CRs) are energetic charged particles that reach the Earth's atmosphere from outer space.
%The distribution in the sky of the arrival direction of CRs is remarkably isotropic.
%The vast majority of CR particles are atomic nuclei, with a contribution from electrons at the percent level, and an even smaller one from antimatter (positrons and antiprotons).
%Amongst CR nuclei, protons largely dominate, with helium and heavier nuclei (metals) amounting to 10\% and 1\% of the total number of particles, respectively \citep[for reviews see][]{gaisser2016,gabici2019,boezio2020}.

\begin{figure*}
% Use the relevant command to insert your figure file.
% For example, with the graphicx package use
\center
  \includegraphics[width=0.7\textwidth]{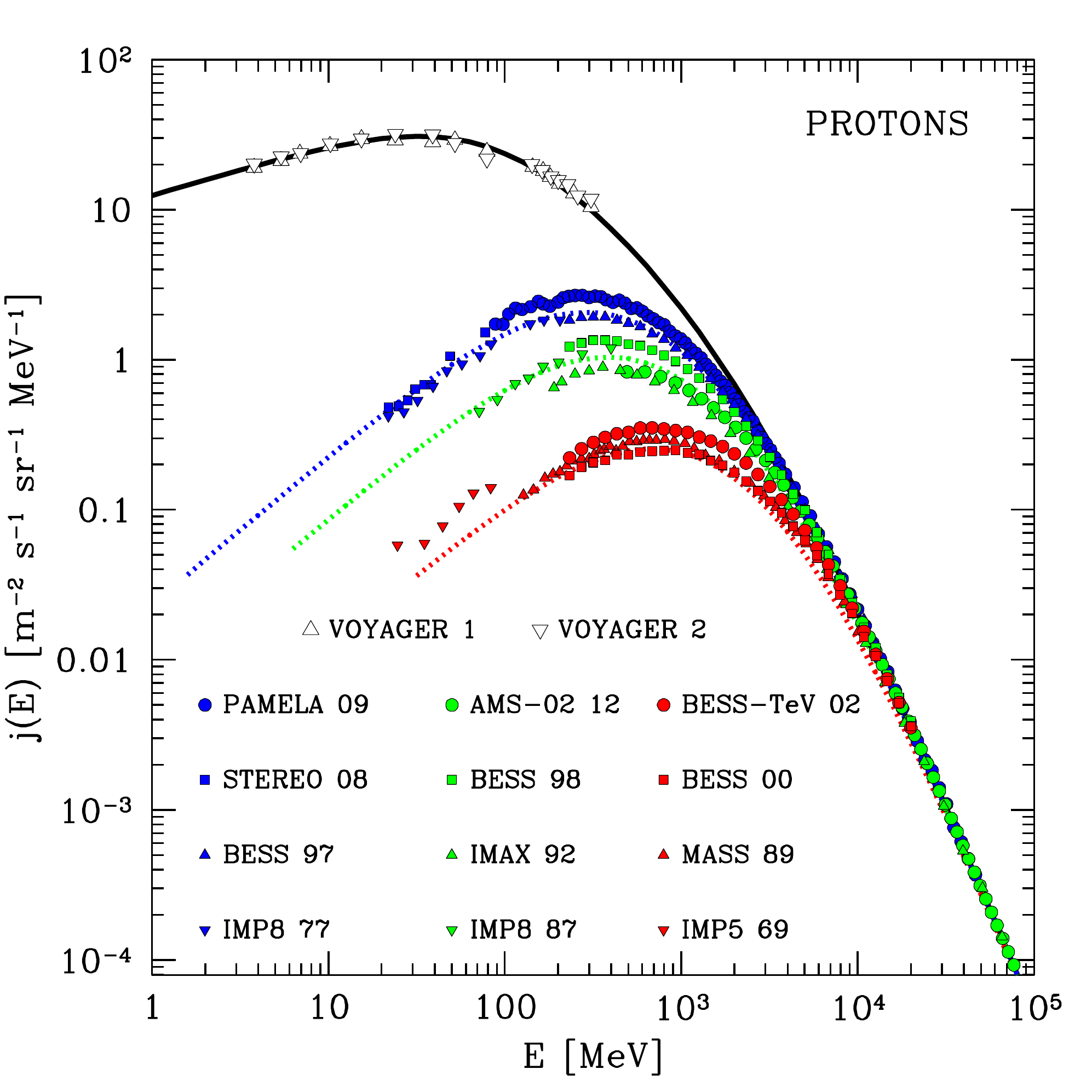}
% figure caption is below the figure
\caption{CR proton spectra measured by the Voyagers (white triangles, \citealt{cummings2016} and \citealt{stone2019}) and by other instruments at different times. Red, green, and blue data points refers to epochs of maximum, intermediate, and minimum solar activity. The black line is an analytic representation of the unmodulated CR spectrum, while the coloured dotted lines have been computed for modulation potentials equal to $\phi =$ 0.6, 0.8, and 1.4 GV. Data are from \citealt{adriani2013} (PAMELA 09); \citealt{zhao2014} (STEREO 08 and IMP8 87); \citealt{shikaze2007} (BESS 97, 98, -TeV 02, 00); \citealt{vonrosenvinge1979} (IMP8 77); \citealt{aguilar2015} (AMS-02 12); \citealt{menn2000} (IMAX 92); \citealt{webber1991} (MASS 89); \citealt{hsieh1971} (IMP5 69). Figure adapted from  \citealt{vos2015}.}
\label{fig:modulation}       % Give a unique label
\end{figure*}

The spectrum of CR nuclei observed at the top of the atmosphere exhibits a broad peak at transrelativistic ($\approx$ GeV) particle energies, and falls very steeply at higher energies (see coloured data points in Fig.~\ref{fig:modulation}).
For this reason, ultra-relativistic particles contribute very little to the CR population both in terms of number of particles and of total energy. 
While the observed intensity of ultra-relativistic CRs is remarkably stable in time, that of lower energy particles is not. The variation in the LECR intensity exhibits a main periodicity of 11 years, indicating a clear link to solar activity. 
This effect is illustrated in Fig.~\ref{fig:modulation}, which shows near-Earth spectra of CR protons measured close to minima (blue), maxima (red) and intermediate levels (yellow data points) of solar activity. The highest (lowest) LECR intensities are measured in correspondence of solar minima (maxima). 
This effect is called solar modulation, and is interpreted as follows: CRs are originated outside of the solar system and, in order to reach the Earth, they have to overcome the solar wind, which is an outward flow of magnetised and turbulent plasma emanated from the Sun.
While high energy CRs can reach virtually undisturbed the Earth, the presence of the magnetised solar wind prevents low energy charged particles of any species to penetrate freely into the inner heliosphere \citep{moraal2013,potgieter2013}.
It can be seen from Fig.~\ref{fig:modulation} that solar modulation is irrelevant for ultra-relativistic particles, and becomes important in the trans-relativistic and non-relativistic energy domain.

% For two-column wide figures use

It follows that the local interstellar spectrum of LECRs cannot be determined by means of near-Earth observations and this has been, until very recently (see Sec.~\ref{sec:voyager}) the main obstacle in the study of these particles. CR spectra measured at the top of the Earth atmosphere need to be demodulated to give the local interstellar spectrum. 
Unfortunately, the demodulation of LECR spectra is not at all a straightforward task due to our limited knowledge of both heliospheric physics \citep[e.g.][]{zank1999} and cosmic ray transport in turbulent magnetic fields \citep[e.g.][]{mertsch2020}. 

\subsection{The transport equation}
\label{sec:transportequation}

Technically, a quantitative understanding of solar modulation can be achieved by solving the CR transport equation, that was first derived by \citet{parker1965}. 
It is a partial differential equation describing how CRs are transported in the six-dimensional phase space $(\bf{x},\bf{p})$. Here, we provide an heuristic derivation of such equation following quite closely the approach by \citet{moraal2013}. 
For more formal derivations of the transport equation the reader is referred to the seminal papers by, e.g., \citet{gleeson1967}, \citet{dolginov1968}, \citet{jokipii1967,jokipii1970}, \citet{skilling1975a}, and \citet{luhmann1976} and to the more recent monographs by \citet{berezinskii1990}, \citet{kirk1994}, \citet{schlickeiser2002}, and \citet{zank2014}.

Consider a volume $V$ containing $N$ CRs. The rate of variation of $N$ with time is given by the continuity equation:
\begin{equation}
\frac{{\rm d}N}{{\rm d}t} = - \oint_V {\bf S} \cdot {\rm d}{\bf a} + Q
\end{equation}
where the first term on the right is the flux of particles ${\bf S}$ integrated across a close surface containing the volume $V$, and $Q$ is a source term representing, for example, an accelerator operating within the volume $V$ and producing CRs at a rate of $Q$ particles per second. The continuity equation can be recast in a more familiar form after using the divergence theorem, $\oint {\bf S} \cdot {\rm d}{\bf a} = \int \nabla \cdot {\bf S} ~ {\rm d} V$, and considering volume densities rather than total number of particles, i.e. $N = \int n ~ {\rm d} V $ and $Q = \int q~{\rm d} V$, to obtain:
\begin{equation}
\label{eq:transport0}
\frac{\partial n}{\partial t} + \nabla \cdot {\bf S} = q
\end{equation}
which can be solved provided ${\bf S}$ and $q$ are known. 
Being interested here in the study of the heliospheric transport of CRs which are produced outside of the Solar system, we can set $q = 0$ and impose that at very large distances from the Sun $n$ has to be equal to the interstellar density of CRs.

In the expressions above, $n$ represents the number density of CRs of {\it any energy}. In fact, what we measure are particle spectra, and so we should rather seek for a transport equation for the particle distribution function $f(p)$, where $p$ is the momentum of the CR particle.  
Since the heliospheric magnetic field is turbulent, CRs are scattered by magnetic irregularities.
The scattering is strong enough to keep the particle distribution function very close to isotropy, so that we can neglect small deviations from isotropy and write $n \sim 4 \pi \int {\rm d} p~ p^2 f(p)$ (a discussion on the accuracy of this approximation can be found in e.g. \citealt{kirk1994}).

Now we have to derive an expression for the {\it differential} flux of particles ${\bf S}(p)$ in the solar wind. To a first approximation, the wind can be treated as a spherically symmetric outflow of plasma moving at a speed ${\bf u_w}$. 
We already said that the plasma is magnetised, and that the magnetic field is turbulent. The magnetic turbulence can be described as a superposition of magnetohydrodynamic (MHD) waves (for example Alfv\'en waves). Such waves move with respect to the plasma at a velocity $v_A \ll u_w$, so that they can be considered to be virtually at rest in the rest frame of the wind. 
CRs are coupled to the outward moving plasma because they scatter off those MHD waves, and this originates an outward advective flux of CR particles ${\bf S_a}(p) = f(p) {\bf u_w}$.
In addition to that, particle scattering also results in a spatial random walk of CRs, which can be described by a diffusive component in the flux ${\bf S_d}(p) = -{\bf D} \cdot \nabla f(p)$, where we made use of Fick's law and introduced the diffusion tensor ${\bf D}$. 

However, we notice that while ${\bf S_a}(p)$ represents the advective flux in the fixed (solar system) rest frame, ${\bf S_d}(p)$ describes the diffusive flux in the wind rest frame, because it is in that frame that the scattering centres are at rest. 
Thus, it is convenient to introduce a {\it mixed frame}, where spatial coordinates are specified in the fixed frame, and particle momenta in the wind frame.
In this mixed frame we can write ${\bf S_{tot}}(p) = {\bf S_a}(p)+{\bf S_d}(p) = f(p) {\bf u_w} -{\bf D} \cdot \nabla f(p)$.

We can now proceed substituting $n$ with $f(p)$ and ${\bf S}$ with ${\bf S_{tot}}(p)$ in Eq.~\ref{eq:transport0}, but this would not result in the correct CR transport equation. 
The reason is that the  additional term $\frac{1}{p^2} \frac{\partial}{\partial p} \left( p^2 \langle \dot{p} \rangle f \right)$ must be added to the left hand side of the transport equation, to account for the fact that particles can gain or lose momentum. 
Here, $\langle \dot{p} \rangle$ represents the average rate of momentum gain/loss.
As pointed out by \citet{moraal2013}, the additional term is the divergence of the particle flux in momentum space, which is analog to the advective part of $\nabla \cdot {\bf S}$ (with the substitutions ${\bf u_w} \rightarrow \langle \dot{p} \rangle$, and $\nabla \rightarrow \frac{1}{p^2} \frac{\partial}{\partial p} p^2$). As noted by Parker, in the heliosphere the most important contribution to $\langle \dot{p} \rangle$ comes from the adiabatic cooling of CRs in the expanding solar wind, at a rate $\langle \dot{p} \rangle = - \frac{p}{3} \nabla \cdot {\bf u_w}$.

After performing these substitutions and additions, and some simple manipulations, the transport equation finally becomes:
\begin{equation}
\label{eq:transport}
\frac{\partial f}{\partial t} + {\bf u_w} \cdot \nabla f - \nabla \cdot \left( {\bf D} \cdot \nabla f \right)  - \frac{p}{3} \left( \nabla \cdot {\bf u_w} \right) \frac{\partial f}{\partial p}  = 0
\end{equation}
and can be further simplified by setting $\frac{\partial f}{\partial t} = 0$, as the typical time scale of variation due to solar modulation (a fraction of the solar cycle) is much longer than the CR propagation time through the heliosphere (less than a year).

\subsection{Solutions of the transport equation}

The simplest solution of the transport equation can be obtained by neglecting adiabatic losses ($\nabla \cdot {\bf u_w} = 0$), by imposing spherical symmetry, and by assuming an isotropic diffusion for CRs (the diffusion coefficient in this case becomes a scalar function of position and momentum). 
Under these simplifying assumptions, and recalling that there are no sources/sinks of particles, the transport equation reduces to the {\it diffusion--advection equation}:
\begin{equation}
\label{eq:diffconv}
u_w f - D \frac{\partial f}{\partial r} = 0
\end{equation} 
where $r$ is the radial coordinate (the distance from the Sun). 
This simplified transport equation describes a situation where an outward advective flux is perfectly balanced by an inward diffusive flux. 
The solution is then:
\begin{equation}
f(r,p) = f_{LIS}(p) ~e^{-M} = f_{LIS}(p) ~ \exp \left[ - \int_r^{r_{LIS}} {\rm d} r^{\prime} \frac{u_w}{D} \right]
\end{equation}
where we introduced a non-dimensional modulation parameter $M$.
The local interstellar spectrum (LIS) of CRs ($f = f_{LIS}$) is recovered at the border of the heliosphere ($r = r_{LIS}$), while an exponential suppression appears for any $r < r_{LIS}$.
Under most circumstances, the diffusion coefficient $D$ is a monotonically increasing function of particle momentum, and it is then possible to define a critical momentum that satisfy $M(r,p_*) \sim 1$.
For any fixed value of $r$, CR particles characterised by momenta $\gg p_*$ are unaffected by the presence of the solar wind, and can penetrate undisturbed into the inner heliosphere. On the contrary, CR particles of low momenta cannot, and for this reason the spectrum of CRs observed at the Earth is strongly suppressed with respect to the local interstellar one for $p \ll p_*$.
 
Let us now proceed a step forward and discuss the most widely used approximate analytic solution of the transport equation in the heliosphere.
To do so, we should remind that $f(r,p)$ is the isotropic part of the particle distribution function computed at a position $r$ (in the fixed frame) and at a particle momentum $p$ (in the wind frame).
When seen from the fixed frame, the particle distribution function is no longer isotropic, because both the energy and the arrival direction of particles are different in that rest frame. 
The anisotropy adds a contribution to the the advective flux in Eq.~\ref{eq:diffconv}, which should be therefore corrected to give:
\begin{equation}
\label{eq:CG}
C u_w f - D \frac{\partial f}{\partial r} = 0
\end{equation}
where $C = -\frac{1}{3} \frac{\partial \ln f}{\partial \ln p}$ is called Compton-Getting coefficient \citep[derivations of $C$ can be found in e.g.][]{compton1935,gleeson1968b,forman1970}.
After substituting $C$ into Eq.~\ref{eq:CG}
%and using particle rigidity $P = pc/Ze$ instead of momentum, 
we obtain an expression which is identical in form to the Liouville equation in a conservative field \citep[e.g.][]{fisk1973,quenby1984}: 
\begin{equation}
\label{eq:forcefield1}
v \frac{\partial f}{\partial r} +  \dot{p} \frac{\partial f}{\partial p} = 0
\end{equation}
where $v$ is the particle velocity.
The effective ``force''
$
\dot{p} = \frac{v p u_w}{3 D}
$
accounts for (in an approximate way) the combined effect of diffusion, advection, and energy losses.
When written in this form, the transport equation is called {\it force field equation} \citep{gleeson1968b}.

Another way to interpret the force field equation is to rewrite Parker's transport equation (Eq.~\ref{eq:transport}) in the equivalent form \citep{moraal2013}:
\begin{equation}
\frac{\partial f}{\partial t} + \nabla \cdot \left[ C {\bf u_w} f - {\bf D} \cdot \nabla f \right] + \frac{1}{3 p^2} \frac{\partial}{\partial p} \left( p^3 {\bf u_w} \cdot  \nabla f \right) = 0
\end{equation}
and notice that the force field equation is obtained after equating to zero the term in square brackets.
This can be obtained by setting, as done above, $\frac{\partial f}{\partial t} = 0$, and  by further imposing that the last term in the left hand side of the equation should be small when compared to both the diffusive and advective fluxes.
The two conditions to be fulfilled are then: {\it i)} $u_w r/D \ll 1$ which is always satisfied provided sufficiently large particle energies are considered ($D$ is a growing function of particle energy), and {\it ii)} $\nabla f/f \ll C/r$, which is satisfied more easily in the inner heliosphere, in the vicinity of the Sun, where $r$ is small.
In other words, the force field approximation applies to particles of high energy, that suffer only moderate solar modulation, and therefore are characterised by a mild spatial gradient \citep{caballerolopez2004}.

Eq.~\ref{eq:forcefield1} can be solved by the method of characteristics, that states that the particle distribution function is constant along the curve %in $(p,r)$ space 
defined by:
\begin{equation}
\label{eq:characteristic}
\frac{{\rm d} p}{{\rm d}r} = \frac{u_w p}{3 D}
\end{equation}
At this point, we should recall that the diffusion coefficient is the product between the particle velocity and the mean free path for spatial scattering. 
A convenient way to write it is: $D = \frac{1}{3} \beta(p) \lambda(p,r)$, where $\beta = v/c$ is the particle speed in units of the speed of light, and $\lambda$ the mean free path multiplied by $c$ (so that it has the same units as $D$). If the latter is separable, i.e. $\lambda(p,r) = \lambda_p(p) \lambda_r(r)$ then we can integrate along the characteristic (Eq.~\ref{eq:characteristic}) and introduce a modulation parameter:
\begin{equation}
\label{eq:potential}
\phi(r) \equiv \int_r^{r_{LIS}} {\rm d}r^{\prime} \frac{u_w}{\lambda_r} = \int_p^{p_{LIS}} {\rm d} p^{\prime} \frac{\beta \lambda_p}{p^{\prime}}
\end{equation}
where $p_{LIS}$ is the momentum a CR particle had when it entered the heliosphere.
The solution of the force field equation can then be written as:
\begin{equation}
\label{eq:ffsolution}
f(p,r) = f_{LIS}(p_{LIS}(p,r)) ~ .
\end{equation}

A simple analytic expression for such solution can be found if the diffusion coefficient is proportional to particle rigidity $\lambda_p \sim R = \frac{p c}{Z e}$, where $Z$ is the charge of the particle in units of the (positive) elementary charge $e$. 
With this assumption, the parameter $\phi$ has the dimensions of an electric potential (V) and is therefore called {\it modulation potential}.
Moreover, the second integral in Eq.~\ref{eq:potential} can be performed analytically and gives $\phi = \frac{E_{LIS}-E}{Ze}$, where $E_{LIS}$ and $E$ are the kinetic energies of the CR particle when it entered the heliosphere and reached the position $r$, respectively.

We notice now that it is more convenient to express Eq.~\ref{eq:ffsolution} in terms of the CR intensity, which is the physical quantity which is actually measured.
It represents the number of particles of a given kinetic energy $E$ passing across a unit surface per unit time and unit solid angle, and is connected to the particle distribution function through $j(E) = p^2 f(p)$. It is a simple exercise to obtain the very well known result \citep{gleeson1968b}:
\begin{equation}
\label{eq:fffsolution}
j(E,r) = \frac{E(E+2 A m_p)}{(E+\Phi)(E+\Phi+2 A m_p)} j_{LIS}(E+\Phi)
\end{equation}
where $A$ is the atomic mass number of the CR particle, $m_p$ is the proton mass, and $\Phi(r) = Z e \phi(r)$.
The expression above has been widely used in the literature, mainly due to its simplicity. 
If we set $r$ equal to the distance between the Sun and the Earth, then only one parameter, $\phi$, is needed in order to connect the local interstellar spectrum of CRs, $j_{LIS}(E)$, to the one observed at the top of the Earth atmosphere, $j_{\Earth}(E)$.

\citet{caballerolopez2004} investigated, among others, the limitations of the force field approximation. 
Interestingly, they found that while this approach provides a good description of the effect of solar modulation suffered by LECRs in the inner heliosphere (see also e.g. \citealt{ghelfi2016}), the diffusion--advection equation is a more appropriate description at large distances from the Sun, where energy losses become less important.
In recent times, continuous observations of CR intensities of unprecedented precision became available, thanks especially to space borne detectors (for reviews see \citealt{bindi2017} and \citealt{boezio2020}). The interpretation of such data calls for more sophisticated approaches, involving either modifications/generalisations of the modulation potential \citep{corti2016,cholis2016,gieseler2017}, or extensions of the force-field model \citep{kuhlen2019}, or fully numerical approaches  \citep{vos2015,boschini2019,corti2019}.

Despite the continuous progresses made in the study of particle transport in the heliosphere, the local interstellar spectrum of LECRs, as derived through a demodulation of the spectra observed at the top of the atmosphere, is inevitably affected by uncertainties. As we will discuss in the following Section, {\it direct} measurements of LECR spectra beyond the heliopause became recently possible, providing us with an unprecedented view on such particles.

\subsection{A breakthrough in the measurement of the local interstellar spectrum of low energy cosmic rays: the Voyager mission}
\label{sec:voyager}

On September 5th 1977, the Voyager 1 probe was launched from the Cape Canaveral Station. 
It was preceded by the twin probe Voyager 2, launched on August 20th.
The goal of the Voyagers was to explore the solar system and, following two distinct paths, reach the outer heliosphere and possibly beyond, before the end of operations.
The study of Galactic CRs and of energetic particles in the heliosphere was one of the scientific objectives of the programme \citep{stone1977,krimigis1977}.

More than 40 years after launch, the Cosmic Ray Systems (CRS) onboard of the Voyagers are still collecting and sending us data.
Each system is composed of three particle detectors: the High-Energy Telescope System, the Low-Energy Telescope System, and the Electron Telescope.
These detectors are sensitive to nuclei of charge Z = 1 to 30 in the energy range 1-500 MeV for protons, 2.5-500 MeV/nucleon for iron, and to electrons of energy 3-110 MeV \citep{stone1977}. %The systems can also detect anisotropies for all components \citep{stone1977}.
These detectors are complemented by the Low Energy Charged Particle instrument (LECP), sensitive to particle energies in the range $15 ~ {\rm keV} \lesssim E \lesssim 40 ~{\rm MeV/nucl}$, and designed to study energetic particles in planetary magnetospheres and interplanetary space \citep{krimigis1977}.

\begin{figure*}
% Use the relevant command to insert your figure file.
% For example, with the graphicx package use
\center
  \includegraphics[width=0.49\textwidth]{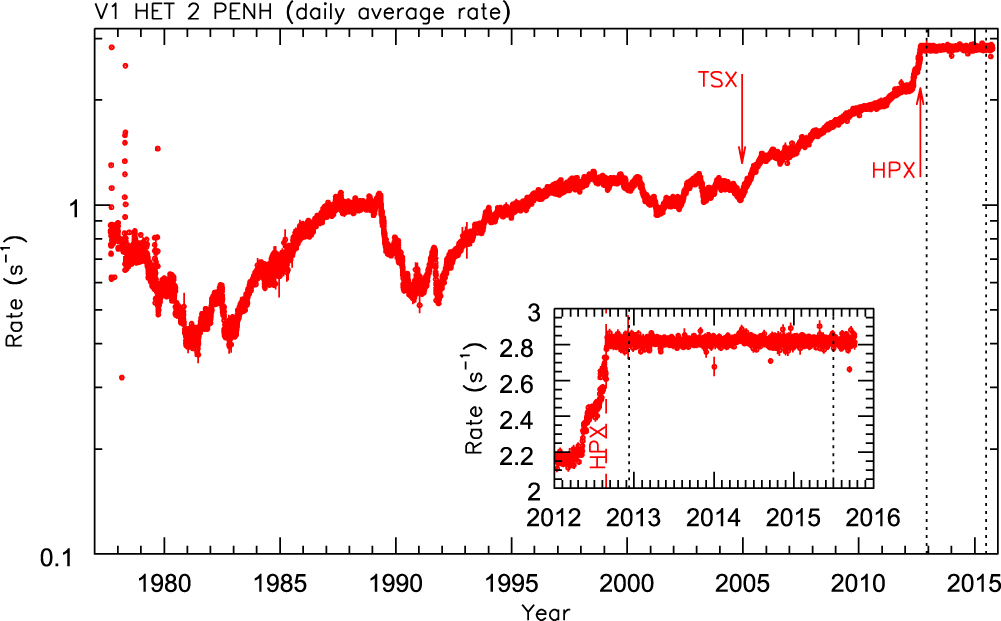}
  \includegraphics[width=0.49\textwidth]{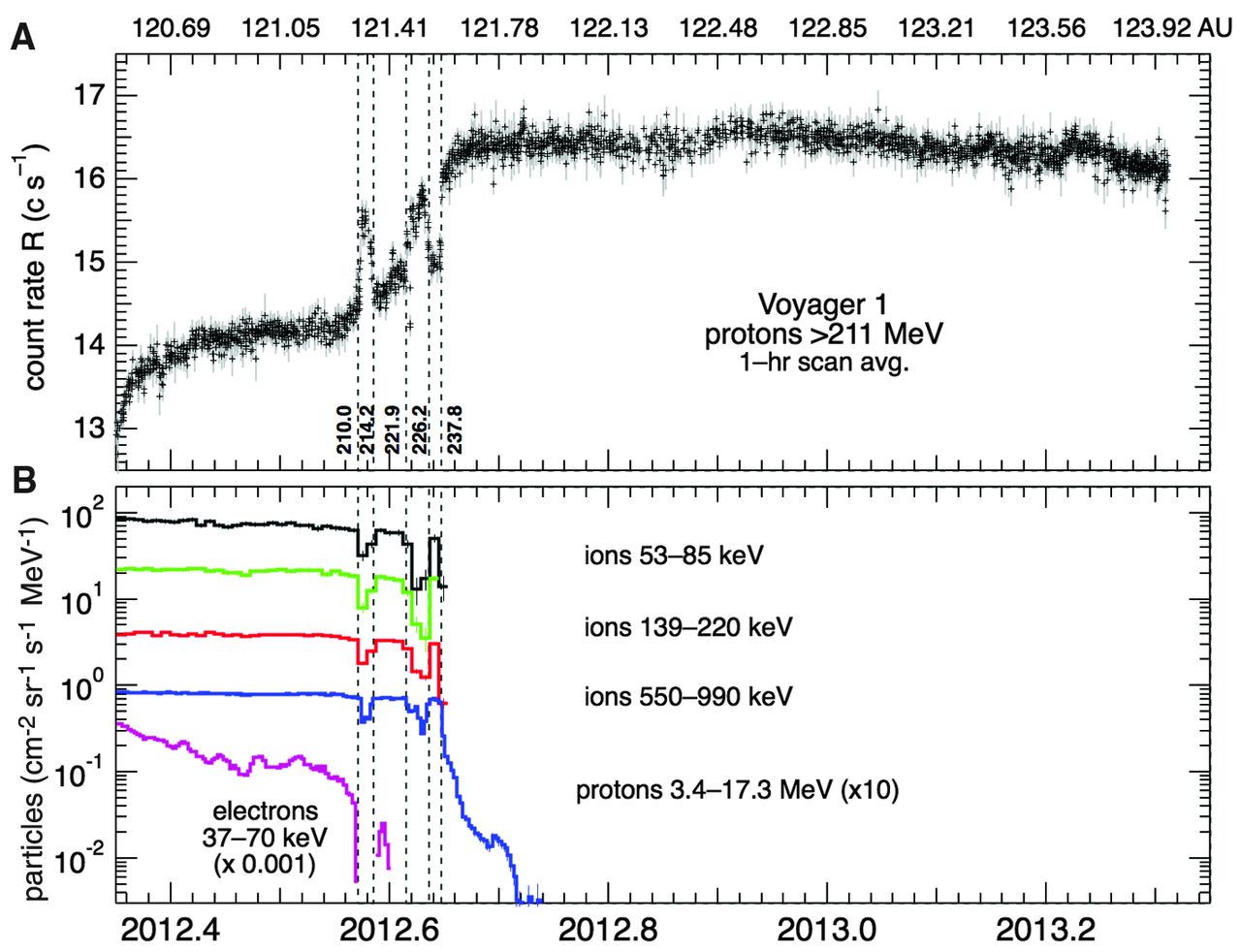}
% figure caption is below the figure
\caption{{\bf Left:} integral count rate of CR nuclei and electrons of energy $\gtrsim$ 70 MeV/nucleon and 15 MeV, respectively, as measured by the CRS onboard Voyager 1. Figure from \citet{cummings2016}. {\bf Right:} integral count rate of CR protons of energy above 211 MeV (A) and lower energy CR nuclei, protons, and electrons (B) measured by the LECP instrument onboard of Voyager 1. Figure from \citet{krimigis2013}. Time is on the x-axis.}
\label{fig:voyager}       % Give a unique label
\end{figure*}

The left panel of Fig.~\ref{fig:voyager} shows the integral (nuclei plus electrons) count rate as measured over decades by the CRS onboard Voyager 1. %Both CR nuclei of energy $\gtrsim$ 70 MeV/nucleon and CR electrons of energy $\gtrsim$ 15 MeV contribute to the total rate. 
Time is on the x-axis, and spans from few days after the launch to 2015. Later times correspond to larger distances between Voyager 1 and the Earth.
The 11 year periodicity induced by solar modulation is clearly visible in earlier data and, as expected, it becomes less evident at later times, when the probe reaches larger distances from the Sun, where the impact of Solar modulation is weaker.
Three milestones can be identified in this epic journey. 
On February 17th 1998, Voyager 1 reached a distance from the Sun of $69.4$ AU and, overtaking Pioneer 10, became the farthest human-made object in space. 
On December 16th 2004 the probe crossed the solar wind termination shock, at a distance of $\sim$ 94 AU from the Sun (label TSX in Fig.~\ref{fig:voyager}), and entered the heliosheath, which is the outermost layer of the heliosphere  \citep{stone2005}. 
Finally, on August 25th 2012 Voyager 1, at a distance of 121.6 AU from the Sun, passed beyond the heliopause (label HPX in Fig.~\ref{fig:voyager}) and entered interstellar space \citep{stone2013,krimigis2013}.
After that, the CR count rate became remarkably stable in time (see inset in Fig.~\ref{fig:voyager}), as one would expect for CRs unaffected by solar modulation.
Few years later, on November 5th 2018, the twin probe Voyager 2 also left the heliosphere \citep{stone2019,krimigis2019}. 

The scenario just described is very widely, though not universally accepted.
\citet{gloeckler2015} claimed that the Voyagers might still be inside the heliosphere. 
Moreover, some residual modulation could possibly affect measurements even outside of the heliosphere, due to the presence of a bow shock or of an adiabatic flow ahead of the heliopause \citep{scherer2011}.
Finally, the data collected by the Voyagers revealed that the structure of the heliospheric boundary is more complex than previously thought \citep{zank2015}.
All these caveats should be kept in mind while reading the rest of this review, where we will assume, as a working hypothesis, that the Voyagers are indeed located beyond the heliopause and are currently measuring the pristine local interstellar spectrum of CRs.

The working hypothesis adopted here was formulated following the unexpected measurements reported by instruments onboard Voyager 1 in the summer of 2012 \citep{stone2013,krimigis2013}. 
In particular, the right panel of Fig.~\ref{fig:voyager} shows the surprising behaviour of particle count rates measured by the LECP instrument for epochs around the presumed heliopause crossing.
The intensity of low energy ions and electrons of solar origin suddenly dropped by more than a factor of $10^3$ on August 25th 2012 and eventually disappeared (bottom panel), while the count rate of higher energy CR protons of Galactic origin simultaneously increased by about 10\% (top panel). 
A very similar behaviour was observed by Voyager 2 on November 5th 2018, when the probe was at a distance of 119 AU from the Sun.
Data may be interpreted by saying that, for an observer that crosses the heliopause, the measured intensity of energetic particles of solar origin drops as such particles stream away in interstellar space, while the intensity of Galactic CRs increases as they become virtually unaffected by solar modulation.

Several analytic fits or representations of the local interstellar spectrum of CR protons can be found in the literature \citep[][]{schlickeiser2014,ivlev2015,phan2018}. 
We adopt here a doubly broken power law function\footnote{Which provides a better representation of data with respect to the singly broken power law spectra often adopted in the literature. However, this is just a convenient descriptive expression, being neither a physically motivated choice, nor a formal fit to data.}: 
\begin{equation}
\label{eq:voyager}
j_{LIS}(E) =  \frac{12.5 \left(\frac{E}{{\rm MeV}}\right)^{0.35}}{\left[ 1 + \left( \frac{E}{80~{\rm MeV}}\right)^{1.3} \right] \left[ 1 + \left( \frac{E}{2.2~{\rm GeV}}\right)^{1.9/s} \right]^s} ~ \rm m^{-2} s^{-1} sr^{-1} MeV^{-1}
\end{equation}
where the parameter $s$ can be tuned to vary the shape of the spectrum around the spectral break at 2.2 GeV/nucleon.
Smaller (larger) values of $s$ make the break sharper (shallower).
The parameter has been introduced because the local interstellar spectrum of CRs is not constrained in the GeV energy domain, as direct near Earth observations are heavily affected by solar modulation.
Eq.~\ref{eq:voyager} is plotted as a solid black line in Fig.~\ref{fig:modulation}, and provides a convenient description of both interstellar measurements of CR protons at low particle energies (Voyager data, white triangles in Fig.~\ref{fig:modulation}), and near-Earth ones (coloured data points in Fig.~\ref{fig:modulation}) at high enough particle energies, where solar modulation has no effect.
To describe the intermediate energy regime, we set $s = 2.1$, a choice that will be motivated in Sec.~\ref{sec:gamma}.

On the contrary, near-Earth proton spectra (coloured data points in Fig.~\ref{fig:modulation}) are shaped by solar modulation for particle energies below few tens of GeV.
As seen in Section~\ref{sec:local}, the effect of solar wind on LECRs can be accounted for in an approximate way by means of Eq.~\ref{eq:fffsolution}. 
In that approach, the modulation potential ($\phi$) alone suffices to account for all of the relevant heliospheric physics.
Before the Voyagers entered interstellar space, the only measured quantity in Eq.~\ref{eq:fffsolution} was the near-Earth spectrum of CRs, $j_{\Earth}(E)$.
Therefore, a value for $\phi$ had to be chosen/estimated in order to derive the local interstellar spectrum of LECRs.
Now, after the two Voyager probes have crossed the heliopause, both the near-Earth and local interstellar spectra of CRs are measured, and Eq.~\ref{eq:fffsolution} can be used to infer $\phi$.
The results of such a procedure are shown in Fig.~\ref{fig:modulation}, where the dotted blue, green, and red curves represent expectations of the near-Earth CR spectra characterised by values of the modulation potential equal to $\phi =$ 0.6, 0.8, and 1.4 GV, respectively.

What said above illustrates one of the many reasons why the LECR data collected by the Voyagers are so important: they provide us, for the very first time, with precious and direct constrains on the physical processes that regulate CR modulation.
More in general, data collected by the Voyagers had a tremendous impact on the study of the heliosphere and of its boundary. 
As this aspect goes beyond the scope of this review, the interested reader is referred to \citet{zank2015}.

\begin{figure*}
% Use the relevant command to insert your figure file.
% For example, with the graphicx package use
\center
  \includegraphics[width=0.7\textwidth]{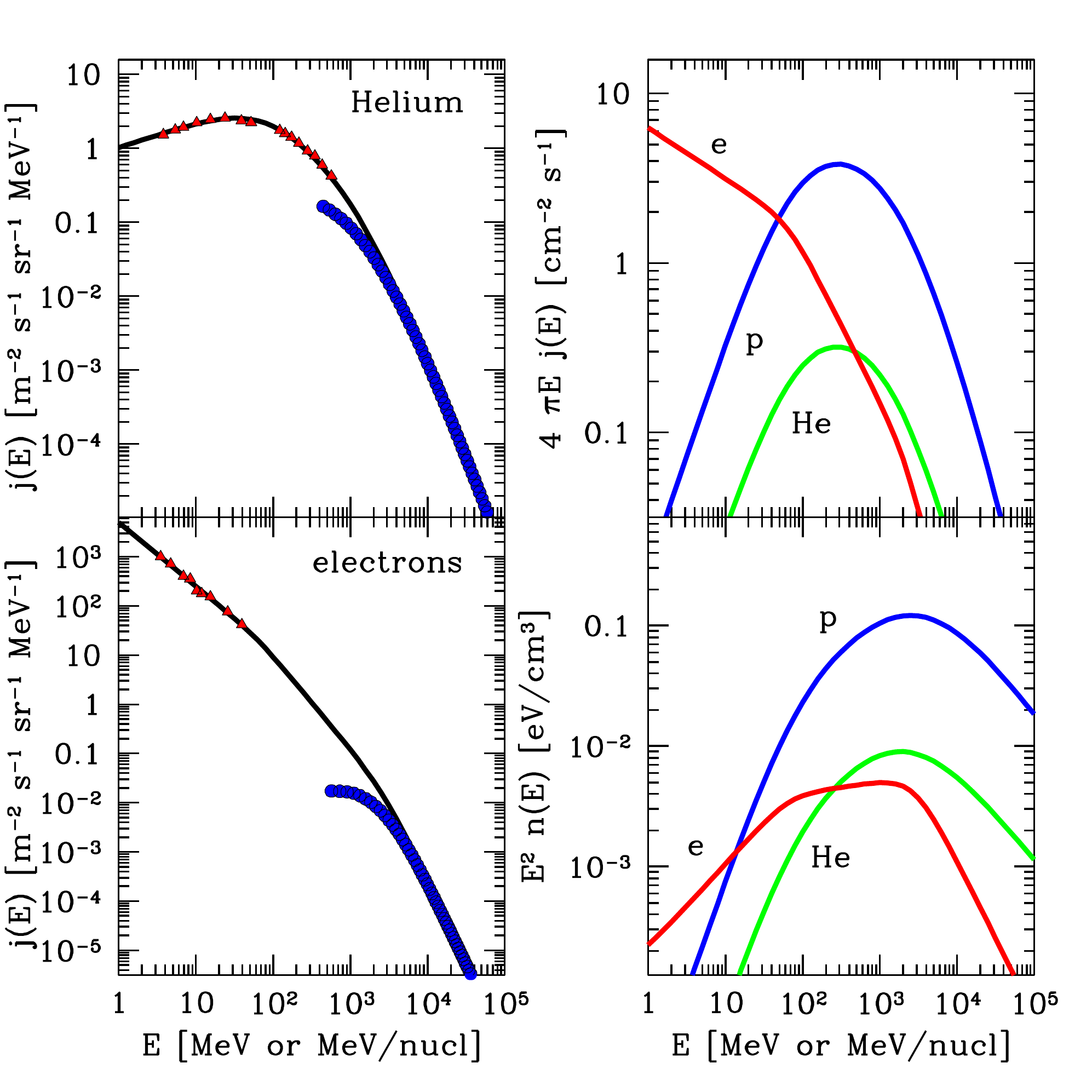}
% figure caption is below the figure
\caption{\textbf{Top left:} CR helium spectrum as measured by Voyager \citep[][red triangles]{cummings2016} and AMS-02 \citep[][blue circles]{aguilar2017}. \textbf{Bottom left:} CR electron spectrum, as measured by Voyager \citep[][red triangles]{cummings2016} and AMS-02 \citep[][blue circles]{aguilar2019}.  \textbf{Top right:} CR intensity for protons (blue), helium (green), and electrons (red curve) in the local ISM. \textbf{Bottom right:} CR energy densities in the local ISM. Same color code as in top right panel.} % for protons (blue), helium (green), and electrons (red curve) in the local interstellar medium.}
\label{fig:Hee}       % Give a unique label
\end{figure*}

The Voyagers measured also the local interstellar spectra of LECR nuclei heavier than hydrogen. 
The most abundant amongst them is helium, whose spectrum is shown in the top-left panel of Fig.~\ref{fig:Hee}.
In the Figure, Voyager and AMS-02 data are shown as red triangles and blue circles, respectively.
\citet{cummings2016} noticed that the spectra measured by Voyager for hydrogen and helium overlap almost perfectly when plotted as a function of the particle energy per nucleon, and when the former is divided by a factor of 12.2.
On the other hand, the spectrum of helium measured by AMS-02 at high energies is appreciably harder than that of hydrogen (both can be described by power laws in energy, the difference between the spectral slopes being $\sim$ 0.07). 
Following these indications, the black line in the top-left panel of Fig.~\ref{fig:Hee} has been obtained by dividing Eq.~\ref{eq:voyager} by 12.2 and substituting $[1+(E/2.2~{\rm GeV})^{(1.9/2.1)}]^{2.1}$ with $[1+(E/1.6~{\rm GeV})^{(1.83/1.8)}]^{1.8}$.
The reason why the spectra of protons and helium nuclei are different in the multi-GeV energy domain is not understood (see discussion in \citealt{gabici2019}).

Finally, the bottom-left panel of Fig.~\ref{fig:Hee} shows the local interstellar CR electron spectrum. 
Also in this case, red triangles and blue circles show Voyager and AMS-02 data, respectively.
As for protons, a doubly broken power law function 
\begin{equation}
\label{eq:voyagere}
j_{LIS}^e(E) = \frac{5 \times 10^3 \left( \frac{E}{\rm MeV} \right)^{-1.3}}{\left[ 1 + \left( \frac{E}{65 ~ {\rm MeV}} \right)^{0.6/\delta_1} \right]^{\delta_1} \left[ 1 + \left( \frac{E}{2.8 ~ {\rm GeV}} \right)^{1.38/\delta_2} \right]^{\delta_2}} ~ \rm m^{-2} s^{-1} sr^{-1} MeV^{-1}
\end{equation}
with $\delta_1 = 0.2$ and $\delta_2 = 0.5$ provides a good description of data and is plotted as a black solid line in the Figure.
We should stress here that Voyager and AMS-02 data alone would not require the presence of two breaks: a singly broken power law would provide an equally good fit to data.
The choice made in Eq,~\ref{eq:voyagere} will become clear in Sec.~\ref{sec:leptonic}, when we will discuss indirect measurements of the local CR electron spectrum.

%It is generally accepted that the impact that LECRs  exert on the interstellar medium is dominated by protons and electrons.
%CR helium nuclei provide a smaller but not negligible contribution, while heavier CR nuclei play only a very minor role.
In order to evaluate the impact that LECRs exert on the ISM, it is instructive to compute and plot the quantities $R_i(E) = 4 \pi E j_i(E)$ and $\omega_i(E) = E^2 n_i(E) = 4 \pi E^2 j_i(E)/v(E)$, where $n_i(E)$ is the particle number density, $v(E)$ the velocity of  a particle characterised by an energy per nucleon $E$, and the suffix $i$ may refer to protons, helium nuclei or electrons.
These two quantities are shown in the top-right and bottom-right panel of Fig.~\ref{fig:Hee}, respectively.

LECRs interact with the interstellar gas in many ways. If $\sigma_j(E) \propto E^{-s}$ is the cross section of a given physical process $j$ (for example, the ionisation cross section), then the quantity $R_i(E) \sigma_j(E)$ represents the rate of interactions per interstellar atom suffered by LECR particles of energy per nucleon within a given logarithmic interval around $E$.
It can be seen from Fig.~\ref{fig:Hee} that the curves for $R_i(E)$ relative to protons and helium are very peaked around a particle energy equal to $E_{peak} \lesssim 1$ GeV/nucleon.
For particle energies below the peak $R_i(E) \propto E^{1.35}$, which implies that for ``well behaved'' cross sections characterised by $s < 1.35$, the total rate of interactions (integrated on all energies) due to the process $j$ is dominated by particles of energy around $\approx E_{peak}$.
As the condition on $s$ is satisfied by ionisation cross sections ($\sigma_{ion} \approx 1/E$, see Sec.~\ref{sec:sigmaion}), the CR ionisation rate in the local ISM is well constrained by Voyager and AMS-02 data for CR nuclei (which are largely dominated by protons and helium).
Unfortunately, this is not the case for electrons, for which $R_i(E)$ decreases monotonically in the energy domain probed by available observations.
The CR ionisation rate is therefore not well constrained for CR electrons, as it depends on the behaviour of their spectrum for energies smaller than few MeV, where no observations have ever been performed.
In this case, only a lower limit for the ionisation rate in the local ISM can be obtained.

On the contrary, as shown in the bottom-right panel of Fig.~\ref{fig:Hee}, the total energy density of CRs in the local ISM is well constrained for both nuclei and electrons.
The largest contribution to the local energy density is provided by particles of energy $\gtrsim$ 1 GeV/nucleon, and integrating over all energies one gets energy densities of $\sim$ 0.55, 0.16, and 0.03 eV/cm$^3$ for CR protons, helium nuclei, and electrons, respectively.
The energy density of nuclei heavier than helium is roughly twice that of electrons \citep{cummings2016}, giving a total CR energy density in the local ISM of the order of $\approx$ 0.8 eV/cm$^3$.

The statements just made are valid under the (very plausible, see Sec.~\ref{sec:questions}) hypothesis that there are no additional components in the CR spectrum that emerge at very low energies and dominate the total CR energy budget in the local ISM.
The presence of such hidden components in the CR spectrum has been sometimes invoked in the past in order to explain some presumed anomaly in the cosmic abundance of light nuclei \citep{meneguzzi1971} or, more recently, as an additional source of ionisation of the interstellar gas \citep{cummings2016}.
Before the Voyagers entered the local ISM, such {\it ad hoc} low energy component was assumed to be present in the MeV energy domain, and to be hidden by solar modulation, while now that measurements of the pristine interstellar spectrum of CRs extend down to few MeV, an hidden component can only show up in the sub-MeV domain. 

Before concluding this Section, few words on heavier CR nuclei are in order.
Here, we will not discuss in great detail the spectra of primary LECRs heavier than helium, but we will just highlight the two most important observational findings.
First of all, the striking similarity of the spectral shape of protons and helium CRs observed by the Voyagers does not extend to heavier nuclei, which exhibit much harder spectra in the MeV domain \citep{cummings2016}.
Second, the spectra of CR carbon and oxygen (the two most abundant nuclei heavier than helium) have been measured by the AMS-02 and found to be very similar to that of helium \citep{aguilar2017}.
Therefore, in the AMS-02 energy domain, the most abundant primary CR nuclei seem to have very similar spectra, with the exception of protons, whose spectrum is slightly softer. 
Conversely, in the lower energy domain probed by the Voyagers, proton and helium have very similar spectra, which are in turn different by those of heavier nuclei.
Both these differences are not understood, and the interested reader is referred to \citealt{gabici2019} and \citealt{tatischeff2021} and references therein for a discussion of this very puzzling issue.
As particle acceleration mechanisms are not expected to distinguish between species when shaping their energy spectra, it is tempting to interpret such differences as the result of a different origin for different species (e.g. acceleration taking place in different phases of the ISM, or in different astrophysical objects, or at different evolutive stages of the same objects, etc.).

Finally, the measurements of the local interstellar spectrum of LECRs reviewed in this Section are also very important in connection with the problem of the origin of CRs.
With this respect, the main goal is to understand how the features observed in the local interstellar spectrum of LECRs could be explained as the result of the injection of CRs in the ISM, followed by their transport in the turbulent interstellar magnetic field \citep{strong2007}.
This is a very complex problem, whose solution requires a knowledge of the nature and spatial distribution of CR sources in the Galaxy, of the topology of the interstellar magnetic field over a huge range of spatial scales, of the details of the transport of CR particles in turbulent fields, and on the spatial distribution of interstellar matter.
In the simple model of CR transport in the Galaxy described in Sec.~\ref{sec:CRs}, the problem was brutally simplified by adopting {\it mean} particle densities or residence times in the Galaxy.
Spatial variations in the CR intensities cannot be neglected in more realistic models, and is therefore important to review the current status of measurements, inevitably indirect, of the intensity of CRs in regions far from the Solar system.
This will be done in the next Section.

\section{Indirect measurements of the remote interstellar spectrum of low energy cosmic rays: gamma-ray and radio observations}
\label{sec:indirect}

%\subsection{Gamma rays as probes of the cosmic ray intensity throughout the Galactic disk} 
\subsection{Hadronic gamma rays}
\label{sec:gamma}

In 1952 Hayakawa \nocite{hayakawa1952} first proposed that, if CRs fill the entire Galaxy, the galactic disk should shine in gamma rays.
Such gamma rays are generated by the decay of light mesons, which are produced in the inelastic collisions between CR nuclei (mainly protons) and nuclei in the ISM (mainly hydrogen).
In this context, the dominant channel for gamma-ray production consists of two steps.
First, an energetic CR proton interacts with a proton at rest in the ISM, to generate a neutral pion, which immediately decays into two gamma rays:
\begin{equation}
\label{eq:pp}
p + p \longrightarrow X + \pi^0 ~~ , ~~~ 
%\end{equation}
%\begin{equation}
%\label{eq:pi0}
\pi^0 \longrightarrow \gamma + \gamma ~ .
\end{equation}
In the expression above, $X$ represents all the other products of the interaction, possibly including more pions.
The process is characterised by an energy threshold $E_{th}$, which can be easily estimated by assuming that the final products of Eq.~\ref{eq:pp} are two protons and a neutral pion at rest, i.e. the minimum energy configuration.
It is straightforward to show that only CR protons of kinetic energy larger than $E_p^{th} = 2 m_{\pi^0} + m_{\pi^0}^2/2 m_p \approx 280$~MeV are able to generate neutral pions (here and in the following of the Section, $c = 1$). 
Neutral pions are also produced in collisions involving nuclei heavier than hydrogen, which are present in both CRs and in the ISM.
Taking this contribution into account enhances the predictions of the gamma-ray signal by a factor of $\epsilon_{N} \lesssim 2$, which depends weakly on the spectrum of the incident CR nuclei \citep{mori2009,kafexhiu2014}.

As gamma rays can pass through large column densities of matter without being absorbed, the implications of Haykawa's proposal are far reaching: gamma-ray observations allow us to observe, indirectly, CR protons and nuclei of energy larger than few hundreds MeV located in remote regions of the Galaxy.
Hayakawa's conjecture was confirmed in the late sixties, when gamma rays of energy exceeding 100 MeV were observed from a band in the sky spatially coincident with the Galactic disk \citep{clark1968}.
The diffuse emission from the Galactic disk is, in fact, the most prominent feature in the gamma-ray sky in the GeV domain, and has been measured over the years with ever-improving accuracy, most recently by the Fermi/LAT space telescope \citep{ackermann2012}.
The spectral shape of the emerging spectrum of gamma rays from inelastic proton proton interactions can be computed from pion production models and/or fits to accelerator data (see e.g. \citealt{stecker1971} and the comprehensive bibliography in \citealt{kafexhiu2014}), and a number of convenient analytic parameterisations of gamma-ray production spectra can be found in the literature \citep[e.g.][]{kelner2006,kamae2006,kafexhiu2014,koldobskiy2021}.
Remarkably, the main features of such gamma-ray spectra can be inferred from the following two simple arguments.

First of all, in the rest frame of the neutral pion, the two gamma rays produced in the decay will have the same energy $E_{\gamma}^* = m_{\pi^0}/2 \sim 70$~MeV and, to conserve momentum, they will move along opposite directions.
After Lorentz transforming to the lab frame, the photon energies are boosted to $E_{\gamma}^{\pm} = \gamma (E_{\gamma}^*\pm\beta p_{\gamma}^* \cos \vartheta^*)$, where $\gamma$ and $\beta$ are the Lorentz factor and velocity of the pion, and $p_{\gamma}^*$ and $\vartheta^*$ the momentum of the photons and the decay angle in the pion rest frame.
In the absence of any preferred direction, the distribution of the decay angles in the pion frame has to be isotropic.
This implies that, for an ensemble of pions of a given energy, the emerging spectrum of gamma rays in the lab frame is flat, i.e. ${\rm d} n_{\gamma}/{\rm d} E_{\gamma} \propto {\rm d} n_{\gamma}/{\rm d} \cos \vartheta^* = 0$, in the energy interval defined by $\cos \vartheta^* = \pm 1$. 
If $E_{\gamma}^{max}$ and $E_{\gamma}^{min}$ are the maximum and minimum energies of the photon distribution, corresponding to $\cos \vartheta^* = $ +1 and -1, respectively, then their geometric mean is always equal to $\sqrt{E_{\gamma}^{max} E_{\gamma}^{min}} = m_{\pi^0}/2$, regardless of the pion energy.
It follows that, for pions characterised by an arbitrary distributions of energies, the gamma ray spectrum will be the superposition of many flat spectra which are all symmetric, when plotted versus $\log ( E_{\gamma} )$, with respect to the logarithm of the geometric mean $\log ( m_{\pi^0}/2 )$.
This symmetry is a distinctive feature of gamma-ray spectra from pion decay, and the peak that invariably appears in the spectrum at $E_{\gamma}^* = m_{\pi^0}/2 \sim 70$~MeV is called \textit{pion bump}.

The second important aspect of proton-proton interactions is that, even though a large number of neutral pions can be produced in each collision, for large enough proton energies ($\gg E_p^{th}$) one leading pion carries away the main fraction (of the order of $\approx$~20\%) of the CR proton kinetic energy.
Moreover, in the same limit, the cross section of inelastic scattering depends quite weakly on the energy of the incident particle.
Therefore, if the energy spectrum of CR protons is a power law $E_p^{-s_p}$, then also the pion and the gamma-ray spectra will be power laws $E_{\pi, \gamma}^{-s_{\pi, \gamma}}$ with similar slopes $s_p \sim s_{\pi} \sim s_{\gamma}$.
In fact, a more accurate analysis shows that the gamma-ray spectrum does not follow a pure power law, but progressively hardens above $\sim 100$ GeV, where $s_{\gamma} \approx s_p-0.1$ \citep{kelner2006}.
Finally, well above the threshold ($E_p \gg E_p^{th}$), the energy of the gamma ray is linked to that of the parent CR proton as $E_{\gamma} \approx 0.1 \times E_p$.

\begin{figure*}
% Use the relevant command to insert your figure file.
% For example, with the graphicx package use
\center
  \includegraphics[width=0.7\textwidth]{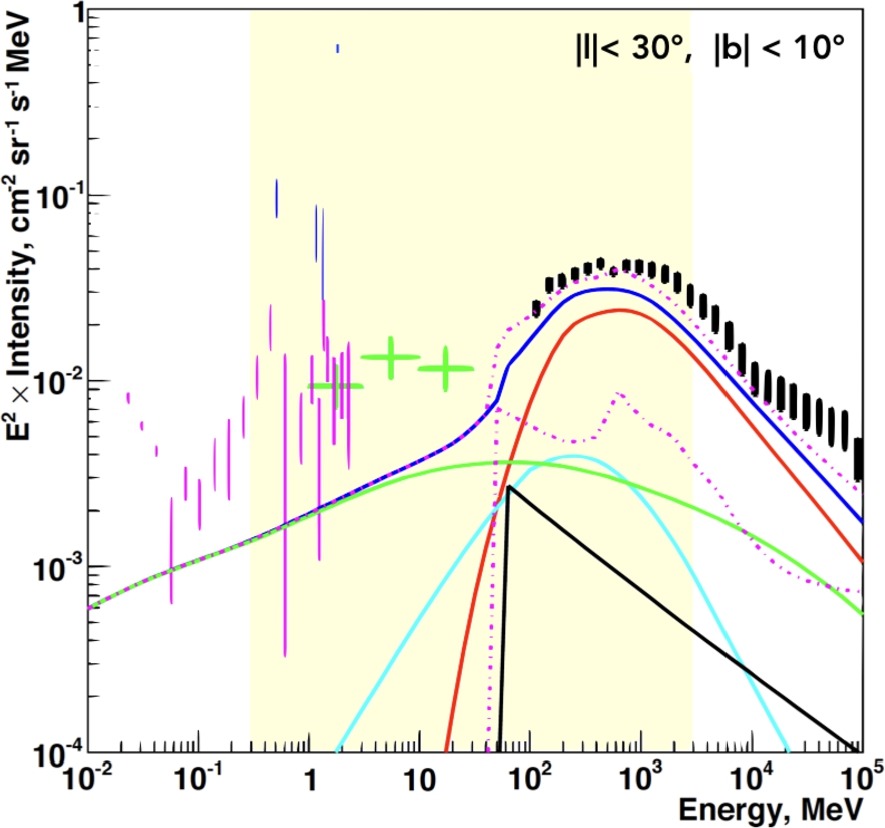}
% figure caption is below the figure
\caption{Hard-X and gamma-ray emission from the inner Galaxy ($|l| < 30^{\circ}$ and $|b| < 10^{\circ}$). Curves show the expected contribution from sources (lower dot-dashed magenta line), diffuse  emission (blue line), and total (upper dot-dashed magenta line). The diffuse emission is the sum of the contributions from $\pi^0$ decay (red), inverse Compton scattering (green), Bremsstrahlung (cyan), isotropic extragalactic (black). Data are from  Fermi (black), INTEGRAL (magenta and blue), and COMPTEL (green). Figure from \citet{deangelis2018}.}
\label{fig:andy}       % Give a unique label
\end{figure*}

%Consider now the diffuse emission from low Galactic latitudes.
Fig.~\ref{fig:andy} shows the diffuse emission observed from the inner part of the Galactic disk (data points) over a very broad energy range, spanning from the hard X-ray to the high-energy gamma-ray domain.
The emission has been extracted from a region around the Galactic centre, defined by Galactic coordinates $|b| < 10^{\circ}$ and $|l| < 30^{\circ}$.
Data points come from observations performed by INTEGRAL and COMPTEL in the hard X-ray/soft gamma-ray domain (magenta, blue, and green points) and by Fermi (black points) in the high energy gamma-ray band \citep{strong2011}.
The observed diffuse emission has been fitted by a multi-component model, that will be described in more detail below.
For the moment, we only anticipate that in the GeV domain the emission is largely dominated by photons from neutral pion decay, represented by the solid red line: a striking validation of Hayakawa's early prediction!
In the Figure, the spectrum has been multiplied by the photon energy squared, to show the {\it spectral energy distribution}.
In this convenient representation, the spectrum of gamma-ray photons from proton-proton interactions is no longer symmetric with respect to $E_{\gamma}^*$, and its peak is shifted to $\lesssim$ GeV energies.

Besides the spectrum, also the morphology of the GeV diffuse emission from the disk carries important information, as it is shaped by the spatial distribution of both CRs and interstellar matter.
As the latter can be derived by a number of observations \citep{ferriere2001}, the former can be constrained by gamma-ray observations.
In the following, we will briefly describe how this can be done, starting with a review of the available estimates of the local (within $\approx 1$ kpc) intensity of CRs, and proceeding then with a discussion of various approaches aimed at probing the remote regions of the Galaxy.

\subsubsection{Constraints on the cosmic ray intensity from diffuse gamma rays}
\label{sec:diffuse}

The expected {\it local} gamma-ray emissivity due to neutral pion decay can be estimated starting from the CR proton spectrum measured in the local ISM (Eq.~\ref{eq:voyager}).
It represents the number of gamma-ray photons emitted per unit energy, time, and per interstellar hydrogen atom.
It is defined as \citep{stecker1971}:
\begin{equation}
\label{eq:qLIS}
q_{\gamma}^{LIS}(E_{\gamma}) = 4 \pi \int_{E_{\gamma}}^{\infty} {\rm d} E_p \sigma(E_{\gamma} | E_p) j_{LIS}(E_p) 
\end{equation}
where $\sigma(E_{\gamma} | E_p)$ is the differential cross section (parameterised in, e.g. \citealt{kafexhiu2014}) describing the probability to have an interaction where a proton of energy $E_p$ produces a gamma ray of energy $E_{\gamma}$.
In general, the intensity of CR protons at a position $\bf{r}$ in the Galaxy, $j(E_p,\bf{r})$, will differ from the local one (${\bf r} = 0$), and the gamma-ray emissivity $q_{\gamma}(E_{\gamma},{\bf r})$ will change accordingly.
The observed intensity of gamma rays from a given direction in the sky is then obtained after integrating along the line of sight:
\begin{equation}
\label{eq:diffuse}
j_{\gamma}(E_{\gamma},l,b) = \frac{\epsilon_N}{4 \pi} \int {\rm d} r ~ q_{\gamma}(E_{\gamma},r) n_{ISM}(r)
\end{equation}
where $l$ and $b$ are the Galactic longitude and latitude, respectively, and $n_{ISM}(r)$ is the hydrogen (atomic plus molecular) density at a distance $r$ along the line of sight.
The spatial distribution of interstellar matter $n_{ISM}(\bf{r})$ can be determined, with some uncertainties, from astronomical observations \citep{ferriere2001,cox2005}.
Then, if the intensity of the diffuse gamma-ray emission $j_{\gamma}(E_{\gamma},l,b)$ is measured, constraints on the spatial distribution of CRs throughout the Galaxy can be obtained.

In order to check the prediction given in Eq.~\ref{eq:qLIS}, it is convenient to observe the diffuse gamma-ray emission at large Galactic latitudes.
This is because, due to the small thickness $h$ of the gaseous Galactic disk (few hundreds parsecs), the gamma-ray emission from such latitudes is dominated by the contribution of CR interactions taking place in the solar neighbourhood, i.e. within a distance $\sim h/(2 \sin b)$.
Under these circumstances, Eq.~\ref{eq:diffuse} can be simplified by making the substitution $q_{\gamma}(E_{\gamma},{\bf r}) \longrightarrow q_{\gamma}^{LIS}(E_{\gamma})$, which is acceptable provided that there are no large spatial gradients in the very local distribution of CRs.
This gives:
\begin{equation}
\label{eq:highlat}
j_{\gamma}(E_{\gamma},l,b) \approx  \frac{\epsilon_N}{4 \pi} q_{\gamma}^{LIS} (E_{\gamma}) \int {\rm d} r ~ n_{ISM}(r) = \frac{\epsilon_N}{4 \pi} q_{\gamma}^{LIS} (E_{\gamma}) N_H(l,b)
\end{equation}
where $N_H$ is the hydrogen column density.
This approximate expression shows how, from the measurement of the diffuse gamma-ray emission and of the gas column density, it is possible to obtain an observational estimate of $q_{\gamma}^{LIS}(E_{\gamma})$, that can be in turn compared with its predicted value (Eq.~\ref{eq:qLIS}).
This can be done to a good accuracy because, as we said above, the neutral pion decay contribution to the diffuse emission from the Galactic disk exceeds that from other radiation mechanisms for photon energies exceeding $\sim 100$ MeV (Fig.~\ref{fig:andy}).

\citet{casandjian2015} analysed Fermi data to extract the diffuse gamma-ray emission at large Galactic latitudes in the range $10^{\circ} < |b| < 70^{\circ}$.
Most of this emission is generated by proton-proton interactions taking place within a distance of roughly a kiloparsecs from the Sun.
The contribution to the diffuse gamma-ray emissivity coming from molecular, atomic, and ionised hydrogen can be separated as the gas column densities of these three components of the ISM are known with fair accuracy.
Once this is done, the derived interstellar gamma-ray emissivity for atomic hydrogen is taken as the most reliable, as the spatial distribution of such component of the ISM, traced by the 21 cm line, is known with high precision.
An estimate of the local spectrum of CRs can then be obtained from such emissivity.
Besides the complications in the determination of the gas column densities, such method also suffers from uncertainties in both the knowledge of hadronic cross sections, and in the determination of the contribution from leptonic processes (see Sec.~\ref{sec:leptonic}) to the gamma-ray emission.
While different values for the gas column densities of the phases of the ISM would shift the overall normalisation of the resulting CR spectrum in an almost energy independent way, the latter two effects have a greater impact at low energies  \citep[see the recent review by][for a more extended discussion of these issues]{tibaldo2021}.

The gamma-ray-based approach just described is important as it probes the local spectrum of CRs in the energy region where direct observations are not available. 
For CR protons, this spans from the highest energy data point obtained by the Voyagers, at about 300 MeV, to few tens of GeV, i.e. the minimum energy for which CR measurements performed inside the heliosphere are unaffected by Solar modulation.
Remarkably, a comparison between direct and indirect measurements of the CR intensity can tell us whether a gradient in the spatial distribution of CRs exists or if CRs are uniformly distributed in a kpc neighbourhood of the Solar system.
A strong spatial gradient would indicate the presence of one or more local sources of CRs, and would put into question the scenario for CR transport and escape from the Galaxy developed in Sec.~\ref{sec:CRs}, which is based on the assumption that the local spectrum of CRs can be taken as a proxy for the typical one in the entire disk.

\begin{figure*}
% Use the relevant command to insert your figure file.
% For example, with the graphicx package use
\center
  \includegraphics[width=0.49\textwidth]{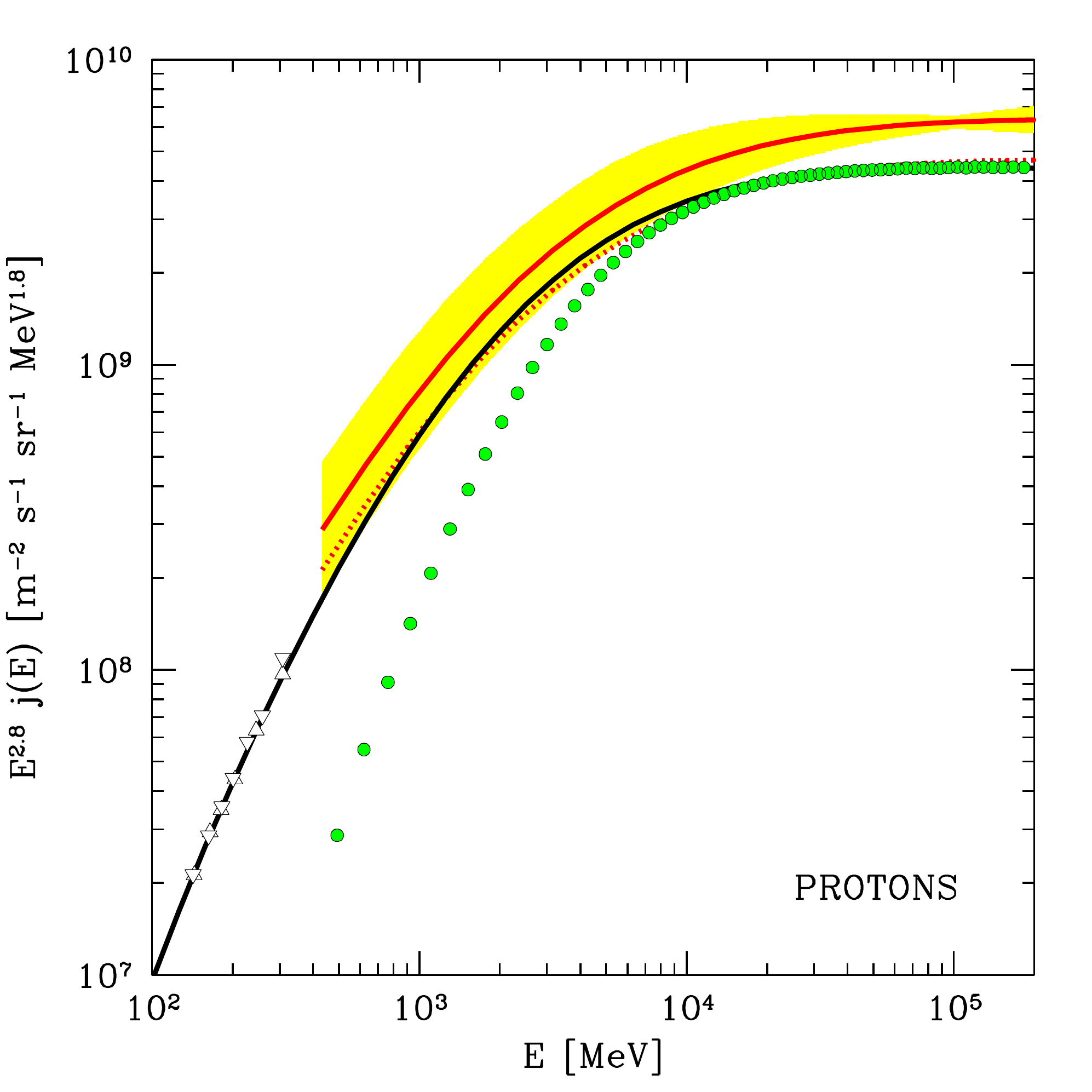}
   \includegraphics[width=0.49\textwidth]{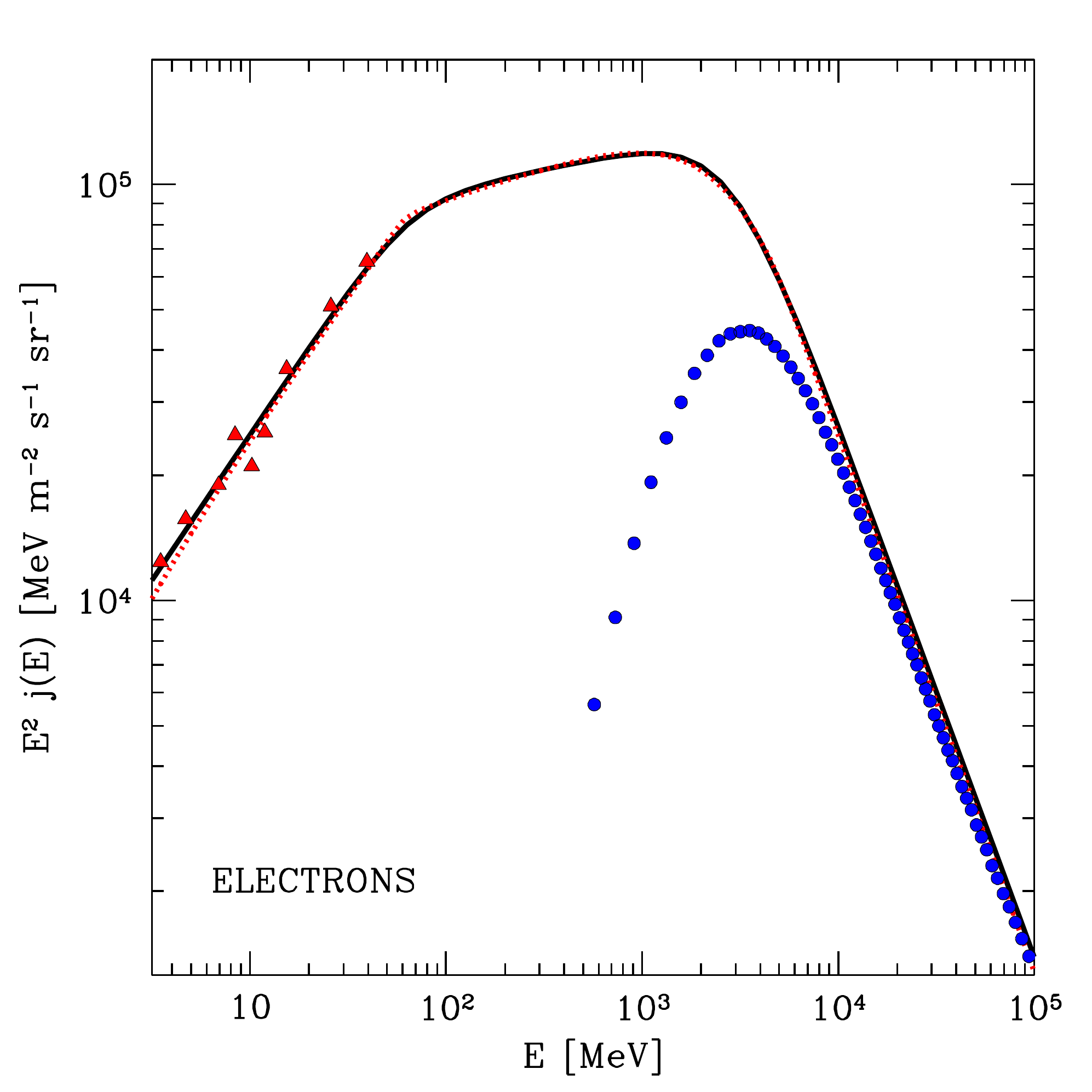}
% figure caption is below the figure
\caption{{\bf Left:} Direct measurements of CR protons (white triangles: Voyager 1 and 2; green circles: AMS-02) are compared with indirect constraints from gamma-ray observations of the local ($\sim$ 1 kpc) ISM (the red solid line, red dotted line, and yellow shaded region are the best fit, the best fit divided by 1.35, and the uncertainty, respectively, as in \citealt{strong2015}). {\bf Right:} Direct measurements of CR electrons (red triangles: Voyager 1; blue circles: AMS-02) are compared with constraints from radio and gamma-ray observations of the local ($\sim$ 1 kpc) ISM and from local CR observations (red dotted line from \citealt{orlando2018}). In both panels, the black line represents the parameterisation adopted in this review.}
\label{fig:LISpe}       % Give a unique label
\end{figure*}

Building on the early work by \citet{casandjian2015}, \citet{strong2015} improved the analysis technique and adopted more updated cross sections, and \citet{orlando2018} revised the estimate of the leptonic contribution to the gamma-ray emission.
The results from these studies indicate that at high enough particle energies, where the effects of Solar modulation are negligible and hadronic models are most accurate, the spectrum derived from gamma-ray observations exceeds by $\approx 30-40$\% direct measurements.
This is shown in the left panel of Fig.~\ref{fig:LISpe}, where direct observations of CRs are compared with indirect ones.
In the plot, the local interstellar spectrum of CRs measured by the Voyagers (white triangles) is shown together with the near Earth observations by AMS-02 (green circles).
The solid red line represents the best fit CR spectrum derived by the local ($\lesssim$ 1 kpc) gamma-ray emissivity, with its uncertainty shown as a yellow shaded region \citep{strong2015}.
If the best fit model is shifted downward by a factor of 1.35 (red dotted curve), it smoothly connects the Voyager data points to the ones from AMS-02.
The black solid line in the Figure represents the expression given in Eq.~\ref{eq:voyager} for $s = 2.1$.
For such a choice of the parameter $s$, the expression constitutes a good description not only of direct measurements at low (Voyagers) and high (AMS-02) energies, but also reproduces the correct shape of the spectrum (though with a slightly different normalisation) in the GeV domain, as derived from gamma-ray observation of the local ISM.

\begin{figure*}
% Use the relevant command to insert your figure file.
% For example, with the graphicx package use
\center
  \includegraphics[width=0.7\textwidth]{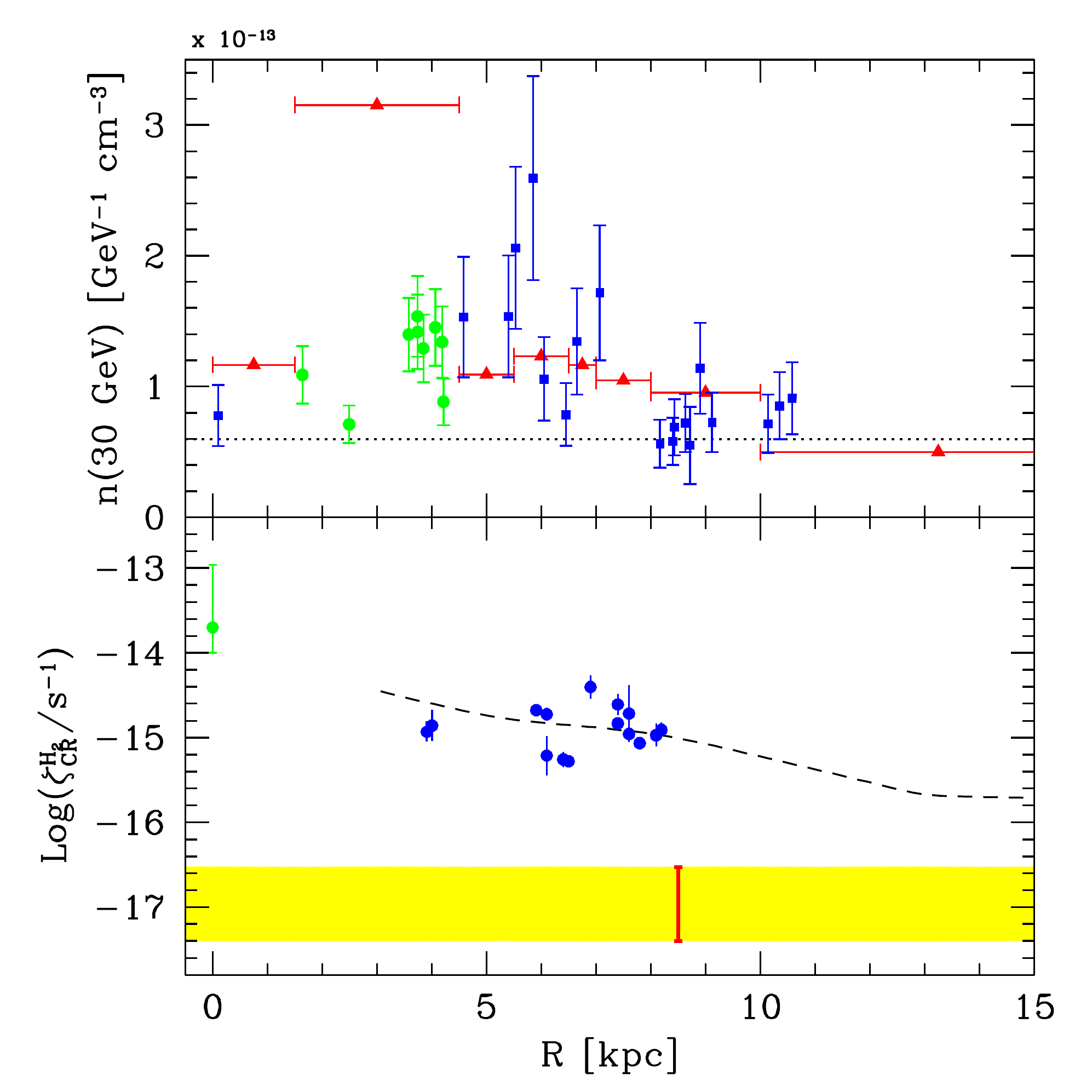}
% figure caption is below the figure
\caption{{\bf Top:} Normalisation of the CR proton spectrum at 30 GeV as a function of Galactocentric distance as derived by gamma-ray observations. Red, green, and blue data points are from \citealt{acero2016}, \citealt{peron2021} and \citealt{aharonian2020}, respectively. The horizontal dotted line indicates the value measured locally by AMS-02 \citep{aguilar2015}. Figure adapted from \citet{peron2021}. {\bf Bottom:} CR ionisation rate versus Galactocentric distance. Data are from \citet{lepetit2016} and \citet{oka2019} (green point), \citet{neufeld2017} (blue points). The yellow shaded region shows the expected value for the local interstellar spectrum of CRs, after the effects of the penetration into a diffuse cloud are taken into account \citep{phan2018}. The red bar shows the position of the Sun. The dashed black line shows the profile predicted by \citet{wolfire2003} renormalised to fit the blue data points.}
\label{fig:profiles}       % Give a unique label
\end{figure*}

At low Galactic latitudes, the interpretation of the observations of the GeV diffuse gamma-ray emission is less straightforward, as in this case the approximation made to derive Eq.~\ref{eq:highlat} is no longer valid, and the full expression given in Eq.~\ref{eq:diffuse} must be adopted.
Nevertheless, both the diffuse emission $j_{\gamma}(E_{\gamma},l,b)$ and the space-dependent interstellar gas density $n_{ISM}({\bf r})$ can be determined from observations, turning Eq.~\ref{eq:diffuse} into an integral equation for the CR intensity $j_p(E_p,{\bf r})$.
The inversion of the integral equation is customarily performed dividing the Galactic disk in concentric rings around the Galactic centre, and imposing that the intensity of CR nuclei is spatially uniform within each ring.
The normalisation and spectral shape of the CR intensity are left as adjustable parameters and may differ in each ring.
Their best fit values are obtained by comparing the expected and  the observed maps of the diffuse gamma-ray emission by means of a likelihood analysis, whose details can be found in, e.g., \citet{acero2016}, \citet{yang2016} and \citet{pothast2018}.
The results obtained by \citet{acero2016} for the density of CR protons as a function of the galactocentric radius $R$ are shown as red data points in the top panel of Fig.~\ref{fig:profiles}.
The values of the CR density obtained in this way for $R \lesssim 15$ kpc differ from each other by less than a factor of $\sim$ 2.5, with the sole exception of the data point centred at $R = 3$ kpc, which stands a factor of $\sim$~3 above neighbouring points.
It has been suggested that this excess might be related to an enhanced CR production in the molecular ring, a region positioned roughly half-way between the Galactic centre and the Sun, where both the distribution of interstellar molecular hydrogen and the star formation rate per unit surface of the disk peak \citep{acero2016}.
Note also that the CR density inferred within the ring that contains the Solar system (that centred at $R = 9$ kpc) exceeds by a factor of $\sim$ 1.6 the value of the local CR density as measured by AMS-02, thereby confirming the earlier claim by \citet{strong2015} that we discussed above.
It seems, then, that large scale variations in the density of CRs are quite mild in the Galactic disk.

\subsubsection{Gamma rays from MCs}
\label{sec:MCs}

The procedure of inversion of Eq.~\ref{eq:diffuse} just described can only provide the intensity of CRs averaged over quite large volumes of the disk, for example a ring centred at a Galactocentric radius $R$ and of width $\Delta R$ %, $\langle j_p(E_p) \rangle (R, \Delta R)$, with $\Delta R$ 
of the order of a kiloparsecs or more (see the red error bars in the top panel of Fig.~\ref{fig:profiles}).
Interestingly, gamma-ray observations of MCs can be used to probe the density of CRs within much smaller volumes, as these objects' typical radii range from parsecs to tens of parsecs \citep{heyer2015}.
The other distinctive characteristic of MCs is their large mass, which provides abundant target for CR proton-proton interactions.  
For this reason, MCs were proposed as potential discrete sources of gamma rays in a seminal paper by \citet{black1973}.
Of particular interest are {\it giant} MCs, i.e. those with masses $M_{cl}$ exceeding $\sim 10^4 M_{\odot}$.
Their masses and radii correlate in such a way that  such objects are characterised by a typical column density $N_H \gtrsim10^{22}$~cm$^{-2}$ \citep{mckee2007}.
This is of the same order of the total average gas column density measured along lines of sight in the direction of the inner Galaxy, which explain why the gamma-ray emission from giant MCs is expected to be seen as an excess above the diffuse interstellar emission, provided that the intensity of CRs is similar in the diffuse ISM and inside clouds.

The gamma-ray luminosity from a MC of mass $M_{cl}$ can be computed by integrating Eq.~\ref{eq:diffuse} over the volume of the cloud, and its flux then reads:
\begin{equation}
\label{eq:MC}
F_{\gamma}^{cl}(E_{\gamma}) = \epsilon_N \frac{q_{\gamma}^{cl}(E_{\gamma})}{4 \pi \mu m_p} \left( \frac{M_{cl}}{d_{cl}^2} \right)
\end{equation}
where $d_{cl}$ is the distance to the cloud, and $\mu \sim 1.4$ accounts for the presence of helium in the cloud.
In deriving the expression above, we implicitly assumed that the spatial distribution of CRs does not vary significantly over the scale of the cloud, that is to say, the intensity of CRs is not attenuated due to ionisation energy losses as they go through the cloud's large column density (note that the entire Sec.~\ref{sec:penetration} will be devoted to the issue of CR penetration into MCs).
We also assumed that the hadronic emission dominates the gamma-ray flux, a condition easily satisfied if photon energies larger than few hundreds of MeV are considered \citep[see e.g.][]{gabici2007,gabici2009}.
Under these conditions, $q_{\gamma}^{cl}(E_{\gamma})$ is the hadronic gamma-ray emissivity at any location inside the MC.

Eq.~\ref{eq:MC} shows that if the mass and distance of a MC are known, the gamma-ray emissivity per hydrogen atom $q_{\gamma}^{cl}(E_{\gamma})$ can be directly estimated from its gamma-ray flux $F_{\gamma}^{cl}(E_{\gamma})$.
The density and energy spectrum of CR protons inside the cloud can be in turn derived from the gamma-ray emissivity. 
This is why MCs have been sometimes referred to as {\it cosmic ray barometers}: we can use them to probe the density and pressure of CRs in remote regions of the Galaxy \citep{issa1981,aharonian1991,casanova2010}.
The uncertainties in the measurements of the cloud mass and distance are the main limitations of this approach.

MCs are almost entirely made of molecular hydrogen, H$_2$, which unfortunately is difficult to detect\footnote{The main problem is that even the lowest (rotational) excited energy level is too far from the ground state to be significantly populated, given the cold environments where H$_2$ is generally found \citep[see][for a detailed discussion]{stahler2004}.}.
For this reason, the mass of a cloud is often estimated from observations of the second most abundant molecule in space, carbon monoxide (CO).
The CO emission line corresponding to the rotational transition $J = 1 \rightarrow 0$ falls in the radio domain ($\lambda = 2.6$ mm) and is the prime tracer of molecular gas \citep{heyer2015}.
As the transition is optically thick, the CO brightness temperature integrated over the line profile, $W_{\rm CO}$, is not proportional to the CO column density, but rather to the cloud velocity dispersion.
For virialized clouds, the velocity dispersion is related to their total mass, which is dominated by H$_2$ \citep{scoville1987}.
The proportionality constant between the H$_2$ column density $N({\rm H}_2)$ and the CO integrated line intensity $W_{\rm CO}$ is usually called the $X_{CO}$ % = N_{H_2}/W_{CO}$ 
factor and has been determined empirically, though with a significant average (mediated over large scales) uncertainty of $\approx 30$\%, with individual clouds fluctuating around that value \citep{bolatto2013}.
The infrared emission from dust grains mixed with the gaseous ISM is another tracer of interstellar matter, providing complementary estimates of the cloud masses. 
The advantage of this approach is that dust emission is optically thin, and therefore directly traces the total gas mass of the cloud.
However, it also suffers from uncertainties and cloud-to-cloud variations \citep[e.g.][]{remy2017,tibaldo2021}. 
Finally, the observed frequency of the CO line (or any other line)  is doppler shifted due to the differential rotation of the Galaxy, a thing that can be used to estimate the distance of a cloud, provided that it is located within the solar circle \citep{scoville1987,ferriere2001}.

A large number of studies on local MCs were aimed at determining their CR content by combining gamma-ray observations and mass estimates (see e.g. \citealt{issa1981} for an early attempt, and the list of recent references compiled by \citealt{tibaldo2021}).
The gamma-ray emissivities derived in this way for clouds located within few hundred parsecs from the Solar system show a moderate scattering around that derived by \citet{casandjian2015} using diffuse gamma rays (see e.g. Fig.~10 in \citealt{remy2017} or  Fig.~2 in \citealt{tibaldo2021}), or around the emissivity computed from AMS-02 data (see blue data points around $R \sim 8.5$ kpc in the upper panel of Fig.~\ref{fig:profiles}, or \citealt{aharonian2020}).
Given the intrinsic uncertainty in the determination of the matter distribution in our neighbourhood, it is not possible to claim whether direct and indirect measurements of the local interstellar spectrum of CR protons agree or not. 
However, if a difference exists, it must be quite small, as data constrain local spatial variations within few tens percent \citep{tibaldo2021}.

The blue and green data points in the upper panel of Fig.~\ref{fig:profiles} show the density of CRs as derived from observations of both local and remote MCs \citep{aharonian2020,peron2021}.
Overall, they broadly agree with the results of studies of the diffuse emission (red data points in the same Figure), except for the region around the molecular ring (green points in the Figure).
Values of the CR density extracted from MC observations are closer to direct AMS-02 measurements with respect to those derived from the diffuse gas.
In fact, this might not be a problem, as MCs probe the intensity of CRs at very specific locations, while studies of the  diffuse gas provide a volume average quantity \citep[for an extended discussion of this issue see][]{peron2021}.

In any case, independently on all these complications, a claim that can be certainly made is that the spatial distribution of CR protons of multi-GeV energies varies very little on large spatial scales.
More quantitatively, variations are constrained to be at most a factor of two from the value measured locally by AMS-02, with the only exception of the molecular ring.
However, also in that case the variation would be of at most a factor of $\approx$ 5.

\subsubsection{The Galactic origin of cosmic rays: the supernova remnant paradigm}
\label{sec:SNR}

Having determined the spatial distribution of CR nuclei in the ISM, we can now estimate their total energy in the Galactic disk.
As seen in the previous Section, the gradient of the large scale distribution of CRs is quite modest.
Therefore, an order of magnitude estimate of the CR energy in the disk is given by $W_{CR} \approx w_{CR}^{LIS} \times V_{disk} \approx 2 \times 10^{55}$~erg, where $w_{CR}^{LIS} \approx 1$~eV/cm$^3$ is the energy density of CR nuclei in the local ISM (Sec.~\ref{sec:voyager}), and $V_{disk} \approx 400$~kpc$^3$ is the volume of the Galactic disk of radius 15 kpc and thickness 500 pc.

The largest contribution to the CR total energy comes from nuclei of energy $\gtrsim$~1 GeV/nucleon (Fig.~\ref{fig:Hee}).
Before escaping the Galaxy, particles having such energy spend $\tau_{ISM} \lesssim$~10 Myr in the ISM (Sec.~\ref{sec:B/C}). 
To compensate for the escape of CRs and maintain stationarity, sources located in the disk have to inject in the ISM a power $P_{CR} \sim W_{CR}/\tau_{ISM} \lesssim 10^{41}$~erg/s in form of energetic particles.
This estimate is admittedly very rough, but provides a result consistent with accurate studies \citep[see e.g.][]{strong2010}.

Following a seminal idea proposed by \citet{baade1934}, it was soon realised that Galactic supernovae could provide the amount of energy needed to explain CRs \citep{terhaar1950}.
The rate $\nu_{SN}$ of supernova explosion in the Galaxy is equal to about three per century, and each explosion releases $E_{SN} \sim 10^{51}$~erg in form of kinetic energy of the stellar ejecta.
Therefore, the total power injected by supernovae in the Galaxy is $P_{SN} = \nu_{SN} E_{SN} \approx 10^{42}$~erg.
 It follows that  about $\lesssim$~10\% of the energy from supernovae has to be converted into energetic particles in order to explain CRs.

With the advent of radio observations it was realised that the remnants of supernova explosions, and not the explosions themselves, were likely to be better candidate as sources of CRs.
Supernova remnants are the expanding shocks that form in the ISM as a result of the stellar explosion, and the synchrotron radio emission detected from these objects implies the presence of magnetic fields and relativistic electrons there.
Finally, in the seventies, the theory of diffusive shock acceleration was developed, independently, by four different teams \citep{krymskii1977,axford1977,blandford1978,bell1978}, showing that strong shocks are indeed expected to accelerate particles with an universal spectral shape: a power law in energy of slope $E^{-2}$.

As the slope predicted from diffusive shock acceleration theory is quite close (though not equal) to what is needed to explain CR observations (slopes in the range 2.1 ... 2.4 are compatible with observations, see Sec.~\ref{sec:diffusive}) a scenario where CRs are accelerated at supernova remnant shocks became so popular that is nowadays considered a paradigm \citep[see][and references therein]{blasi2013}. 
However, two things should be kept in mind: first of all, the supernova remnant paradigm for the origin of CRs remains, in fact, an hypothesis \citep{gabici2019} and, second, the energy-budget argument provided above holds for particles of energy around $\approx$~1 GeV, which are those who dominate the energetic.
It follows that the origin of LECRs of energy significantly smaller than 1 GeV could be different than that of higher energy particles.
We will discuss this issue in Sec.~\ref{sec:questions}.

\subsection{Leptonic gamma rays and synchrotron emission}
\label{sec:leptonic}

\subsubsection{Inverse Compton scattering and relativistic Bremsstrahlung}

Energetic photons produced in interactions of CR electrons in the ISM also contribute to the diffuse gamma-ray emission from the Galactic disk.
The two dominant radiation mechanisms are relativistic Bremsstrahlung and inverse Compton scattering \citep{blumenthal1970}.
%Also CR electrons can produce gamma rays, mainly through two channels: inverse Compton scattering and non-thermal Bremsstrahlung.
Unlike hadronic (neutral pion decay) emission, which dominates the spectral energy distribution in the high energy gamma-ray domain, leptonic processes are characterised by much broader energy spectra, extending down to the soft gamma-ray (MeV) and hard X-ray (keV) domains (Fig.~\ref{fig:andy}).

In inverse Compton scattering, a relativistic electron of mass $m_e$, Lorentz factor $\gamma$, and energy $E_e = \gamma m_e c^2$ scatters off a soft ambient photon of energy $\epsilon$. 
As a result, the latter is boosted to a much larger energy $E_{\gamma}$.
The energy of the soft photon seen from the rest frame of the electron can be computed performing a Lorentz transformation and is of the order of $\epsilon^{\prime} \sim \gamma \epsilon$.
Then, provided that $\gamma \epsilon \ll m_e c^2$, in that rest frame the scattering takes place in the Thomson (elastic) regime, and the energy of the scattered photon remains $\epsilon^{\prime}$.
Lorentz transforming back to the lab frame, one can see that the photon energy after the scattering is boosted to $E_{\gamma} \sim \gamma^2 \epsilon$.
In the expressions above, we neglected all the (small) factors depending on the angle between the velocity of the incident/scattered photons and that of the electron.
Averaging over these angles one gets the exact result for the typical energy of the scattered photon: $E_{\gamma} = (4/3) \gamma^2 \epsilon$ \citep{blumenthal1970}.

In the Thomson regime, a relativistic electron moving through a radiation field made of $n_{rad}$ photons per cubic centimeter would scatter off $\sigma_T n_{rad} c$ photons per second, where $\sigma_T$ is the Thomson cross section.
Taking $\langle \epsilon \rangle$ to be the mean photon energy and $w_{rad} = n_{rad} \langle \epsilon \rangle$ to be the radiation field energy density, the rate at which energy is lost by the electron reads: 
\begin{equation}
\label{eq:ICrate}
P_{IC} = -\frac{{\rm d}E_e}{{\rm d}t} = \left( \sigma_T n_{rad} c \right) \left( \frac{4}{3} \gamma^2 \langle\epsilon\rangle \right) = \frac{4}{3} \sigma_T c \gamma^2 w_{rad}
\end{equation}

The energy spectrum of the inverse Compton emission produced by a population of electrons with a power law spectrum in energy ${\rm d}n_e/{\rm d} E_e \propto E_e^{-\delta}$ (number of particles per unit volume and unit energy) is itself a power law in photon energy $q_{\gamma}^{IC}(E_{\gamma}) \propto E_{\gamma}^{-s}$.
The expected slope of the gamma-ray spectrum can be estimated from the approximate expression:
\begin{equation}
\label{eq:ICspectrum}
E_{\gamma} q_{\gamma}^{IC}(E_{\gamma}) \approx \frac{{\rm d}n_e}{{\rm d} E_e} P_{IC} \frac{{\rm d} E_e}{{\rm d}E_{\gamma}} \propto E_{\gamma}^{-\frac{\delta-1}{2}} ~,
\end{equation}
where we implicitly assumed that electrons of Lorentz factor $\gamma$ interacting with soft photons of mean energy $\langle \epsilon \rangle$ generate gamma-ray photons with a delta function spectrum centred at $E_{\gamma} = (4/3) \gamma^2 \langle \epsilon \rangle$.
From Eq.~\ref{eq:ICspectrum} it follows that $s = (\delta+1)/2$  \citep[see][for a formal derivation of this result]{blumenthal1970}.

The interstellar radiation field pervades the Galactic disk and provides a background of soft photons for inverse Compton scattering.
It is dominated by three components: the starlight emission, characterised by an energy density in the local ISM equal to $w_* \sim 0.43$ eV/cm$^3$ and an average photon energy of $\epsilon_* \sim 0.97$ eV, the emission from dust heated by starlight, with an energy density of $w_d \sim 0.19$ eV/cm$^3$ and average photon energy $\epsilon_d \sim 7.8 \times 10^{-3}$ eV and of course the ubiquitous cosmic microwave background radiation (CMB), whose energy density is $w_{CMB} \sim 0.26$ eV/cm$^3$ and average photon energy $\epsilon_{CMB} \sim 6.3 \times 10^{-4}$ eV \citep[e.g.][and references therein]{vernetto2016}.
If $E_e = \gamma m_e c^2$ is the energy of the relativistic electron, interstellar photons undergoing inverse Compton scattering will be boosted to typical energies
\begin{equation}
\label{eq:ICboost}
E_{\gamma} = 5.1 \left( \frac{E_e}{{\rm GeV}} \right)^2  \left( \frac{\epsilon}{{\rm eV}} \right) ~ \rm MeV ~ .
%&=& 4 \left( \frac{E_e}{10~{\rm GeV}} \right)^2  \left( \frac{\epsilon_d}{7.8 \times 10^{-3}~{\rm eV}} \right) ~ \rm MeV \\
%&=& 0.3 \left( \frac{E_e}{10~{\rm GeV}} \right)^2  \left( \frac{\epsilon_d}{6.3 \times 10^{-4}~{\rm eV}} \right) ~ \rm MeV
\end{equation}
Then, electrons of energy in the range $\approx$ 100 MeV-10 GeV (which is not probed by direct observations) will upscatter photons in a very broad band spanning four decades in energy, and centred around $E_{\gamma} \approx$ 5 MeV, 40 keV, and 3 keV for starlight, dust emission, and CMB target photons, respectively.
%to typical energies of $\sim 5 \times 10^2$, 4, and 0.3 MeV for starlight, dust emission, and CMB target photons, respectively.
This shows that observations in the soft gamma-ray and hard X-ray domains can be used to probe the spectrum of CR electrons in the energy region between those probed by Voyager and AMS-02.

Relativistic Bremsstrahlung radiation is produced when a relativistic electron of Lorentz factor $\gamma$ is accelerated in the Coulomb field of an atomic nucleus of the ISM, that can be taken to be at rest, as thermal velocities are $\ll c$.
If the nucleus belongs to a neutral or partially ionised atom, the screening effect of the atomic electrons has to be taken into account.
It is convenient to look at the interaction from the frame where the incident electron is (initially) at rest, and the nucleus moves at a speed $v \sim c$ towards the electron.
The electron feels the Lorentz-transformed electric potential of the incoming nucleus as a time varying electric field, which is strong for a time $\sim b/\gamma v$, where $b$ is the impact parameter of the collision.
As this time interval is very short, the field of the nucleus appears to the electron as a pulse of radiation: a {\it virtual quantum}.
Relativistic Bremsstrahlung can be seen as Compton scattering of such virtual quanta.
The energy of the emitted photon can be obtained after Lorentz-transforming back to the lab frame. 

In relativistic Bremsstrahlung, emitted photons can have any energy $E_{\gamma}$ smaller than the electron energy $E_e$, with a probability that scales as $\sigma_B \propto 1/E_{\gamma}$ for $E_{\gamma} \ll E_e$.
This fact has two important consequences. First of all, the average energy of the radiated photons is of the order of the electron energy $E_{\gamma} \lesssim E_e$, i.e. Bremsstrahlung is a catastrophic process, where the electron typically loses a significant fraction of its energy in a single encounter with a nucleus.
Second, the energy spectrum of the emitted photons mimics to that of relativistic electrons, as can be easily seen for power law spectra of electrons of slope $\delta$.
The Bremsstrahlung emissivity is then: $q_{\gamma}^B \propto \int_{E_{\gamma}} {\rm d} \gamma \, \sigma_B \, \gamma^{-\delta} \propto E_{\gamma}^{-\delta}$ \citep{blumenthal1970,aharonian2004}.

CR electrons with energies in the range $\approx$ 100 MeV-10 GeV produce Bremsstrahlung photons having approximatively the same energy.
However, in this energy range neutral-pion-decay gamma rays dominate the diffuse emission (Fig.~\ref{fig:andy}).
It follows that Bremsstrahlung emission is a much less effective probe of the CR electron spectrum at energies which are not covered by direct observations.

The expected contribution to the gamma-ray diffuse emission from the disk from the two leptonic processes described above is shown in Fig.~\ref{fig:andy}, together with the hadronic contribution (red line), that was discussed already in the previous Section.
The predictions for the diffuse inverse Compton (green) and Bremsstrahlung (cyan line) emissions have been computed from a model of CR injection and transport in the entire Galaxy \citep{strong2011}.
When comparing predictions with data, one should keep in mind that the former did not include line (blue data points) and positronium emission, nor hard X-ray sources that dominate the signal below 100 keV.
It can be seen that the inverse Compton emission largely dominates the diffuse flux below photon energies of few MeV, while relativistic Bremsstrahlung is subdominant at all energies \citep[see also][]{strong2010}.

%Electrons of energy smaller than several GeV are the main contributors to the emission \citep{bouchet2011}.
%On the other hand, relativistic Bremsstrahlung is predicted to be subdominant at all energies.

For photon energies in the $\approx$~1-10 MeV range, expectations fail to fit COMPTEL measurements (green datapoint).
The observed excess might be due to unresolved discrete sources \citep{strong2011}, but to test this hypothesis observations by superior sensitivity instruments would be badly needed (remember that COMPTEL data are more than 20 years old).
With this respect, the launch of the Compton Spectrometer and Imager (COSI, see \citealt{tomsick2021}), expected in 2025, and the ongoing discussion about future mid-scale missions operating in the MeV domain will hopefully improve the situation in a not too distant future.
Proposals include, among others, e-ASTROGAM \citep{deangelis2018}, AMEGO \citep{mcenery2019}, and GECCO \citep{orlando2021}.

The large number of components (continuum, lines, positronium, sources, and the COMPTEL excess) contributing to the diffuse hard X-ray/soft gamma-ray emission makes it difficult to extract in a reliable way the CR electron spectrum from data.
Luckily, electrons also interact with the interstellar magnetic field to generate synchrotron photons.
The observation of this additional radiation component constrains in a more straightforward way the leptonic spectrum of CRs.

\subsubsection{Synchrotron emission}
\label{sec:synchro}

Synchrotron radiation is produced when a relativistic electron is accelerated due to the presence of a magnetic field.
A relativistic electron of charge $e$ and Lorentz factor $\gamma$ moving in a uniform magnetic field of strength $B$ with pitch angle $\alpha$ (the angle between the electron velocity and the field) will spiral along the magnetic field with a characteristic gyration frequency $\omega_g = e B/\gamma m_e c$, independent on $\alpha$.
In the non-relativistic regime, the spectrum of the radiation emitted by the electron is close to a delta function at the cyclotron frequency $\nu_c = \omega_c/2 \pi = e B/2 \pi m_e c$.
For energetic particles, relativistic corrections apply: the spectrum of the emitted radiation is much broader than a delta function, and extends up to a characteristic frequency \citep{blumenthal1970}:
\begin{equation}
\label{eq:synchro}
\nu_s = \frac{3}{2} \gamma^2 \nu_c  \sin \alpha  \approx 48 \left( \frac{E_e}{\rm GeV} \right)^2 \left( \frac{B}{3~\mu {\rm G}} \right) \rm MHz
\end{equation}½
which does depend on the pitch angle.
For LECR electrons the synchrotron emission falls in the radio domain.
The rightmost equality in the equation above has been computed for the special case $\alpha = \pi/2$ and normalised to the typical value of the interstellar magnetic field \citep{ferriere2001}.
Note that the energy of the synchrotron photon scales as the square of the electron energy, $h \nu_s \propto E_e^2$, as for inverse Compton scattering (see Eq.~\ref{eq:ICboost}). 

The similarity with inverse Compton scattering holds also for the expression of the power radiated by an electron:
\begin{equation}
P_{syn} = -\frac{{\rm d}E_e}{{\rm d}t} = \frac{4}{3} \sigma_T c \gamma^2 w_{B}
\end{equation}
which is identical to Eq.~\ref{eq:ICrate}, except for the fact that the energy density of the radiation field $w_{rad}$ has been substituted with that of the magnetic field $w_B = B^2/8 \pi$.
It follows, then, that the spectrum of synchrotron photons produced by relativistic electrons following a power law distribution in energy (or frequency, as this quantity is customarily used in radio astronomy) ${\rm d}n_e/{\rm d}\gamma = K_e \gamma^{-\delta}$ is also a power law: $\nu \times q(\nu) \propto K_e B^{(\delta+1)/2} \nu^{-(\delta-1)/2} \equiv \nu^{-\beta}$ \citep{blumenthal1970}.

A consequence of Eq.~\ref{eq:synchro} is that observations in the MHz frequency range can be used to constrain the CR electron spectrum in the gap between Voyager and AMS-02 data.
We will proceed in a similar fashion to what done in Sec.~\ref{sec:diffuse} and discuss first the observations of the diffuse synchrotron emission at large Galactic latitudes ($b >$10$^{\circ}$), that can be used to constrain the {\it local} spectrum of CR electrons. Then, we will discuss the difficulties encountered in constraining the spatial distribution of CR electrons throughout the Galactic disk.

The average spectral index of the high latitude diffuse synchrotron emission has been derived from various surveys performed between 22 and 408 MHz \citep[see the compilation of data in][]{strong2011a}.
The values found in this way are all close to $\beta \sim 0.5$, which corresponds to a slope of the electron spectrum equal to $\delta \sim 2$ at $\approx$~GeV energies.
The spectrum of the diffuse synchrotron emission steepens at larger frequencies \citep[see e.g.][]{tartari2008}, reaching $\beta \approx 1$ at tens of GHz, though with significant scatter around this value \citep[][and references therein]{strong2011a}.
Note that $\beta \approx 1$ would correspond to a slope of the electron spectrum of $\delta \approx 3$, consistent with the one measured from direct (AMS-02) observations of CR electrons.

\citet{orlando2018} constrained the local ($\sim 1$~kpc) interstellar spectrum of CR electrons by combining observations of the diffuse radio and gamma-ray emission from intermediate Galactic latitudes with direct observations (Voyager and AMS-02). 
The spectrum she obtained is shown as a red dotted line in the right panel of Fig.~\ref{fig:LISpe}, together with the fitting formula we adopted in Eq.~\ref{eq:voyagere}, shown as a solid black line.
The spectrum has a doubly broken power law shape and is mainly constrained by Voyager and AMS-02 direct observations at low (below tens of MeV) and high (above tens of GeV), respectively, and by radio observations in the intermediate region, where the spectral energy distribution is roughly flat, i.e. $E_e^2 j_{LIS}^e(E_e) \approx const$.

Finally, measuring the large scale spatial variation of the intensity of CR electrons throughout the Galactic disk is not a straightforward task. 
In principle, observations of MCs in the gamma-ray and radio domain could be used to constrain the CR electron spectrum at specific locations in the disk (in a similar fashion to what done in Sec.~\ref{sec:MCs} for CR protons).
Unfortunately, this is problematic, for at least four reasons.
\begin{enumerate}
\item{%First of all, 
As for the diffuse Galactic emission, the leptonic contribution to the gamma-ray flux from MCs is expected to dominate in the MeV domain \citep[e.g.][]{gabici2007,gabici2009} which is, to date, very poorly explored.}
\item{%Second, 
The expected synchrotron radio fluxes from clouds are quite weak, and in several cases only upper limits have been obtained after subtracting the dominant thermal component \citep{jones2008,jones2011}.
Moreover, in the cases where some non-thermal emission has been detected, it was associated to small volumes of the cloud, and its origin was likely unrelated to interstellar CRs (e.g. {\it in situ} acceleration of CRs, \citealt{meng2019}, presence of discrete non-thermal or foreground sources, \citealt{rodriguez2013}, but see also \citealt{jones2014}). 
A possibly unique exception to this is the central molecular zone, a very massive complex of clouds located in the Galactic centre region, exhibiting a radio synchrotron spectrum \citep{yusefzadeh2013} which may be consistent with the emission of interstellar CR electrons penetrating the dense molecular region \citep{dogiel2021}.}
\item{%Second, 
The interactions of CR protons with the gas produce not only neutral pions (Eq.~\ref{eq:pp}), but also charged ones.
This leads to the creation of electrons and positrons (hereafter referred to as {\it secondary} electrons) through the decay chain $\pi^{\pm} \rightarrow \mu^{\pm} \rightarrow e^{\pm}$ \citep{kelner2006}. 
Such secondary electrons are produced in large quantities in dense MCs, and the soft gamma-ray and radio emission that they generate may compete with that from interstellar {\it primary} electrons \citep{brown1977,marscher1978,crocker2007,protheroe2008}.
Ultimately, the predominance of the radiation from primary or secondary electrons depends on the transport properties of these particles and on the cloud density. 
The former regulate the capability of energetic particles to penetrate into or escape from MCs, while the latter determines how many secondary electrons are produced per parent CR proton.
Recent studies seem to indicate that, in fact, the contribution from secondary electrons to the synchrotron emission is subdominant \citep{gabici2009,padovani2018a}.
Given the current observational difficulties, this issue will be hopefully settled by future observations by the Square Kilometer Array \citep{dickinson2015,padovani2018}.}
\item{The synchrotron emissivity depends on both CR electron density and strength of the cloud magnetic field.
The former can be extracted from observations provided that an independent estimate of the field strength is available.
Therefore, the uncertainty in the determination of the field strength \cite{crutcher2012} affects the estimate of the CR electron intensity.}
\end{enumerate}

In the absence of any reliable method to measure the large scale spatial distribution of CR electrons, we resort to theoretical predictions \citep{strong2011a,orlando2013,orlando2018,dibernardo2013}.
Propagation codes describing the transport of CRs in the entire Galaxy can be used to this purpose, as their predictions can be adjusted in order to make them consistent with all available data (both direct and indirect) on CRs.
In this way, the spatial distribution of CR electrons can be predicted and, though results depend on the assumptions made on the spatial distribution of CR sources, they show that large scale fluctuations in the CR electron density are limited to a factor of few \citep[see e.g. Fig.~1 in][]{dibernardo2015}.

We conclude that, as for CR protons, large scale spatial variations of CR electrons in the Galactic disk are likely to be quite mild.

\section{Integral constraints on the remote intensity of low energy cosmic rays: the cosmic-ray ionisation rate}
\label{sec:integral}

In this Section we describe a number of observations that can be used to measure the intensity of CRs in remote locations in the Galaxy, thereby constraining their spatial distribution in the Galactic disk.
The method in based on the measurement of the rate of ionisation of the dense interstellar gas, which under certain conditions is entirely due to CR ionisation, and it differs from that described in the previous Section, as it provides an {\it integral} constrain on the intensity of CRs, rather than their spectral energy distribution.

\subsection{Cosmic ray induced astrochemistry in interstellar clouds: formation and destruction of H$_3^+$}
\label{sec:H3+}

The most important photoionising agent in the ISM is UV radiation emitted by hot stars.
However, UV photons of energy exceeding 13.6 eV (the ionisation potential of the very abundant hydrogen) are very quickly absorbed and do not pervade the interstellar gas.
On the other hand, photons of energy smaller than 13.6 eV can penetrate significant column densities of gas, where they primarily ionise carbon, which is the most abundant species after hydrogen having a smaller ionisation potential ($\sim$ 11.3 eV).
When a column density of gas larger than $N_H \approx 8 \times 10^{21}$ cm$^{-2}$ is traversed, the UV radiation is strongly attenuated and LECRs become the dominant ionising agents \citep{mckee1989}.

Following \citet{snow2006} (see their Fig.~1), we will call diffuse and dense MCs those characterised by a column density smaller than $\approx 2 \times 10^{21}$ cm$^{-2}$ and larger than $\approx 8 \times 10^{21}$ cm$^{-2}$, respectively.
Clouds of intermediate column densities are called translucent, as they can still be studied by means of absorption spectroscopy of background stars \citep{vandishoeck1989}.

In diffuse clouds, fully exposed to stellar radiation, carbon is predominantly singly ionised (C$^+$), and its photoionisation constitutes the main source of free electrons.
The typical ionisation fraction in these objects is of the order of $\approx 10^{-4}$.
Diffuse clouds are called atomic or molecular depending on the predominant state in which hydrogen is found, with atomic hydrogen becoming less and less abundant with increasing column density.
%The fraction of hydrogen in molecular form is an increasing function of the cloud column density.
In translucent clouds, mainly made of molecular hydrogen, photoionisation becomes less important and carbon is found primarily in atomic form (C). 
As a consequence, the electron fraction decreases as the gas column density increases.
Finally, in dense MCs carbon is predominantly locked up in carbon monoxide (CO), the second most abundant interstellar molecule after H$_2$, and the ionisation fraction drops to $\approx 10^{-7}$.
This classification of clouds also serves to describe the different layers of a single object, i.e., dense molecular cores will be surrounded by regions of lower density having the same properties of translucent clouds, which are in turn embedded in envelopes of diffuse gas where hydrogen is present in both atomic and molecular form.

A key point is that molecular hydrogen is primarily ionised by LECRs, and not by stellar photons.
This is because its ionisation potential is 15.4 eV, larger than that of atomic hydrogen.
H$_2$ is therefore well protected from ionising radiation, which is severely absorbed in the outer atomic layers of clouds.
Remarkably, the ionisation of H$_2$ triggers the formation of protonated hydrogen, H$_3^+$, which in turn starts a chain of chemical reactions that leads to the formation of a great variety of molecules.
The importance of CR ionisation in interstellar chemistry was recognised in a number of pioneering papers \citep{watson1973,black1973bis,herbst1973}, showing how the measurements of molecular abundances in clouds could lead to estimations of the ionisation rate and therefore to constraints on the spectrum of the LECRs responsible for the ionisation.

The goal of this Section is to review briefly the available measurements of the CR ionisation rate based on observations of molecular lines from the direction of interstellar clouds.
For more extended discussions the reader is referred to \citet{tielens2013}, \citet{larsson2012}, \citet{oka2006}, and \citet{dalgarno2006}.
A useful database of astrochemical reaction rates can be found in \citet{wakelam2015}\footnote{KIDA: Kinetic Database for Astrochemistry: \url{https://kida.astrochem-tools.org/}}.

Let us start considering the ionisation of molecular hydrogen by a CR particle (either a nucleus or an electron):
\begin{equation}
\label{eq:ionisation}
{\rm H}_2 + {\rm CR} \longrightarrow {\rm H}_2^+ + e^- + {\rm CR}
\end{equation}
The number of ionisations per unit second and unit volume of gas is $\zeta_{CR}^{{\rm H}_2} n({\rm H}_2)$, where $n({\rm H}_2)$ is the number density of molecular hydrogen in the cloud, and $\zeta_{CR}^{{\rm H}_2}$ is the CR ionisation rate per ${\rm H}_2$ molecule, that depends only on the intensity and spectrum of CR nuclei and electrons.
The ionisation is followed by the very fast ion-neutral reaction:
\begin{equation}
\label{eq:H3+}
{\rm H}_2^+ + {\rm H}_2 \longrightarrow {\rm H}_3^+ + {\rm H}
\end{equation}
that leads to the formation of protonated hydrogen, ${\rm H}_3^+$, a pivotal molecule in interstellar chemistry.
As the second reaction (Eq.~\ref{eq:H3+}) is orders of magnitude faster than the first one (Eq.~\ref{eq:ionisation}) for any plausible CR intensity and spectrum, the latter fixes the production rate of protonated hydrogen, which is then: $\zeta_{CR}^{{\rm H}_2} n({\rm H}_2)$.
At equilibrium, the abundance of ${\rm H}_3^+$ in a cloud is obtained by balancing its formation and destruction rates.
The destruction of ${\rm H}_3^+$ proceeds in a different way in diffuse and dense clouds.

In diffuse clouds, the electron fraction is of the order of $x_e = n(e^-)/n_{\rm H} \sim n({\rm C}^+)/n_{\rm H}$, where $n_{\rm H} = n({\rm H}) + 2 n({\rm H}_2)$.
The abundance of C$^+$ and of both atomic and molecular hydrogen have been measured directly for a limited number of clouds thanks to the observations of UV absorption lines in the spectra of background hot stars \citep{cardelli1996,bohlin1978,savage1977}. 
Based on these measurements, a reference value of $x_e \gtrsim 10^{-4}$ is commonly accepted. 
Remarkably, this value is large enough to make dissociative recombination the dominant channel for the destruction of ${\rm H}_3^+$:
\begin{equation}
\label{eq:dissrec}
{\rm H}_3^+ + e^- \longrightarrow {\rm H} + {\rm H} + {\rm H} ~~ {\rm or} ~~ {\rm H}_2 + {\rm H}
\end{equation}
The recombination rate per unit volume is then $k_e n({\rm H}_3^+) n(e^-)$, where $k_e$ is the rate coefficient.

Balancing formation and destruction rates provides the equilibrium density of ${\rm H}_3^+$, which reads:
\begin{equation}
\label{eq:H3+diff}
n({\rm H}_3^+) = \left( \frac{\zeta_{CR}^{{\rm H}_2}}{k_e} \right) ~ \left[ \frac{n({\rm H}_2)}{n(e^-)} \right]
\end{equation} 
Being interested here in the study of LECRs, it is more convenient to rewrite the expression above to obtain the CR ionisation rate as a function of the other physical quantities:
\begin{equation}
\label{eq:zetadiff}
\zeta_{CR}^{{\rm H}_2} = \frac{2  \, k_e  x_e}{f_{\rm H_2}} ~n({\rm H}_3^+) \approx  \frac{2  \, k_e  x_e}{f_{\rm H_2}} ~ \frac{N({\rm H_3^+})}{L}
\end{equation}
where we introduced the electron fraction $x_e$, the fraction of hydrogen in molecular form $f_{\rm H_2} = 2 \, n({\rm H}_2)/n_{\rm H}$, the ${\rm H}_3^+$ column density $N({\rm H_3^+})$, and the depth of the cloud along the line of sight $L$.
Note that the latter quantity is difficult to be constrained observationally.
Imposing the approximate equality $n({\rm H}_3^+) \approx N({\rm H_3^+})/L$, we assumed that ${\rm H}_3^+$ is uniformly distributed in the entire cloud.
Making an assumption on the spatial distribution of H$_3^+$ inside the cloud is needed as column densities, rather than volume densities, are more easily derived from observations.

 At this point one should notice that all the physical quantities on the right side of Eq.~\ref{eq:zetadiff} could be determined from astronomical observations, except for the coefficient $k_e$, which has to be measured in the lab.
Unfortunately, for a very long time the actual value of $k_e$ was so uncertain that its determination was considered one of the most controversial issues in the field of electron-ion recombination, from both an experimental and theoretical point of view \citep{larsson2000,oka2003}.
The breakthrough came when it became possible to perform laboratory measurements of this quantity under nearly interstellar conditions, providing a value of $k_e$ of the order of few times $10^{-7}$~cm$^3$/s for typical cloud temperatures $T$ of the order of tens of degrees, with a mild $\sim~T^{-0.5}$ dependence \citep{mccall2003,mccall2004}. 
Note that, as these measurements revealed that dissociative recombination is faster than previously thought, earlier estimates of the CR ionisation rate in diffuse clouds are often too low \citep[see][for an historical perspective]{dalgarno2006}.
The parameterisation currently adopted in the KIDA database for the rate coefficient is $k_e = 6.7 \times 10^{-8} \, (T/300~{\rm K})^{-0.52}$~cm$^3$/s.

A very similar line of reasoning can be pushed forward also for the case of dense MCs ($f_{\rm H_2} \sim 1$), with one very important difference.
Dense clouds are well shielded from interstellar radiation and therefore the electron fraction is so small ($x_e \approx 10^{-7}$) that dissociative recombination of H$_3^+$ becomes unimportant.
Due to the large abundance of the CO molecule, H$_3^+$ is mainly destroyed through the proton hop reaction:
\begin{equation}
\label{eq:HCO+}
{\rm H}_3^+ + {\rm CO} \longrightarrow {\rm HCO}^+ + {\rm H}_2
\end{equation}
characterised by a rate coefficient $k_{\rm CO} \sim 2 \times 10^{-9}$ cm$^3$/s which is about two orders of magnitude slower than $k_e$ \citep{oka2006}.
Imposing equilibrium between formation and destruction, an expression very similar to Eq.~\ref{eq:H3+diff} can be obtained:
\begin{equation}
\label{eq:H3+dense}
n({\rm H}_3^+) = \left( \frac{\zeta_{CR}^{{\rm H}_2}}{k_{\rm CO}} \right) ~ \left[ \frac{n({\rm H}_2)}{n({\rm CO})} \right]
\end{equation}
or, passing to column densities:
\begin{equation}
\label{eq:zetadense}
\zeta_{CR}^{{\rm H}_2} = 2  \, k_{\rm CO} \left[ \frac{n({\rm CO})}{n_{\rm H}} \right]~n({\rm H}_3^+) \approx  2  \, k_{\rm CO} \left[ \frac{n({\rm CO})}{n_{\rm H}} \right]~ \frac{N({\rm H_3^+})}{L}
\end{equation}
which is analog to Eq.~\ref{eq:zetadiff}.
%The column density of CO can be determined, with the caveats discussed in Sec.~\ref{sec:gamma}, by means of radio observations.
A typical value for $n({\rm CO})/n_{\rm H}$ in dense clouds is $\gtrsim 10^{-4}$ \citep{lee1996}, similar to the value of $x_e \sim n({\rm C}^+)/n_{\rm H}$ found in diffuse clouds.
It follows, then, from Equations \ref{eq:zetadiff} and \ref{eq:zetadense} that the density of H$_3^+$ in interstellar clouds (both diffuse and dense) is mainly set by the CR ionisation rate $\zeta_{CR}^{{\rm H}_2}$.

Reversing this argument, one can see how the determination of the column density of H$_3^+$ is an essential step towards the determination of the CR ionisation rate, as $\zeta_{CR}^{{\rm H}_2} L \propto N({\rm H}_3^+)$. $L$ is in general unknown, but can be estimated when both the column density ($N_{\rm H}$) and the average volume density  ($\bar{n}_{\rm H}$) of hydrogen are known, to give $L \approx N_H/\bar{n}_H$.
It seems therefore useful to review, at this point, the available observations of H$_3^+$ column densities in interstellar clouds.

\subsubsection{${\rm H}_3^+$ in diffuse clouds}
\label{sec:H3+diff}

The presence of H$_3^+$ in interstellar clouds can be revealed in the infrared absorption spectra of background stars. 
Under interstellar conditions, only the two lowest rotational levels of H$_3^+$ are expected to be significantly populated.
Therefore, observations probing the transitions arising from these states can be used to probe the entire content of H$_3^+$ along the line of sight \citep[see e.g. Fig.~1 in][]{mccall1999}.
In particular, the R(1,0) and R(1,1)$^u$ orto-para doublet can be conveniently observed in the IR band, near 3.67~$\mu$m \footnote{For a detailed description of H$_3^+$ spectroscopy notation see \citet{lindsay2001}.}.
After a long history of attempts, H$_3^+$ was finally detected in interstellar space by \citet{geballe1996} (see also \citealt{miller2020} for a recent review on H$_3^+$ astronomy).

As said above, the interpretation of early measurements of H$_3^+$ column densities suffered from the large uncertainty in the knowledge of the dissociative recombination coefficient $k_e$.
After measuring this coefficient in the lab, \citet{mccall2003} could estimate with unprecedented accuracy the CR ionisation rate in a diffuse cloud towards the star $\zeta$~Persei.
The column density of H$_3^+$ was derived from the observations of the R(1,0) and R(1,1)$^u$ absorption lines, resulting in $N({{\rm H}_3^+}) \sim 8 \times 10^{13}$~cm$^{-2}$.
Moreover, a gas temperature of 23 K was estimated from the ratio of the two lines.

Remarkably, not only $N({{\rm H}_3^+})$, but all of the quantities appearing in the right hand side of Eq.~\ref{eq:zetadiff} have been measured along the line of sight towards $\zeta$~Persei.
Absorption lines in the UV domain provided us with an estimate of the column density of molecular hydrogen, $N({{\rm H}_2}) \sim 4.7 \times 10^{20}$~cm$^{-2}$ \citep{savage1977}, atomic hydrogen, $N({{\rm H}}) \sim 6.3 \times 10^{20}$~cm$^{-2}$ \citep{bohlin1978}, and ionised carbon, $N({{\rm C}^+}) \sim 1.8 \times 10^{17}$~cm$^{-2}$ \citep{cardelli1996}.
Such values yield a total hydrogen column density $N_{\rm H} = N({{\rm H}}) + 2 N({{\rm H}_2}) \sim 1.6 \times 10^{21}$~cm$^{-2}$, a fraction of hydrogen in molecular form $f_{{\rm H}_2} = 2 N({\rm H}_2)/N_{\rm H} \sim 0.6$, and an electron fraction $x_e \sim N({\rm C}^+)/N_{\rm H} \sim 10^{-4}$.
Substituting these values in Eq.~\ref{eq:zetadiff}, and using the appropriate value for the dissociative recombination rate coefficient $k_e \sim 2.5 \times 10^{-7}$~cm$^3$/s (from KIDA), gives $\zeta_{CR}^{{\rm H}_2} L \sim 7 \times 10^3$~cm/s.

A measurement of the average hydrogen volume density $\bar{n}_{\rm H}$ is needed to estimate the absorption path length $L$.
This can be obtained from the rotational excitation lines of the molecules CO and C$_2$ \citep{vandishoeck1986}.
The method is quite uncertain and gives values of $\bar{n}_{\rm H}$ of the order of hundreds atoms per cubic centimeter.
Adopting a reference value of $\bar{n}_{\rm H} \approx 250$~cm$^{-3}$ yields a path length $L = N_{\rm H}/\bar{n}_{\rm H} \approx 2$~pc from which one finally gets a CR ionisation rate equal to $\zeta_{CR}^{{\rm H}_2} \approx 10^{-15}$~s$^{-1}$ \citep{mccall2003}.
This was a quite surprising result because, as we will see in the following, such a value for the CR ionisation rate is much larger than that derived from both earlier theoretical predictions and measurements of the H$_3^+$ column density in dense clouds.

\begin{figure*}
% Use the relevant command to insert your figure file.
% For example, with the graphicx package use
\center
  \includegraphics[width=0.7\textwidth]{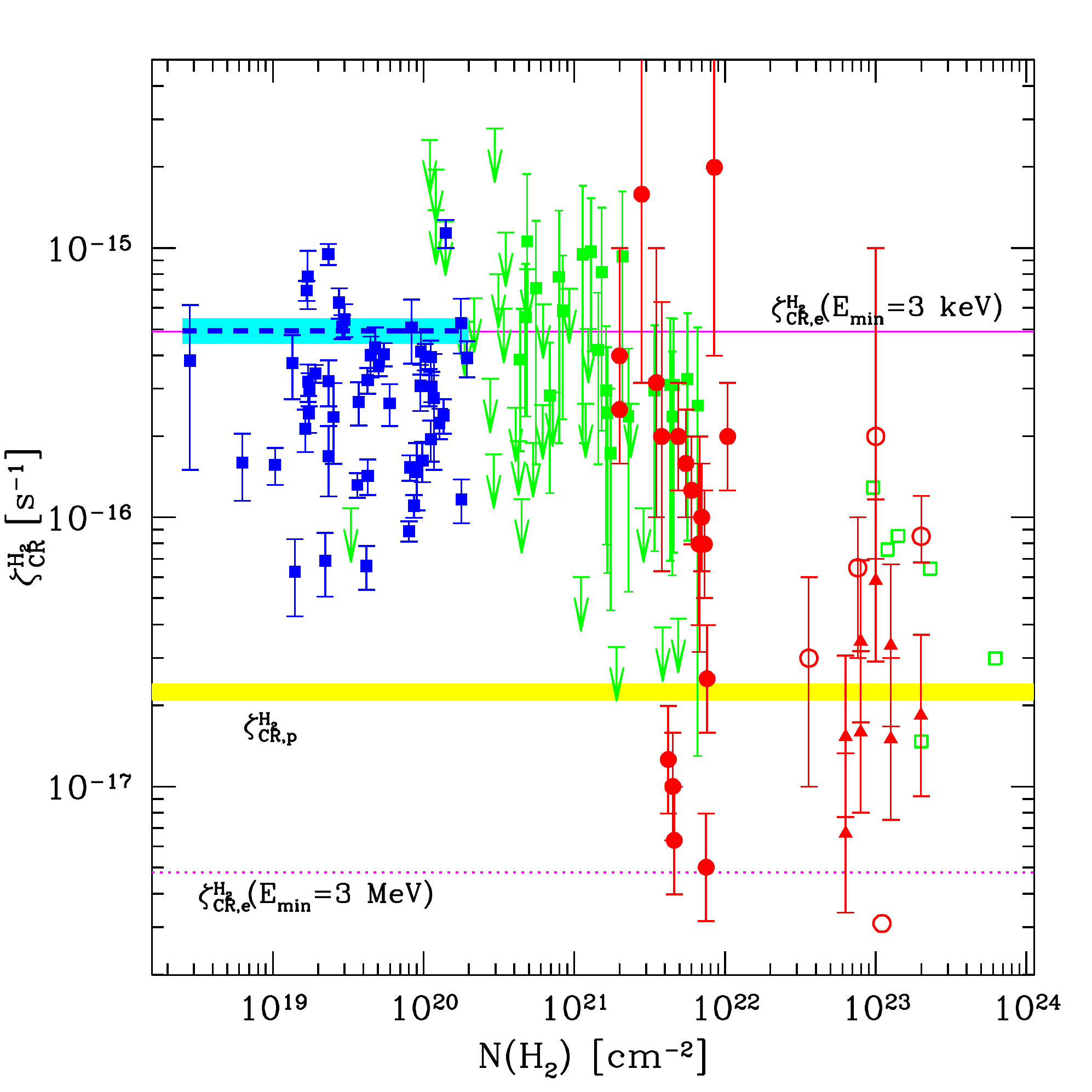}
% figure caption is below the figure
\caption{Measurements of the cosmic ray ionisation rate in diffuse and dense clouds. Data are from \citet{indriolo2015} (blue points), \citet{neufeld2017} (shaded cyan region), \citep{indriolo2012} (filled green points), \citet{vandertak2000} (open green points), \citet{caselli1998} (filled red circles), \citet{sabatini2020} (filled red triangles), and \citet{maret2007}, \citet{hezareh2008}, \citet{moralesortiz2014}, \citet{fuente2016} (open red points). The ionisation rate in the local interstellar medium is shown as a shaded yellow region for CR protons and with magenta lines for CR electrons (dotted and solid line refer to $E_{min} = 3$ MeV and 3 keV, respectively).}
\label{fig:zeta}       % Give a unique label
\end{figure*}

Triggered by this unexpected finding, surveys of H$_3^+$ in diffuse clouds were performed. 
In about a decade, the number of sight lines explored increased from 7 \citep{mccall2002} to 50, resulting in 21 detections \citep{indriolo2007,indriolo2012}.
%\citet{indriolo2012} refined the model of H$_3^+$ astrochemistry, by considering also the destruction of the molecular ion H$_2^+$ due to both dissociative recombination (H$_2$ + $e^- \rightarrow$ H + H) and charge transfer (H$_2$ + H $\rightarrow$ H$_2^+$ + H), and the destruction of H$_3^+$ due to proton transfer to CO (to produce HCO$^+$, see Eq.~\ref{eq:HCO+}, and also the metastable isomer HOC$^+$), O (H$_3^+$ + O $\rightarrow$ H$_2$ +OH$^+$), and N$_2$ (H$_3^+$ + N$_2 \rightarrow$ H$_2$ + HN$_2^+$).
The CR ionisation rates derived in this way are shown in Fig.~\ref{fig:zeta} as filled green square data points, together with 3$\sigma$ upper limits (green arrows).
For some lines of sight, not all of the physical quantities needed to compute $\zeta_{CR}^{{\rm H}_2}$ from Eq.~\ref{eq:zetadiff} could be measured, and typical values where assumed\footnote{Assumed typical values are: $n_{\rm H} = 200$~cm$^{-3}$, $x_3 = 1.5 \times 10^{-4}$, $f_{\rm H_2} = 0.67$, $T = 70$~K.}.
After performing a simple statistical analysis, \citet{indriolo2012} concluded that the mean value of the CR ionisation rate in diffuse clouds is $\sim 3.5 \times 10^{-16}$~s$^{-1}$, and that there is a probability of 68.3\% (1$\sigma$ equivalent) to find a diffuse cloud characterised by a ionisation rate in the range spanning from $5.0 \times 10^{-17}$~s$^{-1}$ to $8.8 \times 10^{-16}$~s$^{-1}$.
In order to investigate the origin of this quite large scattering, correlations between the CR ionisation rate and the position in the sky were searched for, to no avail.
This means that the origin of the scattering is most likely not related to spatial variations of the CR intensity on large spatial scales.
However, it should be kept in mind that all of the clouds in the sample are nearby (background stars are all within a distance of $\approx$ 4 kpc, and 60\% of them are within 1 kpc), suggesting that variations of the CR intensity on quite small spatial scales might occur.

As we will discuss in Sec.~\ref{sec:penetration}, the CR ionisation rate has been predicted to decrease with the cloud column density \citep[e.g.][]{padovani2009}.
This is because LECRs, which are most effective at ionisation, will lose all their energy due to ionisation losses while traversing large column densities of gas \citep[e.g.][]{cravens1978}.
This would make them less abundant inside clouds of larger column density, which would be therefore characterised by a smaller average ionisation rate. 
However, such a trend is not observed in diffuse clouds, as no correlation between $\zeta_{CR}^{{\rm H}_2}$ and $N({\rm H}_2)$ has been found.
The lack of a correlation is indeed expected if LECRs permeate the entire volume of diffuse clouds, which might be a plausible hypothesis, given the characteristic low values of the hydrogen column densities of diffuse clouds.

\subsubsection{${\rm H}_3^+$ in dense clouds}

H$_3^+$ was first detected, and searched for, in dense clouds \citep{geballe1996} and, to date, the number of H$_3^+$ detections in dense clouds amounts to $\lesssim 10$ \citep{mccall1999,brittain2004,gibb2010}.
The reason why dense clouds were (erroneously) thought to be better targets for H$_3^+$ searches can be understood by comparing Equations \ref{eq:H3+diff} and \ref{eq:H3+dense}.
As the ratios $n(e^-)/n({\rm H}_2)$ and $n({\rm CO})/n({\rm H}_2)$ are, in most cases, of the same order (${\cal O}(10^{-4})$), it follows that, for a fixed value of the CR ionisation rate $\zeta_{CR}^{{\rm H}_2}$, the expected density of H$_3^+$ in dense clouds is a factor of $k_e/k_{\rm CO} \approx 100$ larger than that in diffuse clouds \citep{oka2006}.
In fact, as we saw in the previous Section, this is not the case, as observations demonstrated that H$_3^+$ is abundant also in diffuse clouds. %, and that H$_3^+$ column densities roughly of the same order of magnitude are founds in clouds, .
An immediate and important implication emerging from this observational finding is that the CR ionisation rate must be different in diffuse and dense clouds, and not a standard quantity, as previously assumed.

In their pioneer paper, \citet{geballe1996} presented the first detection of interstellar H$_3^+$ from two lines of sight towards the young stellar objects  GL2136 and W33A, both deeply embedded in dense MCs.
Further investigations on these two objects were performed by \citet{mccall1999}, who found H$_3^+$ column densities equal to $N({\rm H}_3^+) \sim 3.8 \times 10^{14}$~cm$^{-2}$ and $\sim 5.2 \times 10^{14}$~cm$^{-2}$, respectively.
They also provided estimated temperatures equal to 47 and 36 K, inferred from H$_3^+$ line ratios.
For such temperatures, the KIDA database suggests a value for the rate coefficient that appears in Eq.~\ref{eq:zetadense} equal to $k_{\rm CO} \sim 2.1 \times 10^{-9}$~cm$^3$/s.
Assuming a typical abundance ratio for dense clouds $n({\rm CO})/n({\rm H}_2) \sim 1.5 \times 10^{-4}$ \citep{lee1996} yields $\zeta_{CR}^{{\rm H}_2} L \approx 120$ and 160 cm/s for the two lines of sight.

The clouds' hydrogen column density can be derived from the measured visual extinction and the dust-to-gas conversion factor, giving in $N_{\rm H} \sim 1.8 \times 10^{23}$~cm$^{-2}$ and $\sim 2.8 \times 10^{23}$~cm$^{-2}$ \citep{mccall1999}.
Taking a typical density $\bar{n}_{\rm H} \approx 10^5$~cm$^{-3}$, as appropriate for the envelopes of deeply embedded stars \citep[e.g.][]{vandertak2000a}, one obtains path lengths equal to $L = N_{\rm H}/\bar{n}_{\rm H} \approx$ 0.6 and 0.9 pc, and values of the CR ionisation rate equal to $\zeta_{CR}^{{\rm H}_2} \approx 7$ and $9 \times 10^{-17}$ s$^{-1}$  for GL2136 and W33A, respectively.
These rates are smaller than the mean value of $\sim 3.5 \times 10^{-16}$~s$^{-1}$ found by \citet{indriolo2012} in diffuse clouds (see previous section).
What we might be witnessing, here, is the predicted decrease of the CR ionisation rate with the cloud column density, due to ionisation losses suffered by LECRs while penetrating interstellar clouds.

In fact, the difference between the values of the CR ionisation rate in dense and diffuse clouds might even be larger, as pointed out by \citet{vandertak2000}.
This is because a significant fraction of the H$_3^+$ observed along the line of sight might reside in the diffuse envelope surrounding the dense cloud cores discussed here, or in foreground diffuse clouds that just happen to be located along the line of sight.
In support of the latter hypothesis, \citet{vandertak2000} provided evidence for a correlation between the measured H$_3^+$ column density and the cloud heliocentric distance (see their Fig. 3), which points to an important role played by intervening clouds.
To overcome this problem, they combined the observations of H$_3^+$ with those of the HCO$^+$ line, which is observed in emission in the millimeter domain.
HCO$^+$ is produced in the dense molecular gas only, and this strongly reduces the importance of intervening foreground diffuse clouds.
Finally, its abundance is proportional to that of H$_3^+$ (see Eq.~\ref{eq:HCO+}) and therefore to the CR ionisation rate.
The revised estimates of $\zeta_{CR}^{{\rm H}_2}$ obtained in this way are shown as empty green squares in Fig.~\ref{fig:zeta} and were found to be a factor of several smaller than those based on H$_3^+$ observations only. 
%Note, however, that such a correlation was not found in the survey of diffuse clouds performed by \citet{indriolo2012}.

As we will see in the next Section, alternative molecular tracers can be used to infer in an independent way the value of $\zeta_{CR}^{{\rm H}_2}$ in dense clouds, confirming that its typical value is significantly smaller than that found in diffuse clouds.
%Especially promising are methods involving molecules that can be observed in emission, rather than absorption spectra of dense clouds, as in this case the importance of intervening diffuse clouds would be strongly diminished.

\subsection{Other molecular tracers of the cosmic ray ionisation rate}
\label{sec:other}

Before the detection of H$_3^+$ in space, constraints on the CR ionisation rate were usually obtained from the abundances of molecules like OH and HD \citep{black1973bis,vandishoeck1986,federman1996}.

Oxygen chemistry begins with the CR ionisation of atomic hydrogen, followed by charge exchange with oxygen to produce O$^+$ in a slightly endothermic reaction. 
This, in turn, starts a chain of exothermic reactions with H$_2$ leading to the production of a number of molecular ions (hydrides, made of one single heavy atom and one or more hydrogen), according to the scheme: O$^+ \rightarrow$ OH$^+ \rightarrow$ H$_2$O$^+ \rightarrow$ H$_3$O$^+$.
In MC interiors, where molecular hydrogen dominates over atomic one, a very similar chain of reactions is initiated by H$_3^+$, that can interact with oxygen to give either OH$^+$ or H$_2$O$^+$ \citep{hollenbach2012}.
Finally, OH is produced as a result of the dissociative recombination of H$_3$O$^+$.

In an analogue manner, the production of HD is the result of CR ionisation of atomic hydrogen, followed by charge exchange first with deuterium, H$^+$ + D $\leftrightarrow$ H + D$^+$, and then with molecular hydrogen, D$^+$ + H$_2 \leftrightarrow$ H$^+$ + HD \citep{black1973bis}.
If LECRs, rather than UV/X-ray photons, dominate the ionisation of hydrogen (as it is the case, unless the cloud is irradiated by an UV flux that largely exceeds the local interstellar one, \citealt{hollenbach2012}), the abundances of OH and HD can be used to constrain the CR ionisation rate.
Early estimates of the CR ionisation rate in diffuse clouds based on the observations of these two molecules fell in the range of a few $10^{-17}$~s$^{-1}$ (see e.g. \citealt{black1977} and \citealt{federman1996}, but see \citealt{vandishoeck1986} who considered also larger values).
As we saw in the previous Section, the detection of H$_3^+$ in diffuse clouds led to an average value of $\zeta_{CR}^{{\rm H}_2}$ about one order of magnitude larger than those early estimates.

It was later recognised that ionised atomic hydrogen can be neutralised via charge exchange with polycyclic aromatic hydrocarbons (PAHs)\footnote{PAHs are molecules made entirely of carbon and hydrogen arranged in cyclic (ring shaped) structures called aromatic rings \citep{tielens2008}.} or small grains.
Therefore, the chain of reactions leading to OH and HD is ``leaky'', as not all ionisations of H lead to the formation of one of such molecules. 
If this effect is not taken into account while interpreting the abundances of OH and HD, the CR ionisation rate is underestimated.
This may explain the discrepancy with the ionisation rates derived from H$_3^+$ observations \citep{wolfire2003,liszt2003,hollenbach2012}.

The opposite might happen deeper into the cloud, when the CR ionisation of molecular hydrogen starts the chemistry.
There, PAHs can efficiently capture electrons, reducing their number and hence diminishing their role in the destruction of H$_3^+$ via dissociative recombination.
Then, the abundance of H$_3^+$ and of the oxygen-bearing molecular ions increases in the presence of PAHs \citep{hollenbach2012}.
In this case, ignoring the presence of PAHs when interpreting the abundance of H$_3^+$ may lead to overestimating the CR ionisation rate, even though an accurate assessment of this effect is prevented due to the uncertainties in our knowledge of the PAH abundance in clouds \citep{shaw2021}.

\subsubsection{Hydride ions: ${\rm OH}^+$, ${\rm H}_2{\rm O}^+$, and {\rm ArH}$^+$}
\label{sec:hydrides}

Not only OH, but also the oxygen-bearing molecular ions formed in the chain of reactions that lead to its formation can be studied in order to constrain the CR ionisation rate.
This fact became of great interest following the detection of the three molecular ions OH$^+$, H$_2$O$^+$, and H$_3$O$^+$ in the submillimeter absorption spectra of sources observed by the HIFI instrument onboard of the Herschel Space Observatory \citep{gerin2010,neufeld2010}.

In a fully molecular gas ($f_{{\rm H}_2} = 1$) each formation of OH$^+$ is followed by the formation of H$_2$O$^+$ through a reaction with H$_2$. 
%Also H$_2$O$^+$, as OH$^+$, is 
Both OH$^+$ and H$_2$O$^+$ are
destroyed in interactions with H$_2$, and the two reactions proceed at a similar rate.
Therefore, the abundance ratio $n({\rm OH}^+)/n({\rm H}_2{\rm O}^+)$ is expected to be of the order of 1 in MCs.
On the other hand, in atomic clouds, where $f_{{\rm H}_2}$ is significantly smaller than 1, OH$^+$ can be destroyed by dissociative recombination before interacting with H$_2$ to produce H$_2$O$^+$.
As a consequence, the abundance of the latter molecular ion is suppressed.
It follows that the abundance ratio $n({\rm OH}^+)/n({\rm H}_2{\rm O}^+)$ can be used to estimate the fraction of hydrogen in molecular form \citep{gerin2010,hollenbach2012}.

\citet{indriolo2015} performed a survey of OH$^+$, H$_2$O$^+$, and H$_3$O$^+$ using Herschel observations along 20 lines of sight.
The first two molecular ions were detected in multiple velocity components along every sightline, while the third was only detected along 7 sightlines.
In the vast majority of cases, the $n({\rm OH}^+)/n({\rm H}_2{\rm O}^+)$ abundance ratio was found to be $\ll 1$, pointing towards very small fractions of hydrogen in molecular form ($f_{{\rm H}_2} < 0.1$). 
This fact makes Herschel observations unique probes of the CR ionisation rate in atomic clouds.

Among the oxygen-bearing molecular ions discussed here,
OH$^+$ is 
%The abundance of OH$^+$ with respect to hydrogen, $n({\rm OH}^+)/n({\rm H})$, seems to be 
the best tracer of CRs, as it is the first one to be produced after the CR ionisation of H, and is therefore less affected by ``leaks'' in the reaction chain.
However, as seen above, not all CR ionisations of H lead to the production of OH$^+$, but only a fraction $\epsilon$ of them.
Then, the ratio $n({\rm OH}^+)/n({\rm H})$ constrains the product $\epsilon \, \zeta_{CR}^{{\rm H}}$, and not the two quantities separately.
\citet{indriolo2012a} were able to calibrate observationally the efficiency parameter, by using data from a sightline were both OH$^+$ and H$_3^+$ were detected.
The CR ionisation rate inferred from the H$_3^+$ column density allowed them to break the degeneracy and determine $\epsilon \sim 0.07$, a value which is in broad agreement with theoretical expectations \citep{hollenbach2012}.
Note that some of the physical parameters needed to determine the CR ionisation rate could not be measured, and typical values had to be assumed \footnote{Assumed typical values: $T = 100$~K, $x_e = 1.5 \times 10^{-4}$, $n_H = 35$~cm$^{-3}$.}.

It should be stressed that what is measured, in this case, is the CR ionisation rate of atomic hydrogen, $\zeta_{CR}^{{\rm H}}$, which is different from that of molecular hydrogen, $\zeta_{CR}^{{\rm H}_2}$.
However, the two quantities differ by a small factor, as suggested by the approximate expression derived by \citet{glassgold1974}: $1.5 \, \zeta_{CR}^{{\rm H}_2} = 2.3 \, \zeta_{CR}^{{\rm H}}$, which was adopted by \citet{indriolo2015}.
The CR ionisation rates obtained in this way are shown as filled blue square data points in Fig.~\ref{fig:zeta}.
We plotted only points for which an estimate of the hydrogen column density is available, and we do not show measurements relative to lines of sight towards the Galactic centre, as this very peculiar region will be examined later in this review.
All the measurements are characterised by quite low column densities of molecular hydrogen, reminding us that hydrides trace LECRs in atomic clouds, whose total column density is $N_{\rm H} \gg N({\rm H}_2)$.
We decided to plot these data versus the molecular hydrogen column density to ease the readability of the Figure (if plotted versus the total hydrogen column density the filled blue and green points would significantly overlap).

Both the mean value and the standard deviation of the CR ionisation rates derived from Herschel observations were found to be in very good agreement with those derived from H$_3^+$ column densities (see Sec.~\ref{sec:H3+diff} and \citealt{indriolo2012}).
A comparison of the two datasets shows that no significant change in the CR ionisation rates is observed in clouds having total hydrogen column density in the range $N_H \approx 0.7 ... 20 \times 10^{21}$ ~cm$^{-2}$ (see Fig.~25 in \citealt{indriolo2015}, where data are plotted versus the total, rather than the molecular hydrogen column density).
The absence of a correlation between $\zeta_{CR}^{{\rm H}_2}$ and $N_{\rm H}$ for diffuse clouds means that they are exposed to the same flux of LECRs, regardless of their column density or, in alternative, that what we observe is in fact the superposition along the line of sight of many small clouds characterised by a very small hydrogen column density.
In this latter scenario, all the gas probed by OH$^+$ or H$_3^+$ observations would be exposed to the very same, almost unattenuated flux of Galactic LECRs.

Finally, the abundances of OH$^+$, and in some cases of OH too, can also be derived from UV absorption spectroscopy, providing alternative estimates of the CR ionisation rate which are in agreement with the Herschel observations described above \citep{porras2014,zhao2015,bacalla2019}.

Before moving on to a discussion of dense clouds, it is worth discussing the role played by argonium (ArH$^+$) in interstellar chemistry.
It was discovered, unexpectedly, by Herschel, first as an ubiquitous and unidentified absorption feature \citep{muller2013}, later identified by \citet{barlow2013}.
As pointed out by \citet{schilke2014}, the existence of this noble gas molecular ions at detectable levels is the result of an astrochemical conspiracy: 
{\it i)} the ionisation potential of argon (15.76 eV) is larger than that of hydrogen, and therefore argon is shielded by the interstellar UV radiation and is mostly neutral in diffuse clouds; {\it ii)} argon is 10 times more easily ionised by LECRs than hydrogen, and this enhances the abundance of Ar$^+$ with respect to that of H$^+$; {\it iii)} while other noble gas cations (He$^+$ or Ne$^+$) are neutralised by a dissociative ionisation reaction with H$_2$, such a reaction is endothermic for argon cations, thus favouring the creation of argonium through Ar$^+$ + H$_2 \rightarrow$ ArH$^+$ + H; {\it iv)} finally, once formed, ArH$^+$ can survive for long times as it has an unusually slow rate of both dissociative recombination and photodissociation.

ArH$^+$ has been detected from a number of the lines of sight in the survey of OH$^+$ and H$_2$O$^+$ performed by \citet{indriolo2015}.
Using a detailed diffuse cloud model, \citet{neufeld2017} re-derived the CR ionisation rate by fitting simultaneously the column densities of OH$^+$, H$_2$O$^+$ and, when available, ArH$^+$.
To do so, they had to assume that two populations of clouds exist. In this scenario, ArH$^+$ absorption lines are produced in small atomic clouds along the line of sight, characterised by very small molecular fractions ($f_{{\rm H}_2} = 10^{-5} ... \,10^{-2}$), while OH$^+$ and H$_2$O$^+$ absorption lines are produced in larger column density clouds, characterised by a larger molecular fraction ($f_{{\rm H}_2} \sim 0.2$).
Small clouds may contribute up to the 50\% of the total hydrogen column density along a line of sight.
Assuming that both populations of clouds are irradiated by the very same LECR flux, they derived an average CR ionisation rate which is a factor of 2.6 larger than that obtained by \citet{indriolo2015}.
This average value and its uncertainty are shown in Fig.~\ref{fig:zeta} as a dashed blue line and shaded cyan region, respectively.
This fits well with the fact that argonium is not protected by large column densities of gas, and therefore the CR ionisation rate derived from ArH$^+$ observations might be (hopefully?) considered a proxy of the actual interstellar value.

\subsubsection{Deuterium fractionation: the {\rm HCO}$^+$/{\rm DCO}$^+$ ratio}

Deuterium fractionation is the process that enriches the abundance of deuterium in interstellar molecules, leading to ratios between the abundance of deuterated molecules and their main isotopologue which are much larger than the cosmic elemental ratio ${\rm D}/{\rm H} \sim 1.6 \times 10^{-5}$.
It was soon recognised that measurements of the abundance of deuterated molecular ions could be used as tools to estimate the ionisation degree in dense clouds \citep{guelin1977}, which is a crucial step in the determination of the CR ionisation rate.
In the following, we provide a brief overview of this process, referring the interested reader to the extended reviews by \citet{caselli2002} and \citet{ceccarelli2014} and references therein.

Deuterium fractionation starts with the formation of H$_3^+$, a product of CR ionisation, followed by an isotope exchange reaction with HD, the major reservoir of deuterium in dense clouds:
\begin{equation}
{\rm H}_3^+ + {\rm HD} \leftrightarrow {\rm H}_2{\rm D}^+ + {\rm H}_2
\end{equation}
The forward reaction proceeds at a rate $k_{\rm D}$, while the reverse one is endothermic, and is therefore inhibited at the cold temperatures typical of dense clouds.
As a consequence, the abundance ratio $n({\rm H}_2{\rm D}^+)/n({\rm H}_3^+)$ becomes much larger than the typical interstellar $n({\rm D})/n({\rm H})$, and its actual level is determined by the destruction rate of H$_2$D$^+$.
Destruction of H$_2$D$^+$ mainly proceeds via dissociative recombination, at a rate $k_e^{\prime}$ (analogue of Eq.~\ref{eq:dissrec}) or through the analogues of Eq.~\ref{eq:HCO+}:
\begin{equation}
\label{eq:DCO+}
{\rm H}_2{\rm D}^+ + {\rm CO} \longrightarrow {\rm DCO}^+ + {\rm H}_2 ~~ {\rm or} ~~ {\rm HCO}^+ + {\rm HD}
\end{equation}
The total rate of this two reactions is $k_{\rm CO}^{\prime} \sim k_{\rm CO}$, the latter being twice faster than the former, i.e., the production of DCO$^+$ proceeds at a rate $k_{\rm CO}/3$ \citep{wootten1979}.

At this point, it is convenient to introduce the ratio between the abundance of DCO$^+$ and HCO$^+$, which after some manipulations can be written as:
\begin{equation}
R_{\rm D} \equiv \frac{n({\rm DCO}^+)}{n({\rm HCO}^+)} \sim \frac{1}{3} \frac{n({\rm H}_2{\rm D}^+)}{n({\rm H}_3^+)} = \frac{1}{3} \frac{k_{\rm D} n({\rm HD})}{k_e^{\prime} n(e^-) + k^{\prime}_{\rm CO} n({\rm CO})}
\end{equation}
where in the first equality we used the fact that HCO$^+$ and DCO$^+$ are destroyed at the same rate due to dissociative recombination, and in the second the equilibrium condition between formation and destruction of H$_2$D$^+$, i.e., $k_{\rm D} n({\rm H}_3^+) n({\rm HD}) = k_e^{\prime} n({\rm H}_2{\rm D}^+) n(e^-) + k^{\prime}_{\rm CO} n({\rm H}_2{\rm D}^+) n({\rm CO})$.

The expression above can be written in a more convenient form introducing abundances relative to H, $x_{\rm X} = n({\rm X})/n_{\rm H}$, to show that the measurement of the deuteration fraction of HCO$^+$, $R_{\rm D}$, can be used to estimate the electron fraction \citep{caselli2002}:
\begin{equation}
\label{xeDCO}
 x_e \sim 10^{-7} \left[ \frac{0.28}{R_{\rm D}} - 1.6 \left( \frac{1}{f_d} + \delta \right) \right] 
\end{equation}
The numerical values in the expression above were obtained by using the typical values for $x_{\rm HD} \sim {\rm H}/{\rm D} \sim 1.6 \times 10^{-5}$ and $x_{\rm CO} \approx 10^{-4}$
and the values of the reaction rates from e.g. \citet{vaupre2014} with T = 10 K.
The depletion factor $f_d$ accounts for the fact that in dense clouds a significant fraction of CO condenses onto grain surfaces ($1/f_d$ is the fraction of CO in gas phase), while $\delta$ is a correction term to account for the destruction of H$_2$D$^+$ with other neutrals (for example atomic oxygen).

Once the electron fraction is constrained, the CR ionisation rate can be derived from the abundance of HCO$^+$. 
The equilibrium between production (Eq.~\ref{eq:HCO+}) and destruction rate (dissociative recombination, at a rate $k_e^{\prime \prime}$) gives
HCO$^+ = k_{\rm CO} n({\rm H}_3^+)/k_e^{\prime \prime} n(e)$, which can be combined with Eq.~\ref{eq:H3+dense} to obtain the observable abundance ratio:
\begin{equation}
R_{\rm H} \equiv \frac{n({\rm HCO}^+)}{n({\rm CO})} = \frac{k_{\rm CO} \left[ \zeta_{CR}^{{\rm H}_2}/n({\rm H}_2) \right]}{4 k_e^{\prime \prime} x_e \left[ k_{\rm CO} x_{\rm CO} \left( 1/f_d + \delta \right)  + k_e x_e \right]}  
\end{equation}
which can be inverted to give the CR ionisation rate \citep{caselli2002}:
\begin{equation}
\label{zetaDCO}
\zeta_{CR}^{{\rm H}_2} \approx 3 \times 10^{-17} \left( \frac{x_e}{10^{-7}} \right) \left[ 4 \left( \frac{1}{f_d} + \delta \right) + \left( \frac{x_e}{10^{-7}} \right) \right] n({\rm H}_2) R_{\rm H} ~ \rm s^{-1} ~ .
\end{equation}

Equations \ref{xeDCO} and \ref{zetaDCO} can be used to compute the CR ionisation rate once the column densities of HCO$^+$, DCO$^+$ and CO are measured from observations of emission lines in the millimeter domain, and provided that an estimate of the depletion factor is also available. 
This was done by \citet{caselli1998}, whose findings are shown in Fig.~\ref{fig:zeta} as filled red circles.

A refined method has been proposed by \citet{bovino2020}, who suggested to use observations of H$_2$D$^+$ (and other deuterated isotopologues of H$_3^+$) together with the DCO$^+$/HCO$^+$ ratio to estimate the column density of H$_3^+$, and then use Eq.~\ref{eq:zetadense} to infer $\zeta_{CR}^{{\rm H}_2}$.
This method has been exploited by \citet{sabatini2020}, whose results are shown in Fig.~\ref{fig:zeta} as filled red triangles.

Finally, the open red circles show the measurement of  by $\zeta_{CR}^{{\rm H}_2}$ by \citet{maret2007}, \citet{hezareh2008}, \citet{moralesortiz2014} and \citet{fuente2016}, also based on the observation of various molecular lines including HCO$^+$.

The main conclusion that can be drawn after inspecting Fig.~\ref{fig:zeta} is that the CR ionisation rate seems to stay quite constant, on average, in diffuse clouds (both atomic and molecular).
The dispersion around the mean value, however, is large (at least one order of magnitude).
On the other hand, the value of $\zeta_{CR}^{{\rm H}_2}$ found in dense clouds is significantly smaller, suggesting a possible correlation with the cloud column density.
The interpretation of these trends is postponed to Sections~\ref{sec:transport} and \ref{sec:questions}.

\subsection{Calculation of the ionisation rate from the cosmic ray spectrum}
\label{sec:sigmaion}

The observations of molecular lines from interstellar clouds that we just reviewed are summarised in Fig.~\ref{fig:zeta}, where the total rate of ionisations due to LECRs per hydrogen molecule $\zeta_{CR}^{{\rm H}_2}$ are plotted versus the cloud column density.
In fact, several definitions of the CR ionisation rate are possible.
In the following, we will list them, and show how these rates are connected to the spectrum of LECRs.

Let us start with the total CR ionisation rate per H$_2$ molecule, representing the number of ionisation suffered by an hydrogen molecule at a given position in space and time (for simplicity the space and time dependences are omitted in the following).
This quantity can be written as \citep[e.g.][]{padovani2009,phan2020}:
%\begin{eqnarray}
%\label{eq:zetatot}
%\zeta_{CR}^{{\rm H}_2} &=& 4 \pi  \sum_k \int_{I({\rm H}_2)}^{E_{max}} {\rm d} E ~j_k(E) \sigma_{k,{\rm H}_2}^{ion}(E) \left[ 1 + \phi_{k,{\rm H}_2}(E) \right] \\ \nonumber
%&+& 4 \pi \int_0^{E_{max}} {\rm d} E ~j_p(E) \sigma_{p,{\rm H}_2}^{e.c.}(E)
%\end{eqnarray}
\begin{equation}
\label{eq:zetatot}
\! \! \! \zeta_{CR}^{{\rm H}_2} \! = \! 4 \pi \!  \left[ \! \sum_k \! \! \! \int\displaylimits_{I({\rm H}_2)}^{E_{max}} \! \! \! {\rm d} E ~j_k(E) \sigma_{k,{\rm H}_2}^{ion} \! (E) \left[ 1 \! + \! \phi_{k,{\rm H}_2} \! (E) \right] 
+ \! \! \! \! \! \int\displaylimits_0^{E_{max}} \! \! \! {\rm d} E ~j_p(E) \sigma_{p,{\rm H}_2}^{e.c.} \! (E) \! \right]
\end{equation}
where $j_k(E)$ is the energy spectrum of CRs (particles per unit surface, solid angle, time and kinetic energy) of species $k$ ($k = p$ for CR protons, $k = e$ for CR electrons, and $k = $~He, etc., for heavier nuclei), $I({\rm H}_2) = 15.4$~eV the ionisation potential of H$_2$ and $\sigma_{k,{\rm H}_2}^{ion}(E)$ its ionisation cross section. Note that throughout this Section, $E$ will represent the total (and not per nucleon) kinetic energy of the CR particle. The function $\phi(E)_{k,{\rm H}_2}$ accounts for the fact that electrons produced during the ionisation of H$_2$ molecules can be energetic enough to ionise more molecules.
The implicit assumptions made in Eq.~\ref{eq:zetatot} is that such secondary ionisations take place in the same place as the primary ones (i.e. the propagation of secondary electrons in space is negligible).
The second term in Eq.~\ref{eq:zetatot} represents the electron capture by a LECR protons in molecular hydrogen, characterised by a cross section $\sigma_{p,{\rm H}_2}^{e.c.}(E)$ (electron capture by heavier LECRs is usually neglected).

Eq.~\ref{eq:zetatot} immediately shows the main limitation of the constrains we can obtain from the measurement of the CR ionisation rate in interstellar clouds.
As we deal with an integral quantity, the information on the spectrum of CR particles is lost.
Also, it is impossible to evaluate from observations how large is the contribution from different species (protons, electrons, nuclei) to the total ionisation rate.

The total CR ionisation rate is often expressed in terms of the CR ionisation rates due to protons, $\zeta_{CR,p}^{{\rm H}_2}$, and electrons, $\zeta_{CR,e}^{{\rm H}_2}$, which are computed by taking only the terms $k = p$ or $k = e$ in the summation in Eq.~\ref{eq:zetatot}, resulting in the expression:
\begin{equation}
\label{eq:heavies}
\zeta_{CR}^{{\rm H}_2} = \eta \zeta_{CR,p}^{{\rm H}_2} + \zeta_{CR,e}^{{\rm H}_2}
\end{equation}
where the enhancement factor $\eta$, of the order of $\approx 1.5$, accounts for the ionisations from CR nuclei heavier than hydrogen (see the appendix in \citealt{padovani2009} for a derivation of $\eta$, or \citealt{chabot2016} for a detailed calculation of the ionisation rate of CR heavy nuclei).
Finally, as we already saw in Sec.~\ref{sec:hydrides}, the CR ionisation rate can also be defined per H atom (and not molecule), $\zeta_{CR}^{\rm H}$.
However, according to the estimate by \citet{glassgold1974}, the two rates are connected through a small multiplication factor, $1.5 \, \zeta_{CR}^{{\rm H}_2} = 2.3 \, \zeta_{CR}^{{\rm H}}$, and therefore in the following we describe in some detail the derivation of $\zeta_{CR}^{\rm H_2}$ only.

\begin{figure*}
% Use the relevant command to insert your figure file.
% For example, with the graphicx package use
\center
  \includegraphics[width=0.49\textwidth]{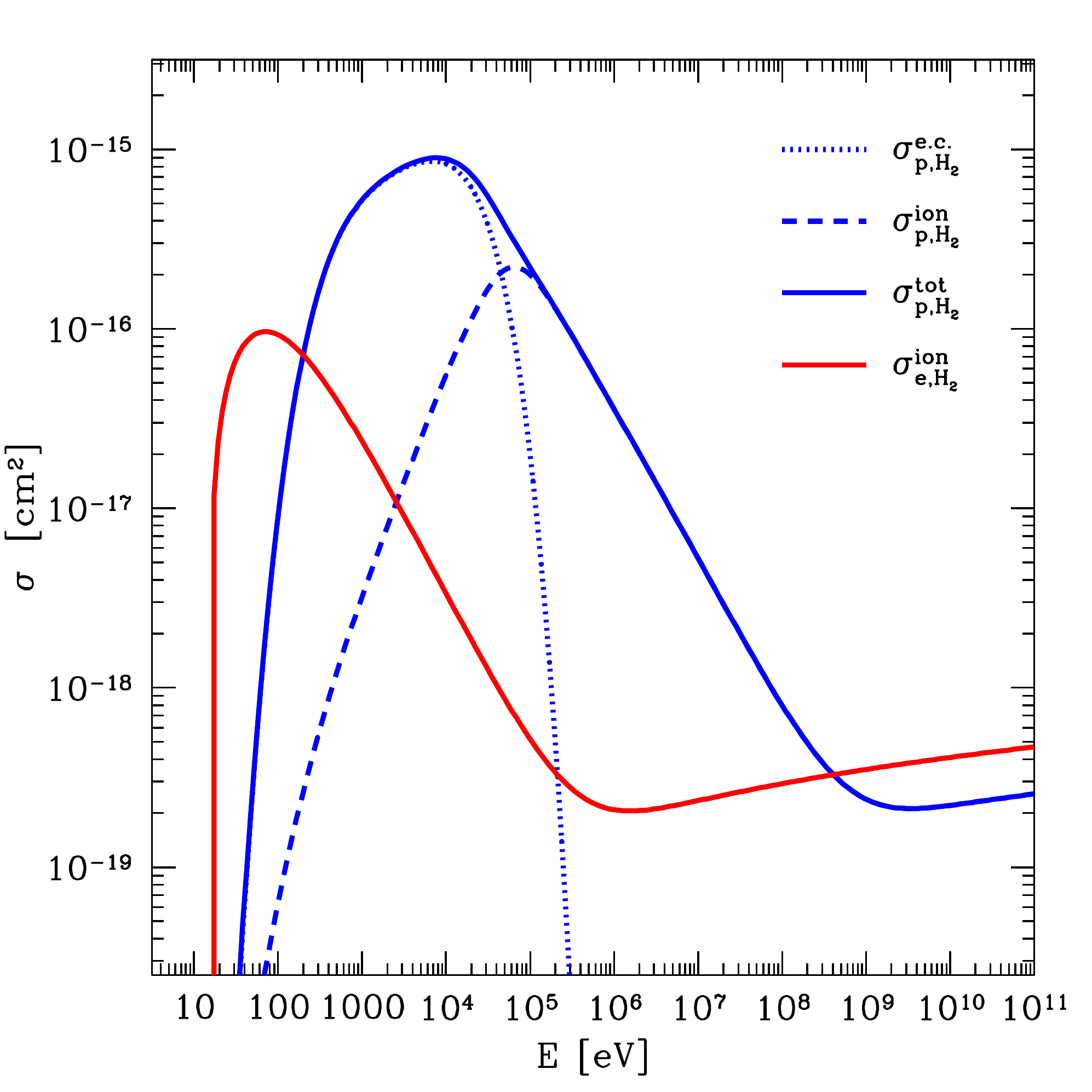}
  \includegraphics[width=0.49\textwidth]{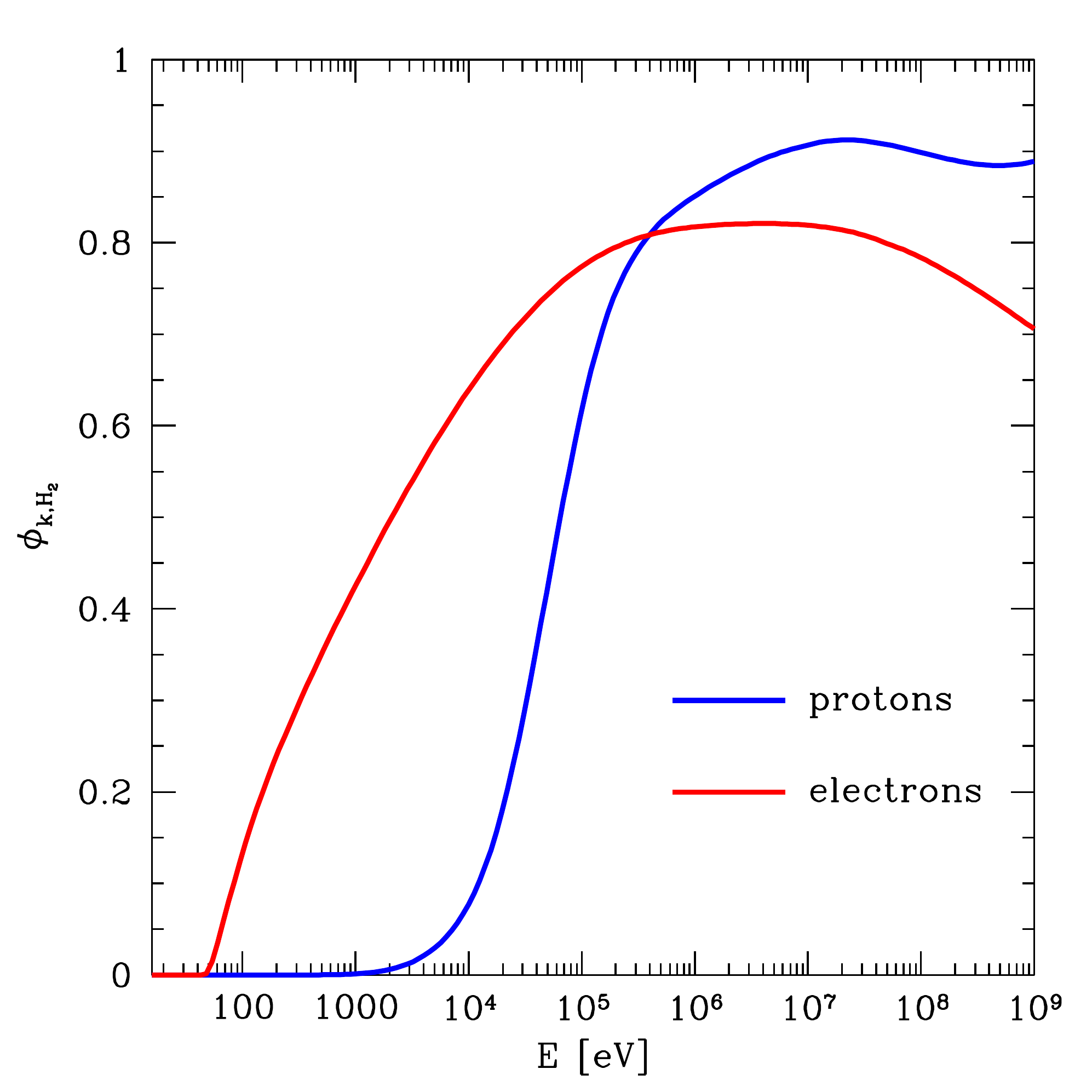}
% figure caption is below the figure
\caption{{\bf Left:} ionisation cross sections for molecular hydrogen: electron impact (e$^-$ + H$_2 \rightarrow$ H$_2^+$ + 2 e$^-$, red curve), electron capture (H$^+$ + H$_2 \rightarrow$ H$_2^+$ + H, dotted blue curve), proton impact (H$^+$ + H$_2 \rightarrow$ H$_2^+$ + H + e$^-$, dashed blue curve), and the sum of the last two (solid blue curve). {\bf Right:} number of secondary ionisations per primary ionisation due to CR protons (blue) and electrons (red curve).}
\label{fig:sigma}       % Give a unique label
\end{figure*}

The cross sections for electron impact (e$^-$ + H$_2 \rightarrow$ H$_2^+$ + 2 e$^-$, red curve, \citealt{kim1994,kim2000,zhong2021}), electron capture (H$^+$ + H$_2 \rightarrow$ H$_2^+$ + H, dotted blue curve, \citealt{padovani2009,phan2021}), and proton impact (H$^+$ + H$_2 \rightarrow$ H$_2^+$ + H + e$^-$, dashed blue curve, \citealt{rudd1988,rudd1992}) on molecular hydrogen are plotted in the left panel of Fig.~\ref{fig:sigma}. The solid blue curve represents the sum of the electron capture and proton impact cross sections. 
In order to obtain the proton impact cross section at relativistic energies, we adjusted the non-relativistic model by \citet{rudd1992} to match the appropriate high energy scaling \citep[see e.g.][]{fano1954}. 

The impact of the secondary ionisations can be estimated from the differential ionisation cross section ${\rm d} \sigma_{k,{\rm H}_2}^{ion}(E)/{\rm d} E_s$ which describes the energy spectral distribution of electrons produced in primary ionisations. Here, $E$ is the energy of the ionising primary particle of species $k$ and $E_s$ that of the ejected secondary electron.
If $j_k(E)$ is the intensity of primary particles, the differential production rate of secondary electrons per unit volume is $q_s(E_s) = 4 \pi n({\rm H}_2) \int j_k(E) ({\rm d} \sigma_{k,{\rm H}_2}^{ion}(E)/{\rm d} E_s) {\rm d}E$, where the integral has to be performed for energies $E$ larger than $I({\rm H}_2)$.
The equilibrium spectrum of secondary electrons can be approximated as $n_s(E_s) \sim q_s(E_s) \tau$, where $\tau$ is the residence time in the region of interest.
If we assume that secondary electrons do not travel away from the production site, the residence time reduces to the energy loss time $\tau_{loss}(E_s)$.
As the energy loss time is short in dense environment, this approximation is in general not too bad (see however \citealt{ivlev2021} for a discussion of the general case).
Finally, the ionisation rate from secondary electrons can be computed as $\zeta_{s,k}^{{\rm H}_2} = 4 \pi \int j_s(E_s) \sigma^{ion}_{e,{\rm H}_2}(E_s) {\rm d}E_s$, where we introduced the intensity $j_s(E_s) = (v_s/4 \pi) n_s(E_s)$, $v_s$ being the velocity of an electron of energy $E_s$. 
Also in this case the integral has to be performed for $E_s > I({\rm H}_2)$.
At this point, after some manipulations and making use of Eq.~\ref{eq:zetatot} one obtains \citep[][]{krause2015}:
\begin{equation}
\label{eq:phi}
\phi_{k,{\rm H}_2}(E) = \frac{1}{\sigma^{ion}_{k,{\rm H}_2}(E)} \int_{I({\rm H}_2)}^{E_{max}} {\rm d} E_s \frac{\sigma^{ion}_{e,{\rm H}_2}(E_s) E_s}{L(E_s)} \frac{{\rm d}\sigma_{k,{\rm H}_2}^{ion}(E)}{{\rm d}E_s} 
\end{equation}
which depends only on cross sections and on an energy loss function $L_e(E_s) = E_s/n({\rm H}_2) v_s \tau_{loss}$, that will be introduced and discussed in Sec.~\ref{sec:losses}.
We anticipate here that the loss function does not depend on the ambient gas density, as in dense environments $\tau_{loss} \propto 1/n({\rm H}_2)$.
The functions $\phi_{k,{\rm H}_2}(E)$ represent the average number of ionisations due to secondary electrons per ionisation of H$_2$ by a primary particle of energy $E$.
They are shown in the right panel of Fig.~\ref{fig:sigma} for CR protons ($k = p$) and electrons ($k = e$).
Note that, formally, Eq.~\ref{eq:phi} is appropriate for a gas made uniquely of molecular hydrogen, but it provides a reasonably accurate approximation also for typical abundances found in clouds.
%For a gas where hydrogen is all in molecular form, but constitutes a fraction $\epsilon_{{\rm H}_2} \lesssim 1$ of the total number of particles, Eq.~\ref{eq:phi} has to be multiplied by $\epsilon_{{\rm H}_2}$, as some of the energy of secondary electrons is lost in the ionisation of species other than H$_2$.

At this point, we can attempt a first prediction of the ionisation rates expected in clouds, and compare them with observations.
The most simplistic prediction can be obtained assuming that the local interstellar spectra of CRs that we discussed in Sections~\ref{sec:voyager} and \ref{sec:indirect} are representative of the entire Galaxy.
The ionisation rate can be computed by plugging in Eq.~\ref{eq:zetatot} the local interstellar spectra of CR protons and electrons, while Eq.~\ref{eq:heavies} can be used to evaluate the contribution of heavier CR nuclei.
In doing so, one should notice that local interstellar spectra of CRs are known only for particle energies larger than few MeV (Fig.~\ref{fig:modulation} and \ref{fig:Hee}), while ionisation cross sections peak at much lower energies (Fig.~\ref{fig:sigma}).
Therefore, an extrapolation of LECR spectra to low energy might be required. 
However, as we saw in Sec.~\ref{sec:voyager} (see the discussion on the top right panel of Fig.~\ref{fig:Hee}), performing a spectral extrapolation is mandatory for CR electrons, as their contribution to $\zeta_{CR,e}^{{\rm H}_2}$ is most likely dominated by sub-MeV particles, for which no observations exist.
On the other hand,  the contribution from CR protons to the total ionisation rate is likely dominated by particles in the sub-GeV domain, which is well constrained by observations, and therefore no extrapolation is needed \footnote{This claim is true unless an unknown and steep spectral component pops up in the sub MeV domain. We will briefly discuss this possibility in Sec.~\ref{sec:questions}.} \citet{padovani2009,phan2018}.
In other words, if we extrapolate the local spectra of CR protons (Eq.~\ref{eq:voyager}) and electrons (Eq.~\ref{eq:voyagere})down to an energy $E_{min}$, the total CR ionisation rate will depend on this minimum energy as as $\zeta_{CR}^{{\rm H}_2}(E_{min}) \sim \zeta_{CR,p}^{{\rm H}_2} + \zeta_{CR,e}^{{\rm H}_2}(E_{min})$.

The ionisation rates estimated in this way are reported in Table~\ref{tab:zeta}, and plotted in Fig.~\ref{fig:zeta} as a yellow shaded region for CR protons, and magenta lines for CR electrons. 
The dotted line refers to the {\it guaranteed} ionisation rate due to CR electrons, obtained setting $E_{min} = 3$~MeV (roughly corresponding to the minimum energy at which observations are available), while the solid line refers to $E_{min} = 3$~keV.
The CR ionisation rate for local CR protons is at the same level of that found in dense clouds, while in order to match the larger ionisation rates observed in diffuse clouds an extrapolation of the CR electron spectrum of 3 orders of magnitude in energy (down to $\sim$~3 keV) is required.
Such a comparison between observations and expectations should be taken with a grain of salt, as we are in fact mixing {\it local} expected ionisation rates with those measured in {\it remote} regions.
Moreover, we are neglecting the suppression of the CR intensity (and therefore of the ionisation rate) induced by energy losses during the penetration of such particles into MCs, a thing that will be discussed in detail in the coming Section.
Nevertheless, the preliminary comparison between observations and expectations shown in Fig.~\ref{fig:zeta} is useful, as it suggests that it might not be easy to explain the large ionisation rates observed in clouds, unless quite wild spectral extrapolations are invoked, or unless we give up the idea that the intensity of CRs is roughly uniform in the Galaxy.

\begin{table}[t]
\label{tab:zeta}
\centering
\begin{tabular}{| c | c | c | c |}
\hline
\multicolumn{4}{| c |}{Cosmic-ray ionisation rates in the local ISM} \\
\hline
\hline
 & & & \\ [-1em]
$E_{min}$ (MeV) & $\zeta_{CR,p}^{{\rm H}_2}$ ($10^{-17}$~s$^{-1}$) & $\zeta_{CR,e}^{{\rm H}_2}$ ($10^{-17}$~s$^{-1}$) & $\zeta_{CR}^{{\rm H}_2}$ ($10^{-17}$~s$^{-1}$) \\
 & & & \\ [-1em]
\hline
3 & 2.1 & 0.49 & 3.6 \\
0.3 & 2.3 & 1.1 & 4.6 \\
0.03 & 2.4 & 4.6 & 8.2 \\
0.003 & 2.4 & 48 & 52 \\
%$I({\rm H}_2)$ & 2.4 & $3.1 \times 10^3$ & $3.1 \times 10^3$  \\
\hline
\end{tabular}
\caption{Cosmic-ray ionisation rates in the local interstellar medium obtained integrating Eq.~\ref{eq:zetatot} above a particle energy $E_{min}$.}
\end{table}

Before concluding we should remind that secondary electrons produced in ionisation events deposit a fraction of their energy into the cloud in form of gas heating.
In molecular gas, CRs are the main heating agents. 
The heating happens through different channels, the most important being the dissociation of molecular hydrogen \citep{stahler2004}.
If, on average, each electron produced in a CR ionisation event releases an amount of heat $Q$, the gas is heated at a rate which is simply proportional to the CR ionisation rate: $\Gamma_h = \zeta_{CR}^{{\rm H}_2} n({\rm H}_2) Q$.
Typically, the value of $Q$ is of the order of several eV.
The balance between heating and cooling determines the thermal properties of the gas, and is therefore a crucial ingredient in the understanding of MCs.
As we will not discuss this topic in this review, we refer the reader to the work by \citet{glassgold2012} and references therein.

\section{The transport of cosmic rays in and around MCs}
\label{sec:transport}

In the previous two Sections (\ref{sec:indirect} and \ref{sec:integral}) we described how the observations of MCs can be used to impose constraints on the intensity and on the spectrum of LECRs inside these objects.
At low particle energies, the severe energy losses suffered by CRs in the dense and neutral gas may prevent them to penetrate freely into the clouds interior.
As a result, the pristine interstellar spectrum of CRs will be modulated by energy losses, and the amount of modulation will depend on the properties of the transport of particles into clouds.
The faster the penetration, the smaller the modulation, as particles might cross the cloud before losing significant energy.
%If the penetration is very fast, the expected amount of modulation might be moderate, as particles may cross the cloud before losing significant energy.
The fastest penetration is obtained assuming that particles penetrate into clouds moving along straight lines.
However, straight line propagation is not likely to happen, due to the ubiquitous presence of turbulent magnetic fields in the ISM, which deflect the trajectories of charged particles.
On one side, this certainly complicates the interpretation of available observations, but on the other hand it provides us with a way to confront with data the theoretical expectations derived from our (quite poor, in fact) knowledge of the transport of particles in turbulent fields.

The transport of CRs in turbulent magnetic fields constitutes by itself an entire field of research.
Therefore, we will limit ourselves to sketch the main aspects of this complex problem, and refer the interested reader to the monographs by \citet{berezinskii1990} and \citet{schlickeiser2002}, or to the recent reviews by \citet{amato2021} and \citet{hanasz2021}.

Following \citet{kulsrud2005}, let us start by considering the simplest problem of a LECR proton moving with a velocity $\vec{v}$ directed at an angle $\vartheta$ (called pitch angle) with respect to an uniform magnetic field $\vec{B}$.
Subject to the Lorentz force, the LECR particle will move at a constant speed around magnetic field lines, following an helical trajectory of radius $r_g= p_{\perp}c/eB$ (called Larmor radius) and gyration frequency $\Omega_g = v_{\perp}/r_g = eB/\gamma m_p c$.
Here, $p_{\perp}$ and $v_{\perp}$ are the components of the particle momentum and velocity orthogonal to $\vec{B}$, while $\gamma$ is the particle Lorentz factor.

Perfectly uniform fields are of course an unrealistic idealisation, and therefore we investigate now the effect of the presence of small scale fluctuations (MHD waves) superimposed to the uniform background field $\vec{B}$.
In particular, we limit our analysis to slab Alfv\'en waves, i.e., transverse, small amplitude magnetic disturbances ($\vec{\delta B} \perp \vec{B}$ and $\delta B \ll B$), which propagate along field lines at the Alfv\'en speed $v_A = B/\sqrt{4 \pi \varrho_i}$, where $\varrho_i$ is the ion mass density in the ISM, and oscillate with a frequency $\omega \sim k v_A$, $k$ being the wavenumber\footnote{See e.g. \citet{hollweg1974} for a formal approach to MHD waves.}.

The fluctuation in the magnetic field will add a $\vec{v} \times \vec{\delta B}$ component to the Lorentz force that will perturb the trajectory of the CR particle.
In fact, under most circumstances the interaction between a CR particle and a wave has very little effect, as $\vec{v} \times \vec{\delta B}$ oscillates many times during one gyration of the particle around the magnetic field, so that the time integrated effect of the Lorentz force on the particle averages to zero.
The particle trajectory is impacted only in the special case of resonant interactions, that occur when the particle interaction with a wave period lasts a time $\tau_c = 2 \pi/v_{\parallel} k$ which is identical to the gyration time $\tau_g = 2 \pi/\Omega_g$. 
When the condition $\tau_c = \tau_g$ is satisfied the Lorentz force does not change direction during the entire interaction, and changes the pitch angle (but not the energy) of the CR particle in a systematic way
(with this respect, Fig.~3 from \citet{wentzel1972} is particularly instructive).

Neglecting factors of order unity, the resonance condition can be written as $r_g \sim 1/k$, indicating that, in first approximation, CRs of a given energy interact only with waves of a given frequency.
For the sake of clarity, we consider the repeated resonant scatterings between a particle and a train of uncorrelated wave packets of length $1/k$ (one period). 
The effect of one of such wave packets is to bend the field line by an angle $\delta B/B$, and the particle pitch angle will be changed by this very same amount after scattering.
As wave packets are uncorrelated (random phases), sometimes the change in the particle pitch angle will be positive, and some other times negative. 
Thus, after a time $t$ the root mean square variation of the pitch angle will be of the order of $\sqrt{\langle \vartheta^2 \rangle} \sim \sqrt{N} (\delta B_k/B)$, where $N = t/\tau_c$ is the number of encounters with wave packets, and the subscript $k$ indicates that only fluctuations at the resonant scale have to be considered.

It follows that an initially anisotropic distribution of particles will be isotropised after a time given by the equation $\sqrt{\langle \vartheta^2 \rangle} \sim 1$ (a large angle), yielding $t_{iso}^{-1} \sim  \Omega_g (\delta B_k/B)^2$, where small numerical factors have been neglected.
The isotropisation time $t_{iso}$ is the time it takes a CR particle to lose memory of its direction of motion, and the distance traveled by the particle %along a field line 
during this time, $\lambda_{mfp} \sim v_{\parallel} t_{iso}$, corresponds then to the mean free path for {\it spatial} scattering along the field line (i.e. after traveling a distance $\lambda_{mfp}$ along a magnetic field line, a particle will have the same probability to continue to move forward or change direction and go backward).
The spatial diffusion coefficient is then:
\begin{equation}
\label{eq:diffusion}
D_{\parallel} = \frac{1}{3} \lambda_{mfp} v_{\parallel}  = \frac{4}{3 \pi} \frac{r_g v}{(\delta B_k/B)^2} 
\end{equation}
where the subscript $\parallel$ indicates that the coefficient refers to the one-dimensional diffusion along magnetic field lines, and the numerical factors come from a more detailed calculation \citep[e.g.][]{blandford1987}.

The derivation of the diffusion coefficient provided above, though admittedly oversimplified\footnote{For a detailed derivation of the spatial diffusion coefficient from the spectrum of magnetic fluctuations based on a state of the art theory of magnetic turbulence see for example \citet{fornieri2021} and the list of references therein.}, highlights a crucial aspect of particle transport: the energy dependence of the spatial diffusion coefficient is determined by the power spectrum of the magnetic perturbations, $(\delta B_k/B)^2$, through the resonance condition $k \sim 1/r_g$. 
This quantity represents the amount of magnetic energy available in form of fluctuations at the scale $k$, and it is unfortunately very difficult to be determined, both observationally and theoretically.
Note that, in this notation, the total magnetic energy in form of fluctuations is $(\delta B_{tot}/B)^2 = \int (\delta B_{k}/B)^2 {\rm d}k/k$.

In addition to parallel spatial diffusion, the scattering of CRs onto MHD waves induces also a diffusion perpendicular to field lines, which is believed to be much less effective than the parallel one in the limit of small amplitude fluctuations \citep[e.g.][]{casse2001}.
Particles can also be transported perpendicularly to the mean magnetic field due to the large scale wandering of magnetic field lines or to plasma motions \citep[e.g.][]{chuvilgin1993}.
In his Figures 1 and 8, \citet{chandran2000} provided a compendium of the various regimes of perpendicular transport, discussing them as a function of few parameters that determine the properties of the turbulent ambient magnetic field. 
He also suggested that magnetic mirroring might play a role when CRs encounter MCs, as magnetic field lines are focussed into their dense cores.

On large Galactic scales, the superposition of all the effects mentioned above might lead to an effective three-dimensional isotropic spatial diffusion of CRs, as it is indeed often assumed in phenomenological studies \citep[e.g.][and references therein]{strong2007}.
However, at the smaller spatial scales of MCs the situation is probably different.
The transport of particles in and around clouds is likely to be one-dimensional, as the size of such objects is roughly of the same order of the coherence length of the interstellar magnetic field.

%The question about the shape of the power spectrum and about the origin of the magnetic turbulence needed to scatter CRs remains debated.
Finally, the origin of the magnetic turbulence needed to scatter CRs remains debated.
However, an appealing scenario was proposed in the late 1960s, based on a plasma instability driven by the (small) anisotropy of CRs \citep{lerche1967,wentzel1968,kulsrud1969}.
As a result of the instability, CRs transfer momentum to MHD waves, enhancing their amplitude $\delta B_k/B$, and reducing correspondingly $D_{\parallel}$.
In other words, the transport of particles might be, in fact, a non-linear process, with CRs producing the turbulence that in turn scatters them.

\subsection{Nonlinear transport of cosmic rays: streaming instability}
\label{sec:nonlinear}

In order to understand how CRs can generate magnetic turbulence, consider a slightly anisotropic distribution of energetic particles in a medium where a uniform background magnetic field is present.
Due to the anisotropy, particles will move along the field lines at an average speed $v_d$, called streaming or drift velocity.
Assuming that energetic particles are protons of typical Lorentz factor $\gamma$ and volume density $n_{CR}$, we can compute the initial total momentum carried by the ensemble of such particles as $P_i = n_{CR} \gamma v_d m_p$.
We further assume that magnetic fluctuations of very small amplitude are superimposed to the background field.
These fluctuations will scatter CRs and lead their angular distribution towards isotropy.
This will reduce both the streaming velocity and the momentum carried by CR particles.

Being a conserved quantity, the momentum lost by CRs has to be transferred to something else.
It can be shown that, due to resonant scattering, the momentum is transferred to resonant waves \citep{kulsrud2005}.
Momentum conservation also implies that only waves traveling in the same direction of the drift velocity will grow (while counterpropagating waves will be damped).
The streaming velocity decreases until the CR particle distribution function becomes isotropic {\it in the frame of the waves}.
Therefore, in the lab frame the final momentum carried by CRs will not be zero, but rather $P_f = n_{CR} \gamma v_A m_p$, as waves move at a speed $v_A$.

The rate at which momentum is transferred from CRs to waves is ${\rm d} P_{CR}/{\rm d}t = -(P_f-P_i)/\tau_{iso}$, where $\tau_{iso}$ is the particle isotropisation time derived in the previous Section.
This has to be equal to the momentum gained by resonant waves ${\rm d} P_{w}/{\rm d}t = \Gamma_g \delta B_k^2/8 \pi v_A$, which is equal to the momentum of waves times a growth rate coefficient $\Gamma_g$.
Equating the two rates yields \citep{kulsrud2005}:
\begin{equation}
\label{eq:kulsrud}
\Gamma_g \sim a \Omega_0 \frac{n_{CR}}{A_i n_i} \left( \frac{v_d-v_A}{v_A} \right) = \Gamma_2 \left( \frac{v_d-v_A}{v_A} \right)
\end{equation}
where $a$ is a numerical factor of order unity, $A_i$ the mass number of the dominant ion in the gas, $\Omega_0 = \gamma \Omega_g$ is the nonrelativistic cyclotron frequency, and $\Gamma_2$ the growth rate corresponding to $v_d/v_A = 2$.
The term in the parenthesis shows that the instability is triggered when CRs stream faster than the Alfv\'en speed, and that it self-regulates, as a very large drift velocity $v_d \gg v_A$ would induce a very fast growth of the waves, that would in turn reduce $v_d$.
Finally, the exponential growth of waves is very fast, being characterised by an e-folding time ($\delta B_k \propto e^{\Gamma_g t}$) which is of the order of $1/\Gamma_g \sim 5 \times 10^3$ yr for typical interstellar conditions ($B \sim 3~\mu$G, $n_{CR}/n_i \sim 10^{-10}$, and $(v_d-v_A)/v_A \sim 2$).  

Eq.~\ref{eq:kulsrud} has been derived in an approximate way, but is in fact quite accurate. 
The exact expression for a CR particle distribution function equal to a power law in momentum, $f(p) \propto p^{-\alpha}$, is recovered after setting $a = (\pi/4) (\alpha-3)/(\alpha-2)$ and substituting $n_{CR}$ with $n_{CR}(>p_{res})$, i.e. the total number of particles per unit volume with momentum exceeding the resonant one \citep{kulsrud1971}.
An equivalent and perhaps more convenient way to express the growth rate was derived by \citet{skilling1975b}, in terms of the spatial gradient of the CR particle distribution function:
\begin{equation}
\Gamma_g = - \frac{16 \pi^2 v_A v p^4}{3 ~\delta B_k^2} \frac{\partial f}{\partial x}
\end{equation}
where $x$ is the coordinate measured along the field line.
This can be rewritten as:
\begin{equation}
\label{eq:work}
2 \left( \frac{\delta B_k}{B} \right)^2 \Gamma_g = - v_A  \frac{\partial P_{CR}}{\partial x}
\end{equation}
where we have introduced the CR partial pressure $P_{CR} = (4 \pi/3) v p^4 f/(B^2/8 \pi)$ normalised to the background field pressure.
Eq.~\ref{eq:work} shows that the rate at which the energy density of Alfv\'en waves increases (left side) is equal to the rate at which CRs do work in scattering off the waves (right side).

The mechanism for the amplification of magnetic fluctuations described here is called {\it streaming instability}, and has been recently reviewed by \citet{marcowith2021}.
In most circumstances, the instability proceeds until equilibrium is reached with competing mechanisms of wave damping.
In this regard, \citet{farmer2004} noticed that Alfv\'en waves generated by the instability may be damped in interactions with waves from interstellar turbulence, and that this is likely to limit the viability of this mechanism to particle energies below $\approx 100$~GeV.
As this is the range of energy considered in this review, we will assume in the following that the transport of CRs is regulated by self-generated turbulence.  

Another process of particular interest is the ion-neutral (or ambipolar) damping of Alfv\'en waves, which operates in partially ionised media \citep{kulsrud1969,zweibel1982,reville2021}.
It is due to the fact that ions follow the motion of the field lines more closely than neutrals.
As a result of their relative motion, momentum-exchanging collisions between ions and neutrals take place at a rate $\nu_{in} \sim m_n/(m_i+m_n) \langle \sigma v \rangle_{mt} n_n$, which depends mainly on the mass of the most abundant ion ($m_i$) and neutral species ($m_n$) in the gas, on the number density of neutrals $n_n$, and on the momentum transfer coefficient for ion-neutral collisions $\langle \sigma v \rangle_{mt}$.
The friction between ions and neutrals subtracts energy to the waves, which are damped.
For high frequency waves (which resonate with low energy particles) the damping rate is $\Gamma_{in} = \nu_{in}/2$ \citep[see e.g.][]{zweibel1982}.

When the growth of waves is balanced by damping ($\Gamma_g = \Gamma_{in}$), making use of Eq.~\ref{eq:kulsrud} one gets $v_d/v_A = 1+(\Gamma_{in}/\Gamma_2)$, which means that CRs can stream much faster than the Alfv\'en waves if $\Gamma_{in} \gg \Gamma_2$.
This can indeed happen in MCs, where the density of neutrals is large.
An estimate of the damping rate for typical conditions found in diffuse clouds\footnote{Assumed values: $n_n = 300$~cm$^{-3}$, $T = 50$~K, $x_e \sim 10^{-4}$, H$_2$ and C$^+$ dominant neutral and ion species, respectively.} was given in \citet{pinto2008} and \citet{recchia2021},  who found a very fast damping time of $1/\Gamma_{in} \approx 1$~ yr. 
It follows that streaming instability is inefficient in clouds, and that CRs stream almost freely in their interiors, unless their number density is of the order of $n_{CR} \approx 10^{-7}$~cm$^{-3}$, which is quite large. 

\subsection{Energy losses of cosmic rays in the interstellar medium}
\label{sec:losses}

An important aspect of the CR transport into clouds is the fact that LECRs lose energy when they interact with the dense molecular gas.
%LECRs lose their energy when they interact with matter or electromagnetic fields in the ISM.
In some cases (e.g. ionisation losses), the energetic particle loses a small fraction of its energy in the interaction, and the process can be described by a continuous function called energy loss rate, ${\rm d}E/{\rm d}t |_i$, where $E$ is the energy of the incident particle and $i$ refers to the interaction responsible for the loss of energy (e.g. $i = ion$ for ionisation losses).
Other processes, such as proton-proton interactions ($i = pp$) or Bremsstrahlung ($i = B$) are catastrophic, i.e. the energetic particle loses a significant fraction of its energy in a single interaction.
However, also in this case it is convenient to introduce a continuous function that has to be considered as an {\it average} energy loss rate \citep{berezinskii1990,aharonian2004}.
This is called the continuous slowing down approximation and has been widely used in the literature \citep[see e.g.][and references therein]{padovani2009}.
The energy loss rate can be defined in an alternative way as ${\rm d}E/{\rm d}x |_i = (1/v) {\rm d}E/{\rm d}t |_i$ to indicate the amount of energy lost in an infinitesimal displacement of the particle ${\rm d}x = v {\rm d}t$, $v$ being the velocity of the energetic particle.

In the dense and neutral environments considered in this review, LECRs lose energy mainly via ionisation losses at low particle energies (both CR electrons and nuclei), and via Bremsstrahlung (CR electrons) or proton-proton interactions (CR nuclei) at large particle energies.
Remarkably, for all of these processes the energy loss rate is proportional to the total (all elements) ambient gas density, ${\rm d}E/{\rm d}x |_i  \propto n$, and is therefore possible to build a quantity, called energy loss function, which is independent on the gas density, and reads \citep[e.g.][]{padovani2009}:
\begin{equation}
L_{k}(E) = - \frac{1}{n} \left( \frac{{\rm d}E}{{\rm d}x} \right)_k = - \frac{{\rm d} E}{{\rm d} N}
\end{equation}
where $k$ refers to the particle species, $({\rm d}E/{\rm d}x )_k$ is the sum of all the relevant loss rates, and $N = \int n {\rm d} x $ is the total gas column density.
From the second equality in the expression above, we see that the energy loss function represents the energy lost per gas column density traversed by the particle.

\begin{figure*}
% Use the relevant command to insert your figure file.
% For example, with the graphicx package use
\center
  \includegraphics[width=0.49\textwidth]{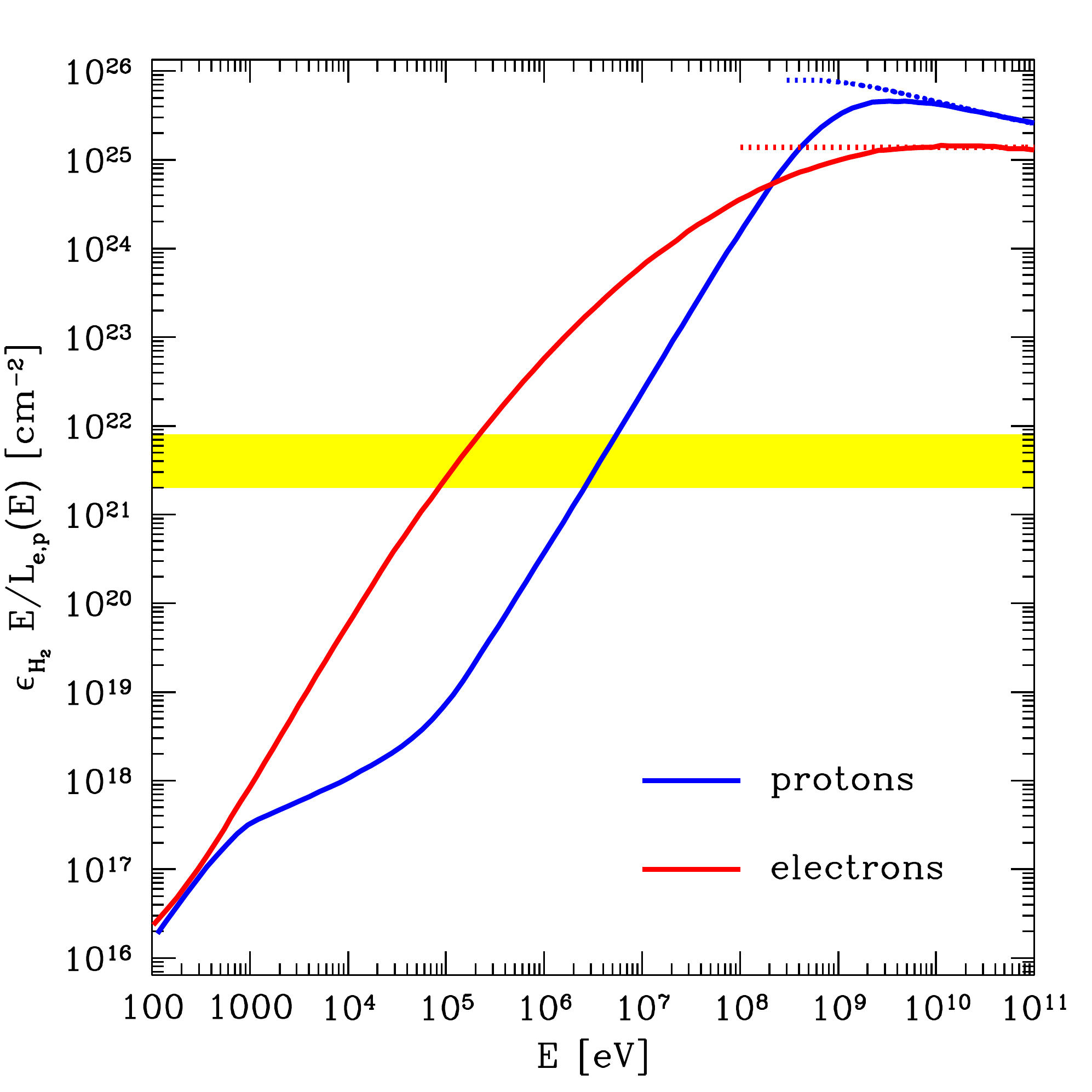}
  \includegraphics[width=0.49\textwidth]{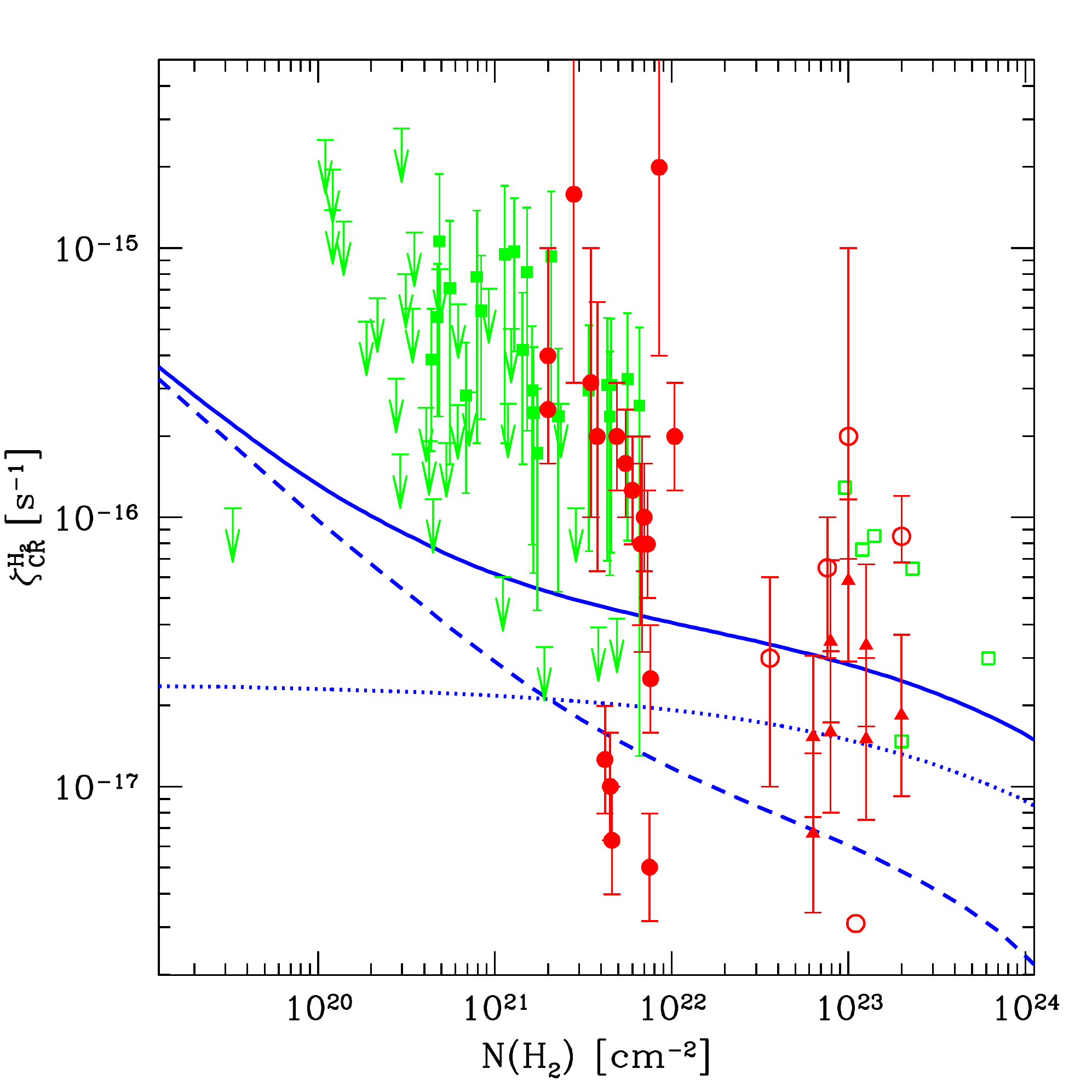}
% figure caption is below the figure
\caption{{\bf Left:} average gas column density crossed by a CR proton (blue curve) or electron (red curve) before losing its initial energy $E$. Energy losses are dominated by ionisation, except at the highest energies where proton-proton interactions (dotted blue line) and Bremsstrahlung (dotted red line) dominate. The yellow shaded region marks the transition between diffuse and dense clouds.  {\bf Right:} CR ionisation rate versus cloud column density. Data as in Fig.~\ref{fig:zeta}. Blue lines show the expected ionisation rates for the free streaming transport model for CR protons (dotted), electrons (dashed), and the sum of the total, including heavy nuclei (solid).}
\label{fig:NHmax}       % Give a unique label
\end{figure*}

\citet{padovani2018a} calculated the energy loss function for CR protons and electrons. 
They considered an ISM composed predominantly by molecular hydrogen and containing traces of heavier elements according to typical interstellar abundances.
Such a gas contains a fraction $\epsilon_{{\rm H}_2} = 0.835$ of molecular hydrogen, and therefore the column density of molecular hydrogen, which has been most often used in the previous Sections, is $N({\rm H}_2) = \epsilon_{{\rm H}_2} N$. %and is made of particles having a mean molecular weight $\bar{A} \sim 2.35$. 
%These two quantities can be used to convert the total column density $N$ into the molecular hydrogen column density $N({\rm H}_2)$ which has been mostly used in the previous Sections.  
Starting from the values of the energy loss function given by \citet{padovani2018a}, we built the quantity $\epsilon_{{\rm H}_2} E/L_k$, which has the dimensions of a column density of H$_2$, and provides an approximate estimate of the amount of interstellar matter a CR particle can traverse before losing most of its energy.
This quantity is shown in Fig.~\ref{fig:NHmax} (left panel) for CR protons (solid blue curve) and electrons (solid red curve).
For most particle energies, ionisation losses provide the dominant contribution to the loss function, with the exception of the highest particle energies, where proton-proton interactions (dotted blue curve) or Bremsstrahlung (dotted red curve) dominate.

The shaded yellow region in Fig.~\ref{fig:NHmax} marks the (blurred) boundary between column densities characteristic of diffuse and dense clouds.
The intersection between the shaded region and the solid curves shows that, even in the most optimistic scenario in which CRs penetrate clouds along straight lines, CR protons (electrons) of energy smaller than $\approx 5$~MeV ($\approx 0.1$~MeV) would be excluded by dense clouds, as they would lose their energy before reaching the cloud's centre.
The exclusion of LECRs from cloud interiors suggests a possible explanation for the difference in the ionisation rates observed in diffuse and dense clouds: electrons characterised by particle energies in the keV domain might penetrate into diffuse clouds and produce there a large ionisation rate. 
On the other hand, they might be excluded from dense clouds, which would be then characterised by a much lower ionisation rate.
The viability of this scenario will be discussed extensively in the coming Section.

\subsection{Modelling the penetration of low energy cosmic rays in molecular clouds}
\label{sec:penetration}

We have now all the elements required to build models describing the transport of LECRs into clouds.
These models can be used to make predictions of the CR ionisation rate as a function of the cloud column density.
Unfortunately, our poor knowledge of the properties of the transport of CRs in turbulent fields will make predictions quite uncertain.
For this reason, it is useful to consider two extreme scenarios for the propagation of LECRs, to obtain the most optimistic and pessimistic predictions on the CR penetration into clouds.
It is very likely that neither scenario is realistic, but reality will certainly be somewhere between these two extremes.

In the scenario that maximises the penetration of LECRs into clouds, and therefore also the CR ionisation rate, LECRs travel along straight lines from the edge to the centre of the cloud.
We call this mode of propagation the {\it free streaming transport}, and we refer the reader to \citet{padovani2009}, who described it in great detail.
In a more conservative scenario, the streaming of LECRs into clouds triggers the plasma instability described in Sec.~\ref{sec:nonlinear}, which in turn strongly suppresses the penetration of CRs as they scatter off self-generated Alfv\'en waves. 
As a consequence, the expected ionisation rate is reduced accordingly.
We call this mode of propagation the {\it self-regulating diffusive transport}.
The latter scenario was first suggested in a number of pioneering papers published in the late seventies/early eighties \citep{skilling1976,cesarsky1978,morfill1982,morfill1982a}, and recently reconsidered by various authors \citep{everett2011,morlino2015,schlickeiser2016,ivlev2018} .

If CRs travel along straight lines, the left panel of Fig.~\ref{fig:NHmax} can be used to estimate the minimum energy $E_{min}^k$ of particles ($k = p,e$ for CR protons and electrons) that can penetrate into a cloud of column density N$_{{\rm H}_2}$.
This can be done by solving the equation $\epsilon_{{\rm H}_2} E_{min}^k/L_k(E^k_{min}) = N_{{\rm H}_2}$.
Once $E_{min}^k(N_{{\rm H}_2})$ is derived, an approximate estimate of the CR ionisation rate can be obtained by means of Eq.~\ref{eq:zetatot}, where the lower limit of the integration should be changed from $I({\rm H}_2)$ to $E_{min}^k(N_{{\rm H}_2})$.
In this Section, we will assume that the intensity of CR nuclei and electrons just outside of the cloud is identical to the one measured in the local ISM.
The CR ionisation rate obtained in this way are plotted as blue lines in the right panel Fig.~\ref{fig:NHmax}, for CR protons (dotted), electrons (dashed), and all particles, including heavy nuclei (solid line).
Note that our approximate result is very close to that from the more accurate calculation from \citet{padovani2022}.
 
The free streaming transport model is an idealisation, but it remains nevertheless quite useful as it provides a way to estimate the maximum possible CR ionisation rate for a given intensity of CRs just outside of a cloud.
A more realistic description can be obtained by assuming that CRs move at a constant speed along magnetic field lines, which are in general turbulent (see e.g. Figures~7 to 12 from \citealt{padovani2013}).
Under these circumstances, the effective gas column density traversed by CRs to reach the cloud's core would be larger than the observed line-of-sight cloud column density, and therefore smaller values of the CR ionisation rate would be expected in this case.
In addition to that, the convergence of magnetic field lines towards cloud cores would focus CRs, enhancing their intensity, but, at the same time, some CRs would be excluded from the cloud's core due to magnetic mirroring.
In fact, it has been shown that the effects of focussing and mirroring cancel each other out almost perfectly, with the exception of magnetic pockets (local minima of the magnetic field strength), where the intensity of CRs can be significantly reduced\citep{padovani2013,silsbee2018}.

In order to highlight the main limitation of the free streaming transport model, consider a cloud of column density $N_{{\rm H}_2}$ embedded in a uniform magnetic field, and assume that the flux of CRs of energy $E$ impinging on each side of the cloud is $F_{in}^k(E)$.
For $E \gg E_{min}^k$, CRs cross the cloud without losing energy, and therefore the flux of CRs getting out from each side of the cloud will be $F_{out}^k(E) = F_{in}^k(E)$.
However, if we consider lower energies, $E \gtrsim E_{min}^k$, energy losses become important, and the beam of CRs will be attenuated while traversing the cloud.
This will result in $F_{out}^k(E) \lesssim F_{in}^k(E)$, i.e. in an anisotropic distribution of CRs at the cloud boundary.
The anisotropy will become of order unity for $E < E_{min}^k$.
As we saw in Sec.~\ref{sec:nonlinear}, anisotropic distributions of particles trigger the streaming instability.
As a result, the amplitude of magnetic fluctuations increases, and the penetration of CRs into clouds becomes a non-linear process.
As $F_{out}^k(E) < F_{in}^k(E)$, the stream of particles is directed towards the cloud, and a converging flow of amplified Alfv\'en waves will form.
Note that under most circumstances the growth of waves takes place just outside of the cloud, in the mostly ionised diffuse ISM, and not inside it, where streaming instability is inhibited by ion-neutral damping %(Sec.~\ref{sec:nonlinear}) 
(\citealt{skilling1976}, \citealt{cesarsky1978}, but see also \citealt{ivlev2018} for a scenario where streaming instability is excited in the diffuse envelope of the cloud).

The emerging picture is that of a self regulating diffusive penetration of LECRs in clouds. 
The intensity of CRs, assumed to be roughly uniform in the Galaxy, is suppressed inside clouds due to severe ionisation energy losses. 
This generates a gradient in the spatial distribution of CRs around clouds, that triggers the amplification of Alfv\'en waves due to streaming instability.
LECRs are advected towards the cloud by a converging flow of Alfv\'en waves, and at the same time undergo spatial diffusion along the background magnetic field due to resonant scattering off waves.
Inside clouds, CRs do not diffuse but they rather free stream along magnetic field lines, due to the absence of efficient scattering (ion-neutral friction is very effective in damping waves).

Following \citet{morfill1982}, we simplify the problem by considering a uniform magnetic field oriented along the $z$ axis, passing through a cloud of total column density $N$ located at $z = 0$. 
In this idealised picture, the cloud has no thickness along $z$, i.e., CRs probe its entire column density every time they cross the point defined by $z = 0$.
The one-dimensional version of the transport equation (Eq.~\ref{eq:transport}) derived in Sec.~\ref{sec:transportequation} can be used to study the penetration of LECRs in clouds.
To do so, the energy loss term $-(1/p^{2}) \partial (p^2 \dot{p} f)/\partial p$ has to be added to its right hand side, to account for ionisation losses in the cloud.
The momentum loss rate can be written as $\dot{p} = L_k N \delta (z)$, where $L_k$ is the loss function defined in the previous Section \citep{jones2001}.
Such a choice of the particle momentum loss rate is equivalent to assuming that outside of the cloud ($z \ne 0$) the gas density is small enough to make ionisation losses negligible.
Finally, the advection velocity $\bf{u_w}$ that appears in Eq.~\ref{eq:transport}  has to be substituted with $v_A$ for $z < 0$ and $-v_A$ for $z > 0$.

The steady-state ($\partial f / \partial t$ = 0) solution of the problem can be obtained by integrating the transport equation first between $z = 0^-$ and $0^+$, and then between $z = -\infty$ and $0^-$.
Then, noticing that the solution must be symmetric with respect to $z = 0$ ($f(z) = f(-z)$) one gets:
\begin{equation}
\label{eq:morfill}
2 v_A \left( f_{\infty} - f_{0} \right) = \frac{2}{3} p ~v_A \frac{\partial f_0}{\partial p} - \frac{N}{p^2} \frac{\partial}{\partial p} \left( p^2 L_k f_0 \right) 
\end{equation}
where $f_0$ is the CR distribution function inside the cloud and $f_{\infty}$ that at $z = \pm \infty$.
The first and second terms in the right hand side of the equation account for the acceleration of particles in the converging flow of Alfv\'en waves, and for ionisation energy losses inside the cloud, respectively.
Remarkably, the diffusion coefficient disappeared from the equation, and only the advective transport term (left hand side of the equation) is present.
We will further discuss this below.

Neglecting numerical factors of order unity, the ratio between the second and third term in Eq.~\ref{eq:morfill} is of the order of \citep{morlino2015}:
\begin{equation}
R \sim  \left( \frac{v_A}{v} \right) \frac{\epsilon_{{\rm H}_2} E}{N_{{\rm H}_2} L_k} \psi(E)
\end{equation}
where $\psi(E) = 1+m_k c^2/(E+ m_k c^2)$ is a slowly varying function of energy whose value changes between 1 and 2.
To simplify the discussion, we set $\psi = 1$ in the following.
When $R < 1$ energy losses dominate, and the intensity of CRs is suppressed inside the cloud ($f_0 < f_{\infty}$).

Note that, as we saw above when discussing the free streaming transport scenario, the condition to be satisfied to have a suppression of the CR intensity in one single cloud crossing is more stringent, and would correspond to $R < v_A/v \ll 1$ (because CRs move much faster than Alfv\'en waves).
This means that in the self regulating diffusive scenario two characteristic energies exist: $E_{min}$ and $E_{br}$, which correspond to the conditions $R = v_A/v$ and $R = 1$, respectively.
Therefore, $E_{br}$ is larger than $E_{min}$, and corresponds to the energy below which the CR intensity is suppressed due to ionisation losses accumulated due to repeated cloud crossings.
Particles of energy in the range $E_{min} < E < E_{br}$ are trapped in the vicinity of the cloud by the converging flow of Alfv\'en waves.
Before losing its energy, a particle of energy $E_{br}$ crosses the cloud $v/v_A$ times.
The number of crossings decreases for lower energy particles, and for $E < E_{br}$ particles lose their energy before a single crossing.
Below $E_{min}$, then, the suppression of the CR intensity becomes more pronounced \citep{morlino2015}.

%This means that the CR intensity is suppressed more effectively in the self-regulating diffusive scenario.
%With this respect, we can proceed as for the free streaming scenario, i.e., impose $R = 1$ to find the minimum energy $E_{min}$ of particles can can penetrate a cloud of a given column density, and estimate then the CR ionisation rate.
%Results are shown as cyan lines in the right panel of Fig.~\ref{fig:NHmax} for CR protons (dotted), electrons (dashed), and all particles, including heavy nuclei (solid line).

\citet{phan2018} performed a detailed modelling of the penetration process, including the effect of streaming instability operating outside of the cloud, and assuming free streaming of CRs inside of it.
The intensity of CRs far away from the cloud was assumed to be identical to the one measured in the local ISM.
The streaming instability is triggered by CR protons, which are most abundant, and CR electrons and nuclei behave as passive particles in the amplified turbulence generated by protons.
Their study was focussed on diffuse clouds, and they found total CR ionisation rates decreasing from $\lesssim 3 \times 10^{-17}$~s$^{-1}$ to $\sim 8 \times 10^{-18}$~s$^{-1}$ when the cloud column density increases from $N_{{\rm H}_2} = 10^{20}$ to $10^{22}$~cm$^{-2}$.
Such ionisation rates are a factor of 4-5 below those predicted within the framework of the free streaming transport.
In a study in preparation (Phan et al. 2022), it will be shown that a similar difference is found also for dense clouds.

The important conclusion of the studies reviewed above is that the standard picture of CRs pervading the Galactic disk with roughly the same intensity fails to reproduce the measurements of the CR ionisation rates in diffuse clouds.
On the other hand, for dense clouds, predictions match observations only if the most optimistic (free streaming) scenario for the CR penetration in clouds is invoked.
To date, this is one of the most puzzling problems in the study of LECRs in clouds, and possible solutions will be discussed in Sec.~\ref{sec:questions}.

We conclude this Section by commenting on the absence of the diffusion coefficient in Eq.~\ref{eq:morfill}, which is used to derive the spectrum of CRs inside the cloud $f_0$.
In fact, the particle diffusion coefficient $D$ does play a role, as it defines a characteristic spatial scale of the problem, $l_d \sim D/v_A$.
When $R < 1$, $l_d$ represents the extension of the region around the cloud where the intensity of CRs is suppressed with respect to the interstellar value $f_{\infty}$.
The value of the typical Galactic diffusion coefficient for sub-GeV protons is not well constrained. 
However, taking a quite plausible value of $D_{ISM} \lesssim 10^{28}$~cm$^2$/s \citep{tatischeff2021}, we get:
\begin{equation}
l_d \sim 3 \times 10^2 \left( \frac{D_{ISM}}{10^{28}~{\rm cm^2/s}} \right) \left( \frac{v_A}{100~{\rm km/s}} \right)^{-1} \left( \frac{\delta B_k}{\delta B_{k,ISM}}\right)^{-2} \rm pc
\end{equation}
where the last term in parenthesis represents the ratio between the energy density of amplified magnetic fluctuations and its typical interstellar value (see Eq.~\ref{eq:diffusion}). 
As pointed out by \citet{morlino2015}, streaming instability is a necessary ingredient of the model, as its effect is to increase $\delta B_k/\delta B_{k,ISM}$ and reduce $l_d$ to values smaller than the typical coherence length of the interstellar magnetic field ($l_B \approx$ 10-100 pc).
Without streaming instability, we would likely have $l_d > l_B$, and the solution of the transport equation obtained in Eq.~\ref{eq:morfill} would no longer be valid.
In that case, indeed, due to the mixing of magnetic field lines on scales larger than $l_B$, the boundary condition $f = f_{\infty}$ should be imposed at $| z | \sim l_B$ and not at $| z | = \infty$.
It can be shown that smaller values of $l_B$ would correspond to a more effective penetration of CRs into the cloud, and that the free streaming scenario can be recovered as a limiting case of the self regulating diffusive scenario for $l_B \rightarrow 0$.

%However, \citet{morlino2015} showed that, if the fraction of neutrals in the ISM surrounding the cloud is small enough to make ion-neutral damping unimportant, amplifications factors in the range $( \delta B_k / \delta B_{k,ISM} )^{2} \approx \mathcal{O}(10)$ are possible, making the self regulating diffusive scenario a viable possibility.
Let us assume, for a moment, that the amplification needed to have $l_d < l_B$ does take place.
In this case, the flux of CR into one side of the cloud can be written then as\footnote{We neglect the small correction due to the Compton-Getting effect.}:
\begin{equation}
- D \frac{\partial f}{\partial z} |_0 + v_A f_0 \sim  D \frac{f_{\infty}-f_0}{l_d} + v_A f_0 = v_A f_{\infty}
\end{equation}
where we assumed that $D$ is uniform in space.
This explain why $D$ disappears from the solution of the transport equation, making the solution of the problem somehow universal.
For $R \ll 1$ losses are so important to make $f_0 \ll f_{\infty}$, and therefore Eq~\ref{eq:morfill} reduces to \citep{skilling1976,cesarsky1978,morlino2015}:
\begin{equation}
2 v_A f_{\infty} = \frac{N}{p^2} \frac{\partial}{\partial p} \left( p^2 L_k f_0 \right) 
\end{equation}
which can be interpreted in a very simple way.
The equilibrium  spectrum of CRs inside the cloud can be derived by balancing the rate at which CRs of a given momentum enter the cloud (left hand side, the factor of 2 indicates that the cloud has 2 sides) with the rate at which they are removed from it due to energy losses (right hand side).

\subsection{Beam meets target: supernova remnant/molecular cloud associations}

The measurements of the CR ionisation rates reported in Fig.~\ref{fig:zeta} refer to {\it passive} clouds.
The term passive is widely used in gamma-ray astronomy to indicate a cloud located in a region where no evidence of current or recent CR acceleration activity has been reported.
The CR ionisation rate and the gamma-ray flux from a passive cloud are therefore ideal probes of the interstellar intensity of CRs which characterises the region of the Galaxy where the cloud is located \citep{aharonian2004}.

The presence of current or recent CR acceleration can be inferred, for example, from an enhanced gamma-ray flux from the cloud, either in the GeV domain, explored by the Fermi space mission, or in the TeV domain, probed by means of ground based observations with Cherenkov telescopes \citep[e.g.][]{gabici2009}.
Interestingly, the CR ionisation rate has been measured in few clouds that also exhibit an enhanced gamma-ray emission.
All these clouds are located in the vicinity of supernova remnants, which are very often invoked as the sources of CRs.

\begin{figure*}
% Use the relevant command to insert your figure file.
% For example, with the graphicx package use
\center
  \includegraphics[width=0.7\textwidth]{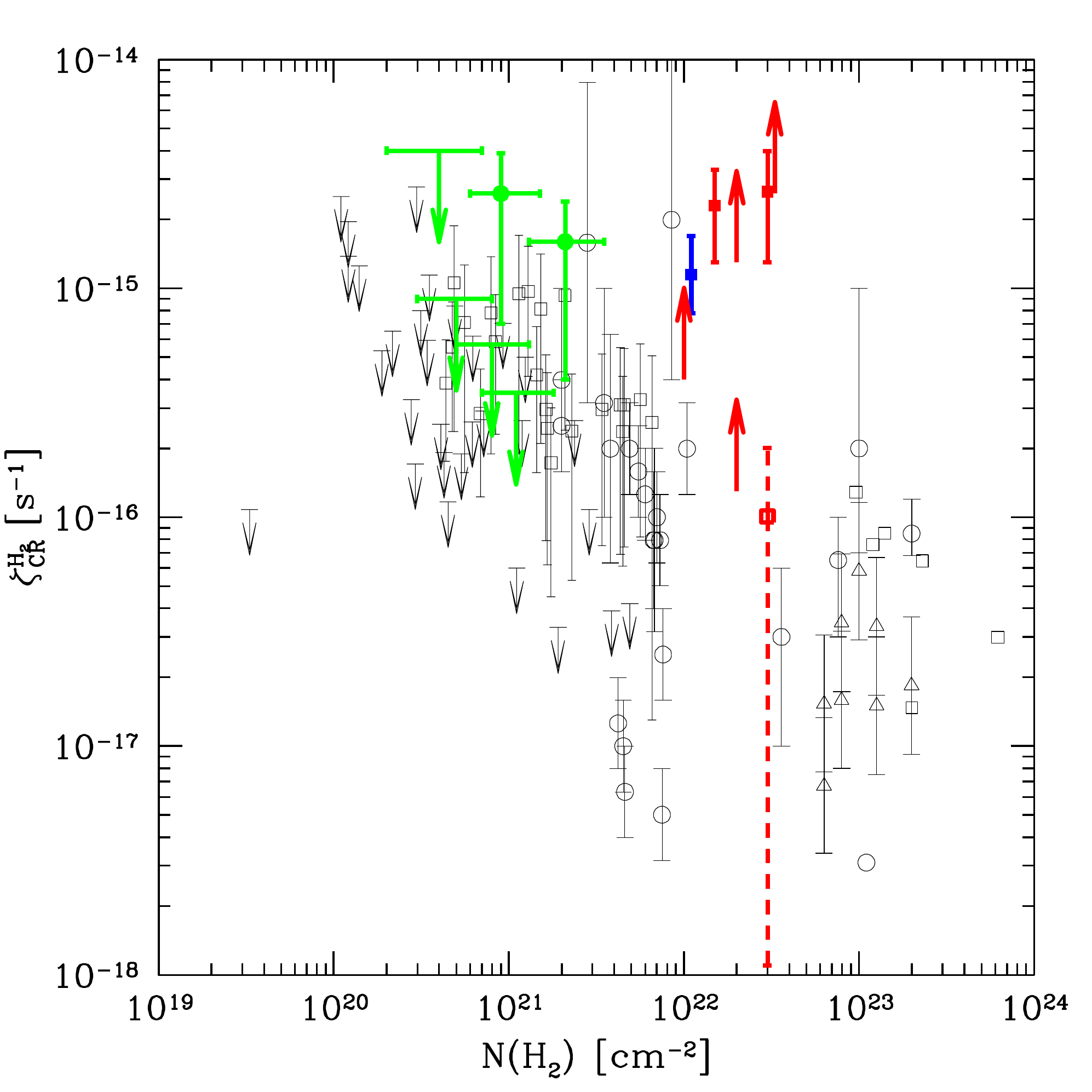}
% figure caption is below the figure
\caption{CR ionisation rates measured in the vicinity of the supernova remnant IC 443 (green points, \citealt{indriolo2010}), W51C (blue point, \citealt{ceccarelli2011}), and W28 (red points, \citealt{vaupre2014}). Black points are as in Fig.~\ref{fig:zeta}.}
\label{fig:SNR}       % Give a unique label
\end{figure*}

The CR ionisation rates measured from these gamma-ray bright clouds are shown in Fig.~\ref{fig:SNR} as coloured data points, together with the ionisation rates measured in passive clouds, shown as black data points.
Green, blue, and red data points refer to measurements taken in the vicinity of the supernova remnants IC 443 \citep{indriolo2010}, W51C \citep{ceccarelli2011}, and W28 \citep{vaupre2014}, respectively.
\citet{indriolo2010} derived the ionisation rates from H$_3^+$ absorption lines, while \citet{ceccarelli2011} and \citet{vaupre2014} from DCO$^+$/HCO$^+$ abundance ratios.

Fig.~\ref{fig:SNR} shows that the values of $ \zeta_{CR,p}^{{\rm H}_2}$ found in clouds located in the vicinity of supernova remnants are systematically larger than the typical values found in passive clouds of the same column density.
They are of the order of $ \zeta_{CR,p}^{{\rm H}_2} \gtrsim 10^{-15}$~s$^{-1}$ and do not show any evident scaling with the cloud column density.
An ionisation rate significantly smaller than $10^{-15}$~s$^{-1}$ was measured only in one case, from a line of sight towards a cloud located at a projected distance of $\sim 10$~pc from the shock of the supernova remnant W28 (red data point with dashed error bar).
This probably means that the cloud is too far from the supernova remnant to be irradiated by the cosmic rays accelerated there.
This hypothesis is corroborated by the fact that all the other red data points in the figure refer to projected distances from the shock of W28 of at most a couple of parsecs.

\citep{phan2020a} studied in more detail the case of the supernova remnant W28.
They constrained the spectrum of GeV and TeV CR protons and electrons in the region, using available gamma-ray and radio observations.
Remarkably, they found that an extrapolation of the CR proton spectrum of only an order of magnitude in particle energy (down to $\approx 30-300$ MeV) would suffice to explain the enhanced ionisation rates measured in the region.
It seems, then, that CR protons accelerated at the supernova remnant shock can explain the enhancement of both the gamma-ray flux and the ionisation rates measured in the region.

\section{Open questions on low energy cosmic rays}
\label{sec:questions}

The right panel of Fig.~\ref{fig:NHmax} illustrates one of the most pressing issues in LECR studies: observed CR ionisation rates in interstellar clouds are very large.
They are much larger than our expectations for diffuse clouds, while for dense clouds expectations match observations only if the most optimistic conditions for CR transport (free streaming) are invoked. 
Thus, the main question we should try to answer is: 
\begin{enumerate}
\item{What induces the large ionisation rates observed in MCs?}
\end{enumerate}

Answering this question is difficult because the measurements of $\zeta_{CR}^{{\rm H}_2}$ provide us with an integral constrain on the CR spectrum, which tells us nothing about the relative contribution from particles of different energies, or different species (CR nuclei or electrons) to the total ionisation rate.
Improving the modelling of the transport of CRs in and around clouds, though certainly possible and desirable, would not help, as we saw that the problem persists also in the idealised scenario where CRs penetrate into clouds moving along straight lines.

\citet{cummings2016} proposed that the large ionisation rates could be due to an unseen component of CRs of very low energies, popping up in the sub-MeV domain, currently unconstrained by observations.
However, we can see from the left panel of Fig.~\ref{fig:NHmax} that energy losses are particularly severe at these very low energies.
For example, in an ISM of density $n_{ISM} \sim 1$~cm$^{-3}$ the ionisation loss time for a proton (electron) of energy smaller than 1 MeV would be shorter than $\tau_{ion}^k \approx 2 \times 10^4$ yr ($10^5$ yr), see e.g. Fig.~2 in \citet{phan2018}.
Moreover, inside clouds of density $n_{cl} \gg n_{ISM}$, the energy loss time would be a factor of $n_{cl}/n_{ISM}$ shorter than that.
It is clear, then, that maintaining such a population of very low energy CRs in the ISM and/or in clouds would require an excessively large amount of energy.
This is because, to balance energy losses, freshly accelerated particles would have to be injected in the ISM at a very fast rate $\propto 1/\tau_{ion}^k$ \citep[see][for quantitative estimates]{recchia2019}.

A thing that should be kept in mind is that the discrepancy between observations and expectations was found under the assumption that the spectra of CRs measured in the local ISM are representative of the entire Galactic disk.
Such an assumption is very well justified for CR protons in the GeV energy domain, as discussed in Sections~\ref{sec:diffuse} and \ref{sec:MCs} and illustrated in the top panel of Fig.~\ref{fig:profiles}, but may not hold for lower energy CRs.
Therefore, a very fundamental question emerges:
\begin{enumerate}
\setcounter{enumi}{1}
\item{Are the spectra of LECRs measured in the local ISM representative of the entire Galaxy? Or, what is the spatial distribution of LECRs throughout the Galactic disk, on both large and small scales?}
\end{enumerate}

In an attempt to answer this questions, \citet{indriolo2015} plotted the CR ionisation rate as a function of the galactocentric distance $R$ (see their Fig.~23).
The values of  $\zeta_{CR}^{{\rm H}_2}$ where obtained from the abundance of oxygen bearing ions in diffuse atomic clouds (see Sec.~\ref{sec:hydrides}).
They found no relation between the ionisation rate and the galactocentric distance for $R > 5$ kpc, and a moderate increase of $\zeta_{CR}^{{\rm H}_2}$ towards the Galactic centre for smaller values of $R$.
Such a trend was predicted by \citet{wolfire2003}, as  a result of the larger concentration of potential CR sources towards the inner Galaxy, coupled with the spatial distribution of the gas density which determines the effectiveness of energy losses.
The prediction by \citet{wolfire2003}, rescaled to match available observations, is shown as a dashed line in the bottom panel of Fig.~\ref{fig:profiles}.
\citet{neufeld2017} refined the analysis by considering only the lines of sights from \citet{indriolo2015} for which also a detection of ArH$^+$ was available.
They then used a detailed diffuse cloud model to fit the abundances of both oxygen bearing ions and argonium.
Their estimates of the CR ionisation rates are shown as blue data points in the bottom panel of Fig.~\ref{fig:profiles}, and show no correlation with galactocentric distance\footnote{The values of the {\it primary} CR ionisation rate per H atom reported by \cite{neufeld2017} were multiplied by 2.3 to obtain $\zeta_{CR}^{{\rm H}_2}$ \citep{glassgold1974}.}.
Note that this is not necessarily in contradiction with the claim made by \citet{indriolo2015}, as only two clouds in the refined sample are located at $R < 5$ kpc.
The yellow shaded region in the Figure indicates the expected level of the CR ionisation rate in diffuse clouds obtained by \citet{phan2018} under the assumption that the intensity of CRs at any location in the Galactic disk is identical to the local one, and assuming  a self regulating diffusive penetration of LECRs into clouds (see Sec.~\ref{sec:penetration}). 
The red bar marks the Sun's location in the Galaxy.
Data are a factor of $\approx 10-100$ above such predictions (the discrepancy would reduce to a factor of $\approx 10$ in the free streaming transport scenario, see e.g. \citealt{padovani2022} or right panel of Fig.~\ref{fig:NHmax}).

At this point, it is worth reminding that for a large number of diffuse clouds only an upper limit for the CR ionisation rate could be derived (see Fig.~\ref{fig:zeta}), implying that we do not know how broad is the distribution of values of $\zeta_{CR}^{{\rm H}_2}$ in such objects.
That is to say: the blue points in the bottom panel of Fig.~\ref{fig:profiles} might represent the upper envelope of the real distribution of CR ionisation rates, which may vary significantly from cloud to cloud.
This suggests that the discrepancy between observations and expectations might be reduced if small scale fluctuations exist in the spatial distribution of LECRs, and the Solar system is located in a minimum of such fluctuating distribution.
Such fluctuations are indeed expected, as LECRs do not travel far from their acceleration sites due to the severe ionisation losses suffered in the ISM.
This idea, pioneered by \citet{cesarsky1975}, was recently investigated in detail by \citet{phan2021}, who found that fluctuations might be relevant if LECRs are accelerated at supernova remnants (the rate of about 3 supernova explosions per century in the Galaxy determines both the average distance between sources and the CR power per source, which are the two key parameters of the problem).
A comparison with measured ionisation rates is in progress (Phan et al. 2022, in preparation).

The search for a dependence of $\zeta_{CR}^{{\rm H}_2}$ on galactocentric radius has been based on data on atomic diffuse clouds.
This is because these objects are characterised by quite small gas column densities, and therefore the suppression of the CR intensity inside the clouds is likely to be quite moderate.
As a consequence, the measured ionisation rate is expected to be quite close to the pristine interstellar one.
However, Fig.~25 from \citet{indriolo2015} shows that there is no appreciable difference in the ionisation rate of atomic and molecular diffuse clouds, for column densities varying over almost two decades \citep[see also][]{neufeld2017}.
Answering the question
\begin{enumerate}
\setcounter{enumi}{2}
\item{Why are diffuse atomic and MCs, despite their different column density, characterised by the same ionisation rate?}
\end{enumerate}
will help us to better understand the transport of LECRs in and around clouds.

Measurements of the CR ionisation rate in the Galactic centre region are shown in the bottom panel of Fig.~\ref{fig:profiles} as a green data point and error bar \citep{lepetit2016,oka2019}.
The ionisation rate is of the order of $\gtrsim 10^{-14}$~s$^{-1}$, which is more than an order of magnitude larger than the ionisation rates measured in diffuse clouds (blue data points), and more than three orders of magnitude larger than the predictions from the self regulating transport scenario (yellow shaded region), leading to the question:
\begin{enumerate}
\setcounter{enumi}{3}
\item{Why is the ionisation rate so large in the Galactic centre region?}
\end{enumerate}

The Galactic centre hosts a supermassive black hole, surrounded by the most massive MC complex in the Galaxy, the Central Molecular Zone, of radius $\approx 100$ pc.
The region is characterised by an intense star formation rate, and a correspondingly enhanced supernova rate.
It also shows signs of nuclear activity in the form of powerful outflows \citep[e.g.][and references therein]{heywood2022}.
Given this rich phenomenology, one might expect to find in that region a largely enhanced intensity of CRs at all energies.
Somewhat surprisingly, gamma-ray observations revealed that the excess in high energy (beyond GeV) CRs is present but moderate, pointing towards an effective rate of escape of CRs from the inner Galaxy \citep{jouvin2017}.
Explaining the extremely large value of the ionisation rate would then require either a very fine tuned mechanism able to confine sub-GeV CRs (most effective in ionising the gas) but not particles of higher energy (constrained by gamma-ray observations), or the presence of other ionising agents (UV/X-ray photons pervading the central molecular zone?).

A possible way to explain the large discrepancy between observations and expectations is to assume that the LECRs responsible for the ionisation of clouds are produced {\it in situ}, inside the clouds.
In this scenario, they would form a distinct population of particles, with no links with local interstellar LECRs.
Ongoing particle acceleration in protostellar environments has been revealed by the presence of non-thermal synchrotron emission (and therefore of relativistic electrons) and of enhanced ionisation rates \citep[e.g.][]{rodriguez2017,favre2018}.
Shocks in protostellar jets seems to be the best places where to accelerate LECRs in these environments \citep{padovani2016}.
Under certain conditions, LECRs accelerated in protostellar environments might contribute to the ionisation rate in the cloud and reduce the discrepancy with observations \citep{gaches2019}.
To test this hypothesis, one should confirm observationally that protostellar particle accelerators are present in all clouds characterised by a large value of $\zeta_{CR}^{{\rm H}_2}$.

In another scenario, ionisation losses suffered by LECRs could be balanced by a continuous acceleration of particles due to their interaction with turbulent reconnecting regions.
As turbulence is ubiquitous in clouds, such mechanism is expected to produce homogeneous distributions of cosmic rays in the entire cloud volume \citep{gaches2021}.
The aim of these lines of research is to answer the question
\begin{enumerate}
\setcounter{enumi}{4}
\item{What are the sites of acceleration of the LECRs responsible for the ionisation of clouds?}
\end{enumerate}

Reversing the argument, instead of invoking an enhanced intensity of CRs inside clouds, we might speculate on the existence of a mechanism able to suppress the {\it local} intensity of LECRs.
In this case, the local CR measurements would have to be demodulated to obtain the {\it true} interstellar spectra.
This scenario is very appealing, as the Solar system is located within a cavity in the ISM, called the local bubble.
Such a cavity has been inflated by the repeated supernova explosions that took place in a star cluster, and is surrounded by a shell of dense and cold gas \citep{breitschwerdt2016,zucker2022}.
It has been suggested that ionisation energy losses experienced by LECRs while crossing the shell could indeed suppress the spectrum of CRs in the MeV domain \citep{silsbee2019,phan2020}.
This raises the question:
\begin{enumerate}
\setcounter{enumi}{5}
\item{Does the observed intensity of LECRs in the local ISM reflects the fact that we live in a special place in the Galaxy?}
\end{enumerate}

Finally, we should also consider the possibility that the uncertainties in our knowledge of chemical reaction networks might impact on the observational estimates of the ionisation rates.
As discussed in Sec.~\ref{sec:other}, ignoring the presence of PAH when defining the chemical network used to interpret H$_3^+$ observations might lead to significant errors in the derivation of $\zeta_{CR}^{{\rm H}_2}$ \citep[][]{wolfire2003,liszt2003,hollenbach2012,shaw2021}. 
We also saw that, in dense clouds, estimates of $\zeta_{CR}^{{\rm H}_2}$ are affected by the poor knowledge of the fraction of carbon and oxygen depleted on dust grains (factors $f_d$ and $\delta$ in Eq.~\ref{xeDCO}).
Moreover, also the assumption of a uniform CR ionisation rate inside clouds introduces errors, and astrochemical models should include the effect of a position (column density) dependent CR ionisation rate \citep{rimmer2012,gaches2021a}.
The last question that we should ask is then:
\begin{enumerate}
\setcounter{enumi}{6}
\item{How reliable are the observational estimates of the CR ionisation rates in interstellar clouds?}
\end{enumerate}

\section{Future perspectives}
\label{sec:conclusions}

During the past two decades, a number of observational breakthroughs in the study of LECRs changed radically our view on these particles.
Remarkably, progresses in this field of research spun across many branches of astrophysics, including space exploration (the Voyagers crossing the boundary of the heliosphere in 2012), laboratory astrophysics (the measurement of the dissociative recombination rate of H$_3^+$ in 2003), gamma-ray astronomy (the unprecedented view of the Galaxy provided by Fermi/LAT, in orbit since 2008), CR detection from space (PAMELA operated until 2016, and AMS-02 since 2011), infrared and millimeter astronomy (the launch of Herschel in 2009, and the numerous ground based observational campaigns).
Theoretical studies, as often happens, are trying to catch up with these observational advancements, but the open questions listed in the previous Section demonstrate that we are still far from a satisfactory understanding of LECRs.

Luckily, the next two decades promise to be equally exciting.
Moving from low to high energies, here is a list of some of the most important advancements we might expect to witness in the next 20 years.

{\it The Square Kilometer Array Observatory (SKAO) --}
Il will be the largest array of radio telescopes ever built.
It is expected to be fully operational in the late twenties.
As discussed in Sec.~\ref{sec:synchro}, radio observations of clouds with instruments of superior sensitivity will help us in better characterise the population of LECR electrons in such objects.

{\it The James Webb Space Telescope (JWST) --} Launched on 2021's Christmas day, it will dramatically improve our view on the infrared sky, which is one of the main sources of information we have on LECRs (see Sec.~\ref{sec:H3+}). 
Moreover, it has been recently proposed that JWST could detect the near-infrared photons produced in the decay of rotovibrational levels of molecular hydrogen. 
In interstellar clouds, the excitation of these levels is dominated by secondary CR electrons impacting on H$_2$.
This implies that the observations of such transitions will provide us with an estimate of $\zeta_{CR}^{{\rm H}_2}$ which is independent on any chemical network \citep{padovani2022}. 
In a similar fashion, emission lines from excited H$_2^+$ might also be observed in the infrared band from clouds characterised by large ionisation rates \citep{becker2011}.

{\it The Athena X-ray observatory --} Expected to be launched in the mid thirties, it will explore with unprecedented sensitivity the X-ray sky.
Of particular relevance for LECR studies is the Fe K$\alpha$ line at 6.4 keV, which is produced when a cold gas is irradiated with either X-rays or energetic particles.
It results from the removal of a K-shell electron followed by a transition from the L shell to fill the vacancy \citep[e.g.][]{tatischeff2012}.
The 6.4 keV line has been detected from clouds near the Galactic centre.
However, its flux varies with time in a way which is not compatible with an origin due to LECRs, but is rather connected to photon irradiation from flares originating from the central supermassive black hole  \citep[e.g.][and references therein]{ponti2010}.
Nevertheless, upper limits on the steady component of such emission can be converted into upper limits on the intensity of LECRs in the Galactic centre region.
The most stringent constraints on the intensity of LECRs obtained in this way are of the same order of the value of the CR proton intensity needed to explain the enhanced ionisation rate measured in the Galactic centre region \citep{rogers2021}.
Future observations are then of paramount importance, as a decrease in the line flux would rule out a CR proton origin of the ionisation rates.

Tentative evidence for the presence of 6.4 keV line emission has been presented also for a handful of clouds interacting with supernova remnants \citep{nobukawa2018}.
The emission, if confirmed, could be due to the presence of LECRs accelerated at the supernova remnant shocks.
Also in this case, future observations will help in clarifying the role played by LECRs close to putative accelerators, and will possibly lead to further detections.

{\it A renaissance of MeV astronomy? --}
Gamma-ray lines in the 0.1-10 MeV energy domain are produced following the collisions of LECRs with interstellar matter, as a result of the de-excitation of the first nuclear level.
They are expected to be particularly intense for interstellar $^{12}$C, $^{16}$O, $^{20}$Ne, $^{26}$Mg, $^{28}$Si, and $^{56}$Fe \citep{ramaty1979}.
Unfortunately, the MeV domain is poorly explored, but the situation might change in a near future.
NASA's Compton Spectrometer and Imager (COSI), expected to be launched in 2025, will reopen the MeV window, and hopefully pave the road to future mid-scale MeV missions.
Remarkably, if the large intensity of LECRs needed to explain the CR ionisation rates of diffuse clouds is a proxy for the typical Galactic value, de-excitation nuclear line emission from the inner Galaxy would be within reach of any of such mid-scale future missions \citep{deangelis2018}.

Theoretical/phenomenological studies will have to focus on both the very local and the remote interstellar medium.
Future observations of clouds from the radio to the gamma-ray domain will help in constraining models of {\it in situ} acceleration of LECRs, while local observations on the composition of LECRs have been recently used to show that such particles are likely accelerate in a diffuse and hot medium, characteristic of superbubbles \citep{tatischeff2021}. The latter result brings us back to the fundamental question: does our location inside the local bubble affects our view on LECRs?

\begin{acknowledgements}
I would like to thank the editors at A\&A Rev, and in particular Luigina Feretti for her advices and infinite patience.
I also thank Phan Vo Hong Minh, Sarah Recchia, and Thibault Vieu, who read and commented on the entire manuscript, Paola Caselli, with whom I had (almost 20 years ago) my first discussion on LECRs, and Cecilia Ceccarelli and Thierry Montmerle who gave me the opportunity to start working in this field.
While writing this review I enjoyed discussing LECRs and related topics with G. Bernardi, D. Breitschwerdt, J. Duprat, C. Evoli, D. Gaggero, D. Galli, R. Giuffrida, D. Grasso, A. Ivlev, D. Maurin, P. Mertsch, M. Miceli, M. Padovani, G. Peron, L. Podio, S. Ravikularaman, J. Raymond, G. Sabatini, F. Schulze, A. Strong, V. Tatischeff, R. Terrier, and F. van der Tak.
Finally, I acknowledge support from Agence Nationale de la Recherche (grants ANR- 17-CE31-0014 and CRitiLISM).
\end{acknowledgements}

% Authors must disclose all relationships or interests that 
% could have direct or potential influence or impart bias on 
% the work: 
%
% \section*{Conflict of interest}
%
% The authors declare that they have no conflict of interest.

% BibTeX users please use one of
%\bibliographystyle{sn-mathphys}
\bibliographystyle{spbasic-FS-etal}      % basic style, author-year citations
\bibliography{biblio}   % name your BibTeX data base

% Non-BibTeX users please use
%\begin{thebibliography}{}
%
% and use \bibitem to create references. Consult the Instructions
% for authors for reference list style.
%
%\bibitem{RefJ}
% Format for Journal Reference
%Author, Article title, Journal, Volume, page numbers (year)
% Format for books
%\bibitem{RefB}
%Author, Book title, page numbers. Publisher, place (year)
% etc
%\end{thebibliography}

\end{document}